\documentclass[a4paper]{article}
\usepackage{amsmath,epsfig}

\numberwithin{equation}{section}

\newcommand{\hs}[1]{\hspace*{#1cm}}

\newcommand{\nn}{\nonumber}

\newcommand{\et}[1]{e^{\mbox{\small $#1$}}}

\newcommand{\ol}[1]{\overline{#1}}

\newcommand{\ket}[1]{| #1 \ra}

\newcommand{\dt}{\! \cdot \!}
\newcommand{\wdg}{\! \wedge \!}
\newcommand{\crs}{\! \times \!}
\newcommand{\scp}{\! \ast \!}
\newcommand{\la}{\langle}
\newcommand{\ra}{\rangle}
\newcommand{\lra}{\leftrightarrow}
\newcommand{\trans}{\; \leftrightarrow \;}
\newcommand{\half}{{\textstyle \frac{1}{2}}}

\newcommand{\qrt}{{\textstyle \frac{1}{4}}}

\newcommand{\tld}{{\textstyle \tilde{\,}}}
\newcommand{\etal}{\textit{et al.}}

\newcommand{\sqhalf}{{\textstyle \frac{1}{\sqrt{2}}}}
\newcommand{\up}{\uparrow}
\newcommand{\down}{\downarrow}

\newcommand{\alp}{\alpha}
\newcommand{\bet}{\beta}
\newcommand{\gam}{\gamma}
\newcommand{\del}{\delta}
\newcommand{\eps}{\epsilon}
\newcommand{\kap}{\kappa}
\newcommand{\lam}{\lambda}
\newcommand{\sig}{\sigma}
\newcommand{\om}{\omega}
\newcommand{\Gam}{\Gamma}

\newcommand{\Om}{\Omega}

\newcommand{\ba}{\mbox{\boldmath $a$}}
\newcommand{\bb}{\mbox{\boldmath $b$}}

\newcommand{\blde}{\mbox{\boldmath $e$}}

\newcommand{\bk}{\mbox{\boldmath $k$}}

\newcommand{\bn}{\mbox{\boldmath $n$}}
\newcommand{\bp}{\mbox{\boldmath $p$}}
\newcommand{\bq}{\mbox{\boldmath $q$}}
\newcommand{\br}{\mbox{\boldmath $r$}}
\newcommand{\bs}{\mbox{\boldmath $s$}}

\newcommand{\bx}{\mbox{\boldmath $x$}}

\newcommand{\bA}{\mbox{\boldmath $A$}}
\newcommand{\bB}{\mbox{\boldmath $B$}}
\newcommand{\bC}{\mbox{\boldmath $C$}}

\newcommand{\bE}{\mbox{\boldmath $E$}}

\newcommand{\bI}{\mbox{\boldmath $I$}}
\newcommand{\bJ}{\mbox{\boldmath $J$}}

\newcommand{\bP}{\mbox{\boldmath $P$}}

\newcommand{\bS}{\mbox{\boldmath $S$}}

\newcommand{\sdot}{\dot{s}}

\newcommand{\vdot}{\dot{v}}
\newcommand{\xdot}{\dot{x}}

\newcommand{\Sdot}{\dot{S}}

\newcommand{\cle}{{\mathcal{E}}}

\newcommand{\clh}{{\mathcal{H}}}

\newcommand{\clj}{{\mathcal{J}}}

\newcommand{\clo}{{\mathcal{O}}}

\newcommand{\dif}[1]{\partial_{#1}}

\newcommand{\dift}{\partial_t}

\newcommand{\si}{\sigma_{1}}
\newcommand{\sj}{\sigma_{2}}
\newcommand{\sk}{\sigma_{3}}

\newcommand{\sr}{\sigma_{r}}
\newcommand{\isi}{i\sigma_{1}}
\newcommand{\isj}{i\sigma_{2}}
\newcommand{\isk}{i\sigma_{3}}
\newcommand{\sigi}{\sig_i}
\newcommand{\sigj}{\sig_j}
\newcommand{\sigk}{\sig_k}

\newcommand{\gi}{\gamma_{1}}
\newcommand{\gj}{\gamma_{2}}
\newcommand{\gk}{\gamma_{3}}
\newcommand{\go}{\gamma_{0}}

\newcommand{\gamum}{\gamma^\mu}
\newcommand{\gamdm}{\gamma_\mu}

\newcommand{\gamdn}{\gamma_\nu}

\newcommand{\isji}{i\sigma_{2}^{1}}
\newcommand{\isjj}{i\sigma_{2}^{2}}
\newcommand{\iski}{i\sigma_{3}^{1}}
\newcommand{\iskj}{i\sigma_{3}^{2}}

\newcommand{\hsi}{\hat{\sigma}_{1}}
\newcommand{\hsj}{\hat{\sigma}_{2}}
\newcommand{\hsk}{\hat{\sigma}_{3}}
\newcommand{\hsigi}{\hat{\sigma}_i}

\newcommand{\hsigk}{\hat{\sigma}_k}
\newcommand{\hsig}{\hat{\sigma}}

\newcommand{\hgo}{\hat{\gamma}_{0}}
\newcommand{\hgamum}{\hat{\gamma}^\mu}
\newcommand{\hgamdm}{\hat{\gamma}_\mu}
\newcommand{\hgam}{\hat{\gamma}}

\newcommand{\Rrev}{\tilde{R}}
\newcommand{\Lrev}{\tilde{L}}
\newcommand{\Rdot}{\dot{R}}
\newcommand{\Rdag}{R^{\dagger}}
\newcommand{\Srev}{\tilde{S}}
\newcommand{\psirev}{\tilde{\psi}}
\newcommand{\psidot}{\dot{\psi}}
\newcommand{\psidag}{\psi^{\dagger}}
\newcommand{\psibar}{\bar{\psi}}
\newcommand{\phirev}{\tilde{\phi}}

\newcommand{\phidag}{\phi^{\dagger}}
\newcommand{\phibar}{\bar{\phi}}
\newcommand{\Phirev}{\tilde{\Phi}}
\newcommand{\Phidot}{\dot{\Phi}}
\newcommand{\Phidag}{\Phi^{\dagger}}

\newcommand{\epsdag}{\eps^{\dagger}}

\newcommand{\ho}{\ol{h}}

\newcommand{\bpht}{\mbox{\boldmath $\hat{p}$}}

\newcommand{\grad}{\nabla}

\newcommand{\dgrad}{\dot{\grad}}
\newcommand{\bgrad}{\mbox{\boldmath $\grad$}}

\newcommand{\deriv}[2]{\frac{\partial #1}{\partial #2}}
\newcommand{\backderiv}[2]{\frac{\overleftarrow{\partial} #1}{\partial #2}}

\newcommand{\Tup}{T^\uparrow}
\newcommand{\Rup}{R^\uparrow}
\newcommand{\Tdn}{T^\downarrow}
\newcommand{\Rdn}{R^\downarrow}
\newcommand{\rup}{r^\uparrow}
\newcommand{\rdn}{r^\downarrow}
\newcommand{\psiup}{\psi^\uparrow}
\newcommand{\psidn}{\psi^\downarrow}
\newcommand{\dn}{\downarrow}

\begin{document}


\thispagestyle{empty}

\vspace*{2cm}

\noindent
\textsf{\textbf{\large SPACETIME ALGEBRA AND ELECTRON PHYSICS}}

\vspace{0.5cm}

\noindent
{\large Chris Doran\footnote{E-mail: c.doran@mrao.cam.ac.uk}, 
Anthony Lasenby\footnote{E-mail: a.n.lasenby@mrao.cam.ac.uk},
Stephen Gull\footnote{E-mail: steve@mrao.cam.ac.uk}, 
Shyamal Somaroo and Anthony Challinor \footnote{E-mail:
  a.d.challinor@mrao.cam.ac.uk}}

\vspace{0.5cm}
\noindent
Astrophysics Group, Cavendish Laboratory, Madingley Road, \\
Cambridge CB3 0HE, UK.

\vspace{1cm}

\begin{abstract}
This paper surveys the application of geometric algebra to the
physics of electrons.  It first appeared in 1996 and is reproduced
here with only minor modifications.  Subjects covered include
non-relativistic and relativistic spinors, the Dirac equation,
operators and monogenics, the Hydrogen atom, propagators and
scattering theory, spin precession, tunnelling times, spin
measurement, multiparticle quantum mechanics, relativistic
multiparticle wave equations, and semiclassical mechanics.
\end{abstract}

\vspace{1cm}

\tableofcontents

\newpage

\section{Introduction}

This paper surveys the application of `geometric algebra' to the
physics of electrons.  The mathematical ideas underlying geometric
algebra were discovered jointly by Clifford~\cite{cli1878} and
Grassmann~\cite{gra-1862} in the late 19th century.  Their discoveries
were made during a period in which mathematicians were uncovering many
new algebraic structures (quaternions, matrices, groups,
\textit{etc.}) and the full potential of Clifford and Grassmann's work
was lost as mathematicians concentrated on its algebraic properties.
This problem was exacerbated by Clifford's early death and Grassmann's
lack of recognition during his lifetime.  This paper is part of a
concerted effort to repair the damage caused by this historical
accident.  We firmly believe that geometric algebra is the simplest
and most coherent language available for mathematical physics, and
deserves to be understood and used by the physics and engineering
communities.  Geometric algebra provides a single, unified approach to
a vast range of mathematical physics, and formulating and solving a
problem in geometric algebra invariably leeds to new physical
insights.
\nocite{DGL93-notreal,DGL93-states,DGL93-lft,DGL93-paths}
In the series of papers \cite{DGL93-notreal}--\cite{DGL93-paths}
geometric algebra techniques were applied to number of areas of
physics, including relativistic electrodynamics and Dirac theory.
In this paper we
extend aspects of that work to encompass a wider range of topics
relevant to electron physics.  We hope that the work presented here
makes a convincing case for the use of geometric algebra in electron
physics.

The idea that Clifford algebra provides the framework for a unified
language for physics has been advocated most strongly by Hestenes, who
is largely responsible for shaping the modern form of the subject.
His contribution should be evident from the number and range of
citations to his work that punctuate this paper.  One of Hestenes'
crucial insights is the role of geometric algebra in the design of
mathematical physics~\cite{hes-unified}.  Modern physicists are
expected to command an understanding of a vast range of algebraic
systems and techniques --- a problem that gets progressively worst if
one is interested in the theoretical side of the subject.  A list of
the some of the algebraic systems and techniques employed in modern
theoretical physics (and especially particle physics) is given in
Table~\ref{tab-list}.  Hestenes' point is that every one of the
mathematical tools contained in Table~\ref{tab-list} can be expressed
within geometric algebra, but the converse is not true.  One would be
hard-pressed to prove that two of the angles in an isosceles triangle
are equal using spinor techniques, for example, but the proof is
simple in geometric algebra because it encompasses vector geometry.
The work of physicists would be considerably simplified if, instead of
being separately introduced to the techniques listed in
Table~\ref{tab-list}, they were first given a firm basis in geometric
algebra.  Then, when a new technique is needed, physicists can simply
slot this into their existing knowledge of geometric algebra, rather
than each new technique sitting on its own, unincorporated into a
wider framework.  This way, physicists are relieved of the burden of
independently discovering the deeper organisational principle
underlying our mathematics.  Geometric algebra fulfills this task for
them.

\begin{table}
\begin{center}
\begin{tabular}{|lcl|} \hline
coordinate geometry & \quad & spinor calculus \\
complex analysis & \quad & Grassmann algebra \\
vector analysis & \quad & Berezin calculus \\
tensor analysis & \quad & differential forms \\
Lie algebras & \quad & twistors \\
Clifford algebra & \quad & algebraic topology \\ \hline
\end{tabular}
\end{center}
\caption{\em Some algebraic systems employed in modern physics}
\label{tab-list}
\end{table}

In the course of this paper we will discuss a number of the algebraic
systems listed in Table~\ref{tab-list}, and demonstrate precisely how
they fit into the geometric algebra framework.  However, the principle
aim here is to discuss the application of geometric algebra to
electron physics.  These applications are limited essentially to
physics in Minkowski spacetime, so we restrict our attention to the
geometric algebra of spacetime --- the \textit{spacetime
algebra}~\cite{hes-sta}.  Our aim is twofold: to show that spacetime
algebra simplifies the study of the Dirac theory, and to show that the
Dirac theory, once formulated in the spacetime algebra, is a powerful
and flexible tool for the analysis of all aspects of electron physics
--- not just relativistic theory.  Accordingly, this paper contains a
mixture of formalism and applications.  We begin with an introduction
to the spacetime algebra (henceforth the STA), concentrating on how
the algebra of the STA is used to encode geometric ideas such as
lines, planes and rotations.  The introduction is designed to be
self-contained for the purposes of this paper, and its length has been
kept to a minimum.  A list of references and further reading is given
at the end of the introduction.

In Sections~\ref{S-spinors} and~\ref{S-opers} Pauli and Dirac column
spinors, and the operators that act on them, are formulated and
manipulated in the STA.  Once the STA formulation is achieved,
matrices are eliminated from Dirac theory, and the Dirac equation can
be studied and solved entirely within the real STA.  A significant
result of this work is that the unit imaginary of quantum mechanics is
eliminated and replaced by a directed plane segment --- a bivector.
That it is possible to do this has many implications for the
interpretation of quantum mechanics~\cite{hes-interp}.

In Sections~\ref{S-scatt}, \ref{S-steps} and~\ref{S-tunn} we turn to
issues related to the propagation, scattering and tunnelling of Dirac
waves.  Once the STA form is available, studying the properties of
electrons via the Dirac theory is no more complicated than using the
non-relativistic Pauli theory.  Indeed, the first-order form of the
Dirac theory makes some of the calculations easier than their
non-relativistic counterparts.  After establishing various results for
the behaviour of Dirac waves at potential steps, we study the
tunnelling of a wavepacket through a potential step.  Indirect timing
measurements for quantum-mechanical tunnelling are now available from
photon experiments, so it is important to have a solid theoretical
understanding of the process.  We argue that standard quantum theory
has so far failed to provide such an understanding, as it misses two
crucial features underlying the tunnelling process.

In Section~\ref{S-spinmeas} we give a relativistic treatment of a
measurement made with a Stern-Gerlach apparatus on a fermion with zero
charge and an anomalous magnetic moment.  As with tunnelling, it is
shown that a disjoint set of outcomes is consistent with the causal
evolution of a wavepacket implied by the Dirac equation.  Wavepacket
collapse is therefore not required to explain the results of
experiment, as the uncertainty in the final result derives from the
uncertainty present in the initial wavepacket.  It is
also argued that the standard quantum theory interpretation of the
measurement performed by a Stern-Gerlach apparatus is unsatisfactory.
In the STA, the anticommutation of the Pauli operators merely
expresses the fact that they represent orthonormal vectors, so cannot
have any dynamical content.  Accordingly, it should be possible to
have simultaneous knowledge of all three components of the spin
vector, and we argue that a Stern-Gerlach apparatus is precisely what
is needed to achieve this knowledge!

Multiparticle quantum theory is considered in Section~\ref{S-multi}.
We introduce a new device for analysing multiparticle states --- the
multiparticle STA. This is constructed from a copy of the STA for each
particle of interest.  The resulting algebraic structure is enormously
rich in its properties, and offers the possibility of a geometric
understanding of relativistic multiparticle quantum physics.  Some
applications of the multiparticle STA are given here, including a
relativistic treatment of the Pauli exclusion principle.  The paper
ends with a brief survey of some other applications of the STA to
electron physics, followed by a summary of the main conclusions drawn
from this paper.

Summation convention and natural units
($\hbar=c=\epsilon_0=1$) are employed throughout, except where
explicitly stated.

\section{Spacetime Algebra}
\label{1-STA}

\textit{`Spacetime algebra'} is the name given to the geometric
(Clifford) algebra generated by Minkowski spacetime.  In geometric
algebra, vectors are equipped with a product that is associative and
distributive over addition.  This product has the distinguishing
feature that the square of any vector in the algebra is a scalar.  A
simple re-arrangement of the expansion
\begin{equation}
(a+b)^2 = (a+b)(a+b) = a^2 + (ab + ba) + b^2
\end{equation}
yields
\begin{equation}
ab + ba = (a+b)^2 - a^2 - b^2,
\end{equation}
from which it follows that the symmetrised product of any two vectors
is also a scalar.  We call this the \textit{inner} product $a\dt b$,
where
\begin{equation}
a \dt b = \half (ab + ba).
\label{1inner}
\end{equation}
The remaining, antisymmetric part of the geometric product is called
the \textit{outer} product $a \wdg b$, where
\begin{equation}
a \wdg b = \half (ab - ba).
\label{1outer}
\end{equation}
The result of the outer product of two vectors is a \textit{bivector}
--- a grade-2 object representing a segment of the plane swept out by
the vectors $a$ and $b$.

On combining equations~\eqref{1inner} and~\eqref{1outer} we see that the
full geometric product of two vectors decomposes as
\begin{equation}
ab = a \dt b + a \wdg b.
\label{1gprod}
\end{equation}
The essential feature of this product is that it mixes two different
types of object: scalars and bivectors.  One might now ask how the
right-hand side of~\eqref{1gprod} is to be interpreted.  The answer is
that the addition implied by~\eqref{1gprod} is that used when, for
example, a real number is added to an imaginary number to form a
complex number.  We are all happy with the rules for manipulating
complex numbers, and the rules for manipulating mixed-grade
combinations are much the same~\cite{DGL93-notreal}.  But why should
one be interested in the sum of a scalar and a bivector?  The reason
is again the same as for complex numbers: algebraic manipulations are
simplified considerably by working with general mixed-grade elements
(multivectors) instead of working independently with pure-grade
elements (scalars, vectors \textit{etc.}).

An example of how the geometric product of two vectors~\eqref{1gprod} is
employed directly is in the description of rotations using geometric
algebra.  Suppose initially that the vector $a$ is reflected in the
hyperplane perpendicular to the unit vector $n$ ($n^2=1$).  The result
of this reflection is the vector
\begin{equation}
a - 2 a \dt n n = a - (an + na)n = -nan.
\end{equation}
The form on the right-hand side is unique to geometric algebra, and is
already an improvement on the usual formula on the left-hand
side.  If one now applies a second reflection in the hyperplane
perpendicular to a second unit vector $m$ the result is the vector
\begin{equation}
-m (-nan) m = mn a (mn)\tld.
\end{equation}
The tilde on the right-hand side  denotes the operation of
\textit{reversion}, which simply reverses the order of the vectors in
any geometric product,
\begin{equation}
(ab \ldots c)\tld = c \ldots ba.
\end{equation}
The combination of two reflections is a rotation in the plane
specified by the two reflection axes.  We therefore see that a
rotation is performed by 
\begin{equation}
a \mapsto R a \Rrev
\label{rot1}
\end{equation}
where 
\begin{equation}
R = mn.
\label{rot-R}
\end{equation}
The object $R$ is called a \textit{rotor}.  It has the fundamental
property that
\begin{equation}
R \Rrev = mnnm = 1.
\end{equation}
Equation~\eqref{rot1} provides a remarkably compact and efficient
formulation for encoding rotations.  The formula for $R$~\eqref{rot-R}
shows that a rotor is formed from the geometric product of two unit
vectors, so does indeed consist of the sum of a scalar and a bivector.
A rotor can furthermore be written as the exponential of a bivector,
$R=\pm\exp (B/2)$, where the bivector encodes the plane in which the
rotation is performed.  This naturally generalises the complex
representation of rotations frequently used in two dimensions.  Rotors
illustrate how mixed-grade objects are frequently employed as
\textit{operators} which act on other quantities in the algebra.  The
fact that both geometric objects and the operators that act on them
are handled in a single unified framework is a central feature of
geometric algebra.

The above discussion applies to vector spaces of any dimension.  We
now turn to the case of specific interest, that of Minkowski
spacetime.  To make the discussion more concrete, we introduce a
set of four basis vectors $\{\gamdm\}$, $\mu = 0 \ldots 3$, satisfying
\begin{equation}
\gamdm \dt \gamdn = \eta_{\mu \nu} 
= \mbox{diag($+$\ $-$\ $-$\ $-$)}.
\end{equation}
The vectors $\{\gamdm\}$ satisfy the same algebraic relations as
Dirac's $\gamma$-matrices, but they now form a set of four independent
basis vectors for spacetime, not four components of a single vector in
an internal `spin space'.  When manipulating (geometric) products of
these vectors, one simply uses the rule that parallel vectors commute
and orthogonal vectors anticommute.  This result is clear immediately
from equation~\eqref{1gprod}.  From the four vectors $\{\gamdm\}$ we can
construct a set of six basis elements for the space of bivectors:
\begin{equation}
\{ \gi \go, \gj\go, \gk \go, \gk\gj, \gi\gk, \gj\gi \}.
\end{equation}
After the bivectors comes the space of grade-3 objects or trivectors.
This space is four-dimensional and is spanned by the basis
\begin{equation}
\{ \gk\gj\gi, \go\gk\gj, \go\gi\gk, \go\gj\gi \}.
\end{equation}
Finally, there is a single grade-4 element.  This is called the
\textit{pseudoscalar} and is given the symbol $i$, so that
\begin{equation}
i = \go\gi\gj\gk.
\end{equation}
The symbol $i$ is used because the square of $i$ is $-1$, but the
pseudoscalar must not be confused with the unit scalar imaginary
employed in quantum mechanics\footnote{In more recent
publications~\cite{gap} we have preferred the symbol $I$ for the
spacetime pseudoscalar, and now recommend this ahead of the older
$i$.}.  The pseudoscalar $i$ is a geometrically-significant entity and
is responsible for the duality operation in the algebra.  Furthermore,
$i$ {\em anti\/}commutes with odd-grade elements (vectors and
trivectors), and commutes only with even-grade elements.

The full STA is spanned by the basis 
\begin{equation}  
1, \qquad \{ \gamdm \}, \qquad \{ \sig_k, \; i \sig_k \} , \qquad \{
i \gamdm \}, \qquad i,
\label{1STA-bas}
\end{equation}
where
\begin{equation}
\sig_k =  \gam_k\go,  \hs{1} k=1,2,3.
\label{1defsig}
\end{equation}
An arbitrary element of this algebra is called a \textit{multivector}
and, if desired, can be expanded in terms of the basis~\eqref{1STA-bas}.
Multivectors in which all elements have the same grade are usually
written as $A_r$ to show that $A$ contains only grade-$r$ components.
Multivectors inherit an associative product from the geometric product
of vectors, and the geometric product of a grade-$r$ multivector $A_r$
with a grade-$s$ multivector $B_s$ decomposes into
\begin{equation} 
A_{r}B_{s} = \la AB \ra_{r+s} +  \la AB \ra_{r+s-2}
\ldots  + \la AB \ra_{|r-s|} . 
\label{1mvlprod}
\end{equation}
The symbol $\la M\ra_r$ denotes the projection onto the grade-$r$
component of $M$.  The projection onto the grade-0 (scalar) component
of $M$ is written as $\la M\ra$.  The scalar part of a product of
multivectors satisfies the cyclic reordering property
\begin{equation}
\la A \ldots B C \ra = \la C A \ldots B\ra.
\end{equation}
The `$ \cdot $' and `$ \wedge$' symbols are retained for the
lowest-grade and highest-grade terms of the series~\eqref{1mvlprod}, so
that
\begin{align}
A_r \dt B_s &= \la AB \ra_{|r-s|} \\
A_r \wdg B_s &= \la AB \ra_{s+r} , 
\end{align}
which are called the inner and outer (or exterior) products
respectively.  We also make use of the scalar product, defined by
\begin{equation}
A \scp B = \la AB \ra,
\end{equation}
and the commutator product, defined by
\begin{equation}
A \crs B = \half (AB - BA).
\label{1comm}
\end{equation}
The associativity of the geometric product ensures that the commutator
product satisfies the Jacobi identity
\begin{equation}
A \crs (B \crs C) + B \crs (C \crs A) + C \crs (A \crs B) =0.
\end{equation}
When manipulating chains of products we employ the operator ordering
convention that, in the absence of brackets,
\textit{inner, outer and scalar products take precedence over
geometric products}.

As an illustration of the working of these definitions, consider
the inner product of a vector $a$ with a bivector $b\wedge c$:
\begin{align}
a \dt (b \wdg c) 
&= \la a b \wdg c \ra_1 \nn \\
&= \half \la abc -a cb \ra_1 \nn \\
&= a \dt b c - a \dt c b - \half \la bac-cab \ra_1.
\end{align}
The quantity $bac-cab$ reverses to give minus itself, so cannot
contain a vector part.  We therefore obtain the result
\begin{equation}
a \dt (b \wdg c) =  a \dt b c - a \dt c b, 
\end{equation}
which is useful in many applications.

\subsection{The Spacetime Split}
\label{1-Stsplt}

The three bivectors $\{\sig_k\}$, where
$\sig_k=\gam_k\go$~\eqref{1defsig} satisfy
\begin{equation}
\half(\sig_j \sig_k + \sig_k \sig_j) = -\half(\gam_j \gam_k +
\gam_k\gam_j) = \delta_{jk}
\label{1pauli}
\end{equation}
and therefore generate the geometric algebra of three-dimensional
Euclidean space~\cite{DGL93-notreal,hes-nf1}.  This is identified as
the algebra for the rest-space relative to the timelike vector $\go$.
The full algebra for this space is spanned by the set
\begin{equation}  
1, \qquad \{ \sig_k \}, \qquad \{ i \sig_k \}, \qquad i,
\label{1pau-bas}
\end{equation}
which is identifiable as the even subalgebra of the full
STA~\eqref{1STA-bas}.  The identification of the algebra of relative space
with the even subalgebra of the STA simplifies the transition from
relativistic quantities to observables in a given frame.  It is
apparent from~\eqref{1pauli} that the algebra of the $\{\sig_k\}$ is
isomorphic to the algebra of the Pauli matrices.  As with the
$\{\gamdm\}$, the $\{\sig_k\}$ are to be interpreted geometrically as
spatial vectors (spacetime bivectors) and not as operators in an 
abstract spin space.  It should be noted that the pseudoscalar
employed in~\eqref{1pau-bas} is the same as that employed in spacetime,
since
\begin{equation}
\si\sj\sk = \gi\go\gj\go\gk\go = \go\gi\gj\gk = i.
\end{equation}

The split of the six spacetime bivectors into relative vectors
$\{\sig_k\}$ and relative bivectors $\{i\sig_k\}$ is a frame-dependent
operation --- different observers determine different relative spaces.
This fact is clearly illustrated using the Faraday bivector $F$.  The
`spacetime split'~\cite{hes-sta,hes74a} of $F$ into the $\go$-system
is made by separating $F$ into parts which anticommute and commute
with $\go$.  Thus
\begin{equation}
F = \bE + i \bB 
\label{1Fsplit}
\end{equation}
where
\begin{align}
\bE &= \half(F - \go F \go) \\
i\bB  &= \half(F + \go F \go).
\end{align}
Both $\bE$ and $\bB$ are spatial vectors in the $\go$-frame, and
$i\bB$ is a spatial bivector.  Equation~\eqref{1Fsplit} decomposes $F$
into separate electric and magnetic fields, and the explicit
appearance of $\go$ in the formulae for $\bE$ and $\bB$ shows how this
split is observer-dependent.  Where required, relative (or spatial)
vectors in the $\go$-system are written in bold type to record the
fact that in the STA they are actually bivectors.  This distinguishes
them from spacetime vectors, which are left in normal type.  No
problems arise for the $\{\sigk\}$, which are unambiguously spacetime
bivectors, and so are left in normal type.

When dealing with spatial problems it is useful to define an operation
which distinguishes between spatial vectors (such as $\bE$) and
spatial bivectors (such as $i\bB$).  (Since both the $\{\sig_k\}$ and
$\{i\sig_k\}$ are spacetime bivectors, they behave the same under
Lorentz-covariant operations.)  The required operation is that of
spatial reversion which, as it coincides with Hermitian conjugation for
matrices, we denote with a dagger.  We therefore define
\begin{equation}
M^\dagger = \go \tilde{M} \go,
\label{1defdag}
\end{equation}
so that, for example,
\begin{equation}
F^\dagger = \bE - i\bB.
\end{equation}
The explicit appearance of $\go$ in the definition~\eqref{1defdag} shows
that spatial reversion is not a Lorentz-covariant operation.

When working with purely spatial quantities, we often require that the
dot and wedge operations drop down to their three-dimensional
definitions.  For example, given two spatial vectors $\ba$ and $\bb$,
we would like $\ba\wdg\bb$ to denote the spatial bivector swept out by
$\ba$ and $\bb$.  Accordingly we adopt the convention that,
\textit{in expressions where both vectors are in bold type, the dot
and wedge operations take their three-dimensional meaning}.  Whilst
this convention may look clumsy, it is simple to use in practice and
rarely causes any confusion.

Spacetime vectors can also be decomposed by a spacetime split, this
time resulting in a scalar and a relative vector.  The spacetime split
of the vector $a$ is achieved via
\begin{equation}
a\go = a \dt \go + a \wdg \go = a_0 + \ba,
\label{1stsplit}
\end{equation}
so that $a_0$ is a scalar (the $\go$-time component of $a$) and $\ba$
is the relative spatial vector.  For example, the 4-momentum $p$
splits into
\begin{equation}
p \go = E + \bp
\end{equation}
where $E$ is the energy in the $\go$ frame, and $\bp$ is the
3-momentum.  The definition of the relative vector~\eqref{1stsplit}
ensures that
\begin{align}
a \dt b
&= \la a\go \go b\ra \nn \\
&= \la (a_0 +\ba)(b_0 - \bb) \ra \nn \\
&= a_0b_0 - \ba \dt \bb,
\end{align}
as required for the inner product in Minkowski spacetime.

\subsection{Spacetime Calculus}

The fundamental differential operator on spacetime is the derivative
with respect to the position vector $x$.  This is known as the
\textit{vector derivative} and is given the symbol $\grad$.  The
vector derivative is defined in terms of its directional derivatives,
with the derivative in the $a$ direction of a general multivector $M$
defined by
\begin{equation}
a \dt \grad M(x) = \lim_{\eps \rightarrow 0} \frac{M(x+\eps a) -
M(x)}{\eps}. 
\label{1dderiv}
\end{equation}
If we now introduce a set of four arbitrary basis vectors $\{e_j\}$,
with reciprocal vectors $\{e^k\}$ defined by the equation $e_j\dt
e^k=\delta_j^k$, then the vector derivative assembles from the
separate directional derivatives as
\begin{equation}
\grad = e^j e_j \dt \grad.
\label{1vecdiv}
\end{equation}
This definition shows how $\grad$ acts algebraically as a vector, as
well as inheriting a calculus from its directional derivatives.

As an explicit example, consider the $\{\gamdm\}$ frame introduced
above.  In terms of this frame we can write the position vector $x$ as
$x^\mu\gamdm$, with $x^0=t$, $x^1=\mathsf{x}$ \textit{etc}. and
$\{\mathsf{x,y,z}\}$ a usual set of Cartesian components for the
rest-frame of the $\go$ vector.  From the definition~\eqref{1dderiv} it is
clear that
\begin{equation}
\gamdm \dt \grad = \deriv{}{x^\mu} 
\end{equation}
which we abbreviate to $\partial_\mu$.  From the
definition~\eqref{1vecdiv} we can now write
\begin{equation}
\grad = \gamum \partial_\mu = \gam^0 \partial_t + 
\gam^1 \partial_{\mathsf{x}} + \gam^2 \partial_{\mathsf{y}} +
\gam^3 \partial_{\mathsf{z}}
\label{1stvecdiv}
\end{equation}
which, in the standard matrix language of Dirac theory, is the
operator that acts on Dirac spinors.  It is not surprising, therefore,
that the $\grad$ operator should play a fundamental role in the STA
formulation of the Dirac theory.  What is less obvious is that the
same operator should also play a fundamental role in the STA
formulation of the Maxwell equations~\cite{hes-sta}.  In tensor
notation, the Maxwell equations take the form
\begin{equation}
\partial_\mu F^{\mu\nu} = J^\nu, \hs{1} \partial_{[\alp} F_{\mu\nu]}
=0,
\label{1max}
\end{equation}
where $[\ldots]$ denotes total antisymmetrisation of the indices
inside the bracket.  On defining the bivector
\begin{equation}
F = \half F^{\mu\nu} \gamdm \wdg \gamdn
\end{equation}
and the vector $J= J^\mu\gamdm$ the equations~\eqref{1max} become
\begin{equation}
\grad \dt F = J
\label{1max1}
\end{equation}
and
\begin{equation}
\grad \wdg F = 0.
\label{1max2}
\end{equation}
But we can now utilise the geometric product to combine these separate
equations into the single equation
\begin{equation}
\grad F = J
\label{1maxga}
\end{equation}
which contains all of the Maxwell equations.  We see from~\eqref{1maxga}
that the vector derivative plays a central role in Maxwell theory, as
well as Dirac theory.  The observation that the vector derivative is
the sole differential operator required to formulate both Maxwell and
Dirac theories is a fundamental insight afforded by the STA.  Some
consequences of this observation for propagator theory are discussed
in~\cite{DGL93-paths}.

The vector derivative acts on the object to its immediate right unless
brackets are present, when it acts on everything in the brackets.
Since the vector derivative does not commute with multivectors, it is
useful to have a notation for when the derivative acts on a
multivector to which it is not adjacent.  We use overdots for this,
so that in the expression $\dgrad A\dot{B}$ the $\grad$ operator acts
only on $B$.  In terms of a frame of vectors we can write
\begin{equation}
\dgrad A\dot B = e^j A e_j \dt \grad B.
\end{equation}
The overdot notation provides a useful means for expressing Leibniz'
rule via
\begin{equation}
\grad (AB) = \dgrad \dot{A} B + \dgrad A \dot{B}.
\end{equation}
The spacetime split of the vector derivative requires some care.  We
wish to retain the symbol $\bgrad$ for the spatial vector
derivative, so that
\begin{equation}
\bgrad = \sig_k \partial_k, \hs{1} k=1 \ldots 3.
\end{equation}
This definition of $\bgrad$ is inconsistent with the
definition~\eqref{1stsplit}, so for the vector derivative we have to
remember that
\begin{equation}
\grad \go = \dif{t} - \bgrad.
\end{equation}

We conclude this introduction with some useful results concerning the
vector derivative.  We let the dimension of the space of interest be
$n$, so that the results are applicable to both space and spacetime.
The most basic results are that
\begin{equation}
\grad x = n
\end{equation}
and that
\begin{equation}
\grad \wdg \grad \psi = 0
\end{equation}
where $\psi$ is an arbitrary multivector field.  The latter result
follows from the fact that partial derivatives commute.  For a
grade-$r$ multivector $A_r$ the following results are also useful:
\begin{align}
\dgrad \xdot \dt A_r &= r A_r \\
\dgrad \xdot \wdg A_r &= (n-r) A_r \\
\dgrad A_r \xdot &= (-1)^r (n-2r) A_r.
\end{align}
More complicated results can be built up with the aid of Leibniz'
rule, for example
\begin{equation}
\grad x^2 = \dgrad \xdot \dt x +  \dgrad x \dt \xdot = 2x.
\end{equation}

This concludes our introduction to the spacetime algebra.  Further
details can be found in `\textit{Space-Time Algebra}' by
Hestenes~\cite{hes-sta} and `\textit{Clifford Algebra to Geometric
Calculus}' by Hestenes and Sobczyk~\cite{hes-gc}.  The latter is a
technical exposition of geometric algebra in general and does not deal
directly with spacetime physics.  A number of papers contain useful
introductory material including those by
Hestenes~\cite{hes74,hes74a,hes-unified} and the series of
papers~\cite{DGL93-notreal,DGL93-states,DGL93-lft,DGL93-paths} written
by three of the present authors.  Baylis \etal~\cite{bay92} and
Vold~\cite{vol93,vol93a} have also written good introductory pieces,
and the books \textit{`New Foundations for Classical Mechanics'} by
Hestenes~\cite{hes-nf1} and \textit{`Multivectors and Clifford Algebras
in Electrodynamics'} by Jancewicz~\cite{jan-mul} provide useful
background material.  Further work can be found in the three
conference proceedings~\cite{cliffconf1,cliffconf2,cliffconf3}, though
only a handful of papers are directly relevant to the work reviewed in
this paper.  Of greater interest are the proceedings of the conference
entitled `\textit{The Electron}'~\cite{elecconf}, which contains a
number of papers dealing with the application of the STA to electron
physics.

%
%

\section{Spinors and the Dirac Equation}
\label{S-spinors}

In this section we review how both the quantum states and matrix
operators of the Pauli and Dirac theories can be formulated within the
real STA.  This approach to electron theory was initiated by
Hestenes~\cite{hes67,hes71} and has grown steadily in popularity ever
since.  We start with a review of the single-electron Pauli theory and
then proceed to the Dirac theory.  Multiparticle states are considered
in Section~\ref{S-multi}.

Before proceeding, it is necessary to explain what we mean by a
\textit{spinor}.  The literature is replete with discussions about
different types of spinors and their inter-relationships and
transformation laws.  This literature is highly mathematical, and is
of very little relevance to electron physics.  For our purposes, we
define a spinor to be an element of a linear space which is closed
under left-sided multiplication by a rotor.  Thus spinors are acted on
by rotor representations of the rotation group.  With this in mind, we
can proceed directly to study the spinors of relevance to physics.
Further work relating to the material in this section is contained
in~\cite{DGL93-states}.

\subsection{Pauli Spinors}
\label{3-Pau}

We saw in Section~\ref{1-Stsplt} that the algebra of the Pauli
matrices is precisely that of a set of three orthonormal vectors in
space under the geometric product.  So the Pauli matrices are simply a
matrix representation of the geometric algebra of space.  This
observation opens up the possibility of eliminating matrices from the
Pauli theory in favour of geometrically-significant quantities.  But
what of the operator action of the Pauli matrices on spinors?  This
too needs to be represented with the geometric algebra of space.  To
achieve this aim, we recall the standard representation for the
Pauli matrices
\begin{equation}
\hsi = \begin{pmatrix}
        0 & 1 \\
        1 & 0
      \end{pmatrix} , \mbox{\hspace*{.2in}}
\hsj = \begin{pmatrix}
        0 & -j \\
        j & 0
      \end{pmatrix} , \mbox{\hspace*{.2in}}
\hsk = \begin{pmatrix}
        1 & 0 \\
        0 & -1
      \end{pmatrix} .
\end{equation}
The overhats distinguish these matrix operators from the $\{\sig_k\}$
vectors whose algebra they represent.  The symbol $i$ is reserved for
the pseudoscalar, so the symbol $j$ is used for the scalar unit
imaginary employed in quantum theory.  The $\{\hsigk\}$ operators act
on 2-component complex spinors
\begin{equation}
\ket{\psi} = \begin{pmatrix} 
 \psi_1  \\
 \psi_2 
\end{pmatrix} ,
\end{equation}
where $\psi_{1}$ and $\psi_{2}$ are complex numbers.  Quantum states
are written with bras and kets to distinguish them from STA
multivectors.  The set of $\ket{\psi}$'s form a two-dimensional complex
vector space.  To represent these states as multivectors in the STA we
therefore need to find a four-dimensional (real) space on which the
action of the $\{\hsigk\}$ operators can be replaced by operations
involving the $\{\sig_k\}$ vectors.  There are many ways to achieve
this goal, but the simplest is to represent a spinor $\ket{\psi}$ by
an element of the even subalgebra of~\eqref{1pau-bas}.  This space is
spanned by the set $\{1,i\sig_k\}$ and the column spinor $\ket{\psi}$
is placed in one-to-one correspondence with the (Pauli)-even
multivector $\psi=\go\psi\go$ through the
identification~\cite{DGL93-states,DGL-polspin}
\begin{equation}
\ket{\psi} = \begin{pmatrix} 
 a^0 + ja^3  \\
-a^2 + ja^1
\end{pmatrix} \trans \psi = a^0 + a^k i\sigma_k . 
\label{3ptrans}
\end{equation}
In particular, the spin-up and spin-down basis states become
\begin{equation}
\begin{pmatrix} 
1  \\
0 \end{pmatrix} 
\trans 1
\end{equation}
and 
\begin{equation}
\begin{pmatrix} 
0  \\
1 \end{pmatrix} 
\trans -i \sj.
\end{equation}
The action of the quantum operators $\{\hat{\sigma}_{k}\}$ and $j$ is
now replaced by the operations
\begin{align}
\hsigk \ket{\psi} & \lra  \sig_k \psi \sk \hs{0.8} (k=1,2,3) \\
\mathrm{and} \hs{0.5}  j \ket{\psi} & \lra  \psi \isk .
\label{3ptrans1}
\end{align}
Verifying these relations is a matter of routine computation; for
example 
\begin{equation}
\hsi \ket{\psi} = \begin{pmatrix}
-a^2 + ja^1 \\
 a^0 + ja^3  
\end{pmatrix} 
\trans 
-a^2 + a^1 i\sk -a^0 i\sj + a^3 i\si  
 = \si \psi \sk.
\end{equation}

We have now achieved our aim.  Every expression involving Pauli
operators and spinors has an equivalent form in the STA and all
manipulations can be carried out using the properties of the
$\{\sig_k\}$ vectors alone, with no need to introduce an explicit
matrix representation.  This is far more than just a theoretical
nicety.  Not only is there considerable advantage in being able to
perform the computations required in the Pauli theory without
multiplying matrices together, but abstract matrix algebraic
manipulations are replaced by relations of clear geometric
significance.

\subsubsection*{Pauli Observables}

We now turn to a discussion of the observables associated with Pauli
spinors.  These show how the STA formulation requires a shift in our
understanding of what constitutes scalar and vector observables at the
quantum level.  We first need to construct the STA form of the spinor
inner product $\la\psi\ket{\phi}$.  It is sufficient just to consider
the real part of the inner product, which is given by
\begin{equation}
\Re \la \psi \ket{\phi} \trans \la \psidag \phi \ra ,
\end{equation}
so that, for example,
\begin{align}
\la \psi \ket{\psi} \trans \la \psidag \psi \ra &= \la(a^{0} -i
a^{j}\sigj)(a^{0} + ia^{k}\sigk)\ra \nonumber \\
&= (a^{0})^{2} + a^{k} a^{k} .
\end{align} 
(Note that no spatial integral is implied in our use of the bra-ket
notation.)  Since
\begin{equation}
\la \psi | \phi \ra = \Re \la \psi | \phi \ra  - j \Re\la \psi | j\phi \ra, 
\end{equation}
the full inner product becomes
\begin{equation}
\la \psi \ket{\phi} \trans (\psi, \phi)_S = 
\la\psidag\phi\ra  - \la\psidag\phi \isk \ra \isk .
\label{3pinner}
\end{equation}
The right hand side projects out the $\{1, \isk\}$ components from the
geometric product $\psidag \phi$.  The result of this projection on a
multivector $A$ is written $\la A \ra_S$.  For Pauli-even multivectors
this projection has the simple form 
\begin{equation}
\la A \ra_S = \half(A - \isk A \isk).
\end{equation}

As an application of~\eqref{3pinner}, consider the expectation value of
the spin in the $k$-direction,
\begin{equation}
\la \psi | \hat{\sigma}_{k} \ket{\psi} \trans \la \psidag \sigk
\psi \sk \ra - \la \psidag \sigk \psi i\ra \isk.
\label{3pspin1.1}
\end{equation}
Since $\psidag i\sigk\psi$ reverses to give minus itself, it has zero
scalar part.  The right-hand side of~\eqref{3pspin1.1} therefore reduces
to
\begin{equation}
\la \sigk \psi \sk \psidag \ra = \sigk \dt \la \psi \sk \psidag \ra_v,
\label{3pspin}
\end{equation}
where $\la\ldots\ra_v$ denotes the relative vector component of the
term in brackets.  (This notation is required because $\la\ldots\ra_1$
would denote the spacetime vector part of the term in brackets.)  The
expression~\eqref{3pspin} has a rather different interpretation in the STA
to standard quantum mechanics --- it is the $\sig_k$-component of the
vector part of $\psi\sk\psidag$.  As $\psi\sk\psidag$ is both
Pauli-odd and Hermitian-symmetric it can contain only a relative
vector part, so we define the spin-vector $\bs$ by
\begin{equation}
\bs = \psi \sk \psidag.
\label{3defspin}
\end{equation}
(In fact, both spin and angular momentum are better viewed as spatial
bivector quantities, so it is usually more convenient to work with
$i\bs$ instead of $\bs$.)  The STA approach thus enables us to work
with a single vector $\bs$, whereas the operator/matrix theory treats
only its individual components.  We can apply a similar analysis to
the momentum operator.  The momentum density in the $k$-direction is
given by
\begin{equation}
\la\psi | -j\partial_k |\psi\ra \trans - \la\psidag \sig_k\dt\bgrad
\psi \isk\ra -  \la\psidag \sig_k\dt\bgrad \psi\ra \isk,
\end{equation}
in which the final term is a total divergence and so is ignored.
Recombining with the $\{\sig_k\}$ vectors, we find that the momentum
vector field is given by
\begin{equation}
\bp = - \dot{\bgrad} \la \psidot i\sk \psidag \ra.
\end{equation}

It might appear that we have just played a harmless game by redefining
various observables, but in fact something remarkable has happened.
The spin-vector $\bs$ and the momentum $\bp$ are both legitimate
(\textit{i.e}. gauge-invariant) quantities constructed from the spinor
$\psi$.  But standard quantum theory dictates that we cannot
simultaneously measure all three components of $\bs$, whereas we can
for $\bp$.  The `proof' of this result is based on the
non-commutativity of the $\{\hsigk\}$ operators.  But, in the STA
formulation, this lack of commutativity merely expresses the fact that
the $\{\sigk\}$ vectors are orthogonal --- a fact of geometry, not of
dynamics!  Furthermore, given a spinor $\psi$ there is certainly no
difficulty in finding the vector $\bs$.  So how then are we to
interpret a spin measurement, as performed by a Stern-Gerlach
apparatus for example?  This problem will be treated in detail in
Section~\ref{S-spinmeas}, but the conclusions are straightforward.  A
Stern-Gerlach apparatus is \textit{not} a measuring device --- it
should really be viewed as a spin
\textit{polariser}.  When a spinor wavepacket with arbitrary initial
vector $\bs$ enters a Stern-Gerlach apparatus, the wavepacket splits
in two and the vector $\bs$ rotates to align itself either parallel or
anti-parallel to the $\bB$ field.  The two different alignments then
separate into the two packets.  Hence, in the final beams, the vector
$\bs$ has been polarised to point in a single direction.  So, having
passed through the apparatus, all three components of the spin-vector
$\bs$ are known - not just the component in the direction of the $\bB$
field.  This is a major conceptual shift, yet it is completely
consistent with the standard predictions of quantum theory.  Similar
views have been expressed in the past by advocates of Bohm's `causal'
interpretation of quantum theory~\cite{boh55,vig-causal,dew88}.
However, the shift in interpretation described here is due solely to
the new understanding of the role of the Pauli matrices which the STA
affords.  It does not require any of the additional ideas associated
with Bohm's interpretation, such as quantum forces and quantum
torques.

\subsubsection*{Spinors and Rotations}

Further insights into the role of spinors in the Pauli theory are
obtained by defining a scalar 
\begin{equation}
\rho = \psi \psidag,
\end{equation}
so that the spinor $\psi$ can be decomposed into 
\begin{equation}
\psi = \rho^{1/2} R.
\label{3pau-decomp}
\end{equation}
Here $R$ is defined as
\begin{equation}
R= \rho^{-1/2} \psi 
\end{equation}
and satisfies
\begin{equation}
R \Rdag = 1.
\label{3rotnorm}
\end{equation}
In Section~\ref{1-STA} we saw that rotors, which act double-sidedly to
generate rotations, satisfy equation~\eqref{3rotnorm}.  It is not hard to
show that, in three dimensions, all even quantities
satisfying~\eqref{3rotnorm} are rotors.  It follows from~\eqref{3pau-decomp}
that the spin-vector $\bs$ can now be written as
\begin{equation}
\bs = \rho R \sk R^{\dagger} ,
\end{equation}
which demonstrates that the double-sided construction of the
expectation value in equation~\eqref{3pspin} contains an instruction
to rotate the fixed $\sk$ axis into the spin direction and dilate it.
The decomposition of the spinor $\psi$ into a density term~$\rho$ and
a rotor~$R$ suggests that a deeper substructure underlies the Pauli
theory.  This is a subject which has been frequently discussed by
Hestenes~\cite{hes71,hes751,hes79,hes-interp}.  As an example of the
insights afforded by this decomposition, one can now `explain' why
spinors transform single-sidedly under rotations.  If the vector $\bs$
is to be rotated to a new vector $R_0\bs\Rdag_0$ then, according to
the rotor group combination law, $R$ must transform to $R_0 R$.  This
induces the spinor transformation law
\begin{equation}
\psi \mapsto R_0 \psi
\end{equation}
which is the STA equivalent of the quantum transformation law
\begin{equation}
\ket{\psi} \mapsto \exp\{ \frac{j}{2} \theta n_k \hsigk \} \ket{\psi}
\end{equation}
where $\{n_k\}$ are the components of a unit vector.

\begin{table}
\renewcommand{\arraystretch}{1.2}
\begin{center}
\begin{tabular}{lll}
\hline \hline
\\
&  \begin{tabular}{l}
Pauli \\ Matrices 
\end{tabular} &
\begin{minipage}[c]{8.2cm}
\fbox{
\( 
\hsi = \begin{pmatrix}
        0 & 1 \\
        1 & 0
      \end{pmatrix}  
\quad
\hsj = \begin{pmatrix}
        0 & -j \\
        j & 0
      \end{pmatrix}
\quad
\hsk = \begin{pmatrix}
        1 & 0 \\
        0 & -1
      \end{pmatrix} 
\)
}
\end{minipage} \\
\\
&  \begin{tabular}{l}
Spinor \\ 
Equivalence 
\end{tabular} &
\begin{minipage}[c]{7cm}
\fbox{ 
\(
\ket{\psi} = \begin{pmatrix} 
 a^0 + ja^3  \\
-a^2 + ja^1
\end{pmatrix} \trans \psi = a^0 + a^k i\sigma_k 
\)}
\end{minipage} \\
\\
&  \begin{tabular}{l} 
Operator \\
Equivalences \end{tabular} &
\begin{minipage}[c]{5cm}
\fbox{ 
\begin{tabular}{rcl}
$\hsig_k \ket{\psi} $ & $\lra$ & $ \sig_k \psi \sk $ \\
$ j \ket{\psi} $ & $\lra$ & $ \psi i\sk $ \\
$ \la \psi | \psi' \ra $  & $\lra$ & $ \la \psidag \psi' \ra_S$
\end{tabular} }
\end{minipage} \\
\\
&  \begin{tabular}{l}
Observables 
\end{tabular} &
\begin{minipage}[c]{4.1cm}
\fbox{ 
\begin{tabular}{l}
\( \rho = \psi \psidag \) \\
\( \bs = \psi \sk \psidag \)
\end{tabular} }
\end{minipage} \\
\\
\hline \hline
\end{tabular}
\end{center}
\caption[dummy1]{\sl Summary of the main results for the STA
representation of Pauli spinors}
\label{3tab-Pauli}
\end{table}

We can also now see why the presence of the $\sk$ vector on the
right-hand side of the spinor $\psi$ does not break rotational
invariance.  All rotations are performed by left-multiplication by a
rotor, so the spinor $\psi$ effectively shields the $\sk$ on the right
from the transformation.  There is a strong analogy with rigid-body
mechanics in this observation, which has been discussed by
Hestenes~\cite{hes-interp,hes90}.  Similar ideas have also been
pursued by Dewdney, Holland and Kyprianidis~\cite{dew86,hol-qtm}.  We
shall see in the next section that this analogy extends to the Dirac
theory.  The main results of this section are summarised in
Table~\ref{3tab-Pauli}.

\subsection{Dirac Spinors}
\label{3-Dirac}

The procedures developed for Pauli spinors extend simply to Dirac
spinors.  Again, we seek to represent complex column spinors, and the
matrix operators acting on them, by multivectors and functions in the
STA.  Dirac spinors are four-component complex entities, so must be
represented by objects containing 8 real degrees of freedom.  The
representation that turns out to be most convenient for applications
is via the 8-dimensional even subalgebra of the
STA~\cite{hes71,hes75}.  If one recalls from Section~\ref{1-Stsplt}
that the even subalgebra of the STA is isomorphic to the Pauli
algebra, we see that what is required is a map between column spinors
and elements of the Pauli algebra.  To construct such a map we begin
with the $\gamma$-matrices in the standard Dirac--Pauli
representation~\cite{bjo-rel1},
\begin{equation}
\hat{\gamma}_0 = \begin{pmatrix}
        I & 0 \\
        0 & -I
      \end{pmatrix},\hs{0.6}
\hat{\gamma}_k = \begin{pmatrix}
        0 & -\hat{\sigma}_k \\
        \hat{\sigma}_k & 0
      \end{pmatrix} \hs{0.4} \mathrm{and} \hs{0.4} 
\hat{\gamma}_5 = \begin{pmatrix}
        0 & I \\
        I & 0
      \end{pmatrix},
\end{equation}
where $\hgam_5 = \hgam^5 = -j \hgam_0\hgam_1\hgam_2\hgam_3$ and
$I$ is the $2\times 2$ identity matrix.  A Dirac column spinor $|\psi
\ra$ is placed in one-to-one correspondence with an 8-component even
element of the STA via~\cite{DGL-polspin,gul-steps}
\begin{equation}
\ket{\psi} = \begin{pmatrix} 
 a^0 + ja^3  \\
-a^2 + ja^1  \\
b^0 + jb^3  \\
-b^2 + jb^1
\end{pmatrix} \trans \psi = a^0 + a^k i\sigma_k + (b^0 + b^k
i\sigma_k) \sk.
\label{3dtrans}
\end{equation}
With the spinor $|\psi \ra$ now replaced by an even multivector, the
action of the operators $\{ \hgamdm, \hat{\gamma}_{5}, j\}$ becomes
\begin{align}
\hgamdm \ket{\psi} & \trans  \gamma_{\mu} \psi \gamma_0
\quad (\mu= 0,\ldots,3) \\
j\ket{\psi} & \trans  \psi \, i\sigma_3 \\
\hat{\gamma}_{5} \ket{\psi} & \trans  \psi \sk .
\label{3dDPops}
\end{align}
To verify these relations, we note that the map~\eqref{3dtrans} can be
written more concisely as
\begin{equation}
\ket{\psi} = \begin{pmatrix} 
\ket{\phi} \\
\ket{\eta}
\end{pmatrix} \trans \psi = \phi + \eta \sk,
\label{3dtrans1}
\end{equation}
where $\ket{\phi}$ and $\ket{\eta}$ are two-component spinors, and
$\phi$ and $\eta$ are their Pauli-even equivalents, as defined by the
map~\eqref{3ptrans}.  We can now see, for example, that
\begin{equation}
\hat{\gamma}_k \ket{\psi} = \begin{pmatrix}
-\hsigk \ket{\eta} \\ 
\hsigk \ket{\phi} 
\end{pmatrix} 
\trans 
- \sigk \eta \sk + \sigk \phi = \gam_k (\phi + \eta \sk) \go,
\end{equation}
as required.  The map~\eqref{3dtrans1} shows that the split between the
`large' and `small' components of the column spinor $\ket{\psi}$ is
equivalent to splitting $\psi$ into Pauli-even and Pauli-odd terms in
the STA.

\subsubsection*{Alternative Representations}

All algebraic manipulations can be performed in the STA without ever
introducing a matrix representation, so equations~\eqref{3dtrans}
and~\eqref{3dDPops} achieve a map to a representation-free language.
However, the explicit map~\eqref{3dtrans} between the components of a
Dirac spinor and the multivector $\psi$ is only relevant to the
Dirac--Pauli matrix representation.  A different matrix representation
requires a different map so that that the effect of the matrix
operators is still given by~\eqref{3dDPops}.  The relevant map is easy to
construct given the unitary matrix $\hat{S}$ which transforms between
the matrix representations via
\begin{equation}
\hgamdm' = \hat{S} \hgamdm \hat{S}^{-1}.
\label{4dreps1}
\end{equation}
The corresponding spinor transformation is
$\ket{\psi}\mapsto\hat{S}\ket{\psi}$, and the map is constructed by
transforming the column spinor $|\psi\ra'$ in the new representation
back to a Dirac--Pauli spinor $\hat{S}^{\dagger}|\psi\ra'$.  The spinor
$\hat{S}^{\dagger}|\psi\ra'$ is then mapped into the STA in the usual
way~\eqref{3dtrans}.  As an example, consider the Weyl representation
defined by the matrices~\cite{itz-quant}
\begin{equation}
\hat{\gamma}_0^\prime = \begin{pmatrix}
        0 & -I \\
       -I  & 0
      \end{pmatrix}  \quad \mbox{and} \quad
\hat{\gamma}_k^\prime = \begin{pmatrix}
        0 & -\hat{\sigma}_k \\
        \hat{\sigma}_k & 0
      \end{pmatrix} .
\label{3dweyl}
\end{equation}
The Weyl representation is obtained from the Dirac--Pauli
representation by the unitary matrix
\begin{equation}
\hat{U} = \frac{1}{\sqrt{2}}\begin{pmatrix}
        I &  I \\
       -I  & I
      \end{pmatrix}.
\end{equation}
A spinor in the Weyl representation is written as 
\begin{equation}
|\psi \ra^\prime = \begin{pmatrix}
      | \chi \ra \\
      | \bar{\eta} \ra
 \end{pmatrix} ,
\label{3wchi}
\end{equation}
where $ | \chi \ra $ and $| \bar{\eta} \ra$ are 2-component spinors.
Acting on $|\psi\ra^\prime$ with $\hat{U}^\dagger$ gives
\begin{equation}
\hat{U}^\dagger |\psi \ra^\prime = \sqhalf \begin{pmatrix}
      | \chi \ra - | \bar{\eta} \ra \\   
      | \chi \ra + | \bar{\eta} \ra
  \end{pmatrix} .
\end{equation}
Using equation~\eqref{3dtrans}, this spinor is mapped onto the even
element
\begin{equation}
\hat{U}^\dagger |\psi \ra^\prime = \sqhalf \begin{pmatrix}
      | \chi \ra - | \bar{\eta} \ra \\   
      | \chi \ra + | \bar{\eta} \ra
  \end{pmatrix}
\trans \psi = \chi \sqhalf (1+\sk) - \bar{\eta} \sqhalf (1-\sk),
\label{3dweylmap}
\end{equation}
where $\chi$ and $\bar{\eta}$ are the Pauli-even equivalents of the
2-component complex spinors $| \chi \ra$ and $|\bar{\eta} \ra$, as
defined by equation~\eqref{3ptrans}.  The even multivector
\begin{equation}
\psi = \chi \sqhalf (1+\sk) - \bar{\eta} \sqhalf (1-\sk)
\end{equation}
is therefore our STA version of the column spinor
\begin{equation}
|\psi \ra^\prime = \begin{pmatrix}
      | \chi \ra \\
      | \bar{\eta} \ra
 \end{pmatrix} ,
\end{equation}
where $|\psi \ra^\prime$ is acted on by matrices in the Weyl
representation.  As a check, we observe that
\begin{equation}
\hgo^\prime |\psi \ra^\prime =  \begin{pmatrix}
      - |\bar{\eta} \ra \\
      - |\chi \ra 
 \end{pmatrix}
\trans - \bar{\eta} \sqhalf(1+\sk) + \chi \sqhalf(1-\sk) = \go \psi \go 
\end{equation}
and
\begin{equation}
\hat{\gamma}_k \ket{\psi} =  \begin{pmatrix}
      - \hat{\sigma}_{k}|\bar{\eta} \ra \\
       \hat{\sigma}_{k} |\chi\ra
 \end{pmatrix}
\trans - \sigk \bar{\eta} \sk \sqhalf(1+\sk) - \sigk \chi \sk
\sqhalf(1-\sk) =  \gamma_k \psi \go . 
\end{equation}
(Here we have used equation~\eqref{3ptrans1} and the fact that $\go$
commutes with all Pauli-even elements.)  The map~\eqref{3dweylmap} does
indeed have the required properties.

While our procedure ensures that the action of the
$\{\hgamdm,\hat{\gamma}_5\}$ matrix operators is always given
by~\eqref{3dDPops}, the same is not true of the operation of complex
conjugation.  Complex conjugation is a representation-dependent
operation, so the STA versions can be different for different
representations.  For example, complex conjugation in the Dirac--Pauli
and Weyl representations is given by
\begin{equation}
|\psi \ra^{\ast} \trans - \gj \psi \gj ,
\label{3dcong}
\end{equation}
whereas in the Majorana representation complex conjugation leads to
the STA operation~\cite{DGL93-states}
\begin{equation}
|\psi \ra^{\ast}_{\mathrm{Maj}} \trans \psi \sj.
\label{3dmajcong}
\end{equation}
Rather than think of~\eqref{3dcong} and~\eqref{3dmajcong} as different
representations of the same operation, however, it is simpler to view
them as distinct STA operations that can be performed on the
multivector $\psi$.

\subsection{The Dirac Equation and Observables}
\label{Ss3-Deqn}

As a simple application of~\eqref{3dtrans} and~\eqref{3dDPops}, consider the
Dirac equation
\begin{equation}
\hgamum(j \partial_{\mu} - e A_{\mu}) \ket{\psi} = m\ket{\psi}. 
\label{3ddeqnold}
\end{equation}
The STA version of this equation is, after postmultiplication by $\go$,
\begin{equation}
\grad \psi i \sk - eA \psi = m \psi \go,
\label{3Dirac}
\end{equation}
where $\grad = \gamum \partial_\mu$ is the spacetime vector
derivative~\eqref{1stvecdiv}.  The STA form of the Dirac
equation~\eqref{3Dirac} was first discovered by Hestenes~\cite{hes-sta},
and has been discussed by many authors since; see, for example,
references~\cite{gul-steps,ham84,bou-dual,kru-sols,dav91}.  The
translation scheme described here is direct and unambiguous and the 
resulting equation is both coordinate-free and representation-free.
In manipulating equation~\eqref{3Dirac} one needs only the algebraic rules
for multiplying spacetime multivectors, and the equation can be solved
completely without ever introducing a matrix representation.
Stripped of the dependence on a matrix representation,
equation~\eqref{3Dirac} expresses the intrinsic geometric content of the
Dirac equation.

In order to discuss the observables of the Dirac theory, we must first
consider the spinor inner product.  It is necessary at this point to
distinguish between the Hermitian and Dirac adjoint.  These are
written as
\begin{equation}
\begin{array}{rcl}
\la \bar{\psi} | & - & \mbox{Dirac adjoint} \\ 
\la \psi | & - & \mbox{Hermitian adjoint} ,
\end{array}
\end{equation}
which are represented in the STA as follows,
\begin{equation}
\begin{array}{rcl}
\la \bar{\psi} | & \trans & \psirev \\
\la \psi | & \trans & \psidag = \go \psirev \go.
\end{array}
\end{equation}
One can see clearly from these definitions that the Dirac adjoint is
Lorentz-invariant, whereas the Hermitian adjoint requires singling out
a preferred timelike vector.

The inner product is handled as in equation~\eqref{3pinner}, so that
\begin{equation}
\la \psibar \ket{\phi} \trans \la \psirev \phi \ra - \la \psirev \phi
\isk \ra \isk =  \la \psirev \phi \ra_S,
\label{3dinner}
\end{equation}
which is also easily verified by direct calculation.  By utilising
\eqref{3dinner} the STA forms of the Dirac spinor bilinear
covariants~\cite{itz-quant} are readily found.  For example,
\begin{equation}
\la \psibar| \hgamdm \ket{\psi} \trans \la\psirev \gamdm \psi \go\ra -
\la\psirev \gamdm \psi i\gk\ra \isk = \gamdm \dt \la \psi \go \psirev
\ra_{1} 
\end{equation}
identifies the `observable' as the $\gamdm$-component of the vector
$\la\psi\go\psirev\ra_1$.  Since the quantity $\psi\go\psirev$ is odd
and reverse-symmetric it can only contain a vector part, so we can
define the frame-free vector $J$ by
\begin{equation}
J = \psi \go \psirev.
\label{3DiracJ}
\end{equation}

The spinor $\psi$ has a Lorentz-invariant decomposition which
generalises the decomposition of Pauli spinors into a rotation and a
density factor~\eqref{3pau-decomp}.  Since $\psi \psirev$ is even and
reverses to give itself, it contains only scalar and pseudoscalar
terms.  We can therefore define
\begin{equation} 
\rho e^{i \beta} = \psi \psirev,
\label{3beta}
\end{equation}
where both $\rho$ and $\beta$ are scalars.  Assuming that $\rho \neq
0$, $\psi$ can now be written as
\begin{equation}
\psi = \rho^{1/2}e^{i\beta/2}R
\end{equation}
where
\begin{equation}
R = (\rho e^{i\beta})^{-1/2} \psi .
\end{equation}
The even multivector $R$ satisfies $R\Rrev=1$ and therefore defines a
spacetime rotor.  The current $J$~\eqref{3DiracJ} can now be written as
\begin{equation}
J = \rho v
\label{3Dj2}
\end{equation}
where
\begin{equation}
v = R \go \Rrev.
\label{3Diracv}
\end{equation}
The remaining bilinear covariants can be analysed likewise, and the
results are summarised in Table~\ref{3tab-bicov}.  The final column of
this Table employs the quantities
\begin{equation}
s = \psi \gk \psirev, 
\hs{0.4} \mathrm{and} \hs{0.4} 
S = \psi i\sk \psirev.
\end{equation}

\begin{table}[t!]
\begin{center}
\begin{tabular}{|c|c|c|c|}
\hline
Bilinear  & Standard &    STA     & Frame-Free \\ 
Covariant &   Form   & Equivalent &   Form     \\
\hline
Scalar & $\la\psibar\ket{\psi}$ & $\la\psi\psirev\ra$ &
$\rho\cos\!\beta$ \rule{0cm}{5mm} \\ 
Vector & $\la\psibar | \hgamdm\ket{\psi}$ & $\gamdm\dt
(\psi\go\psirev)$ & $\psi\go\psirev = J$ \\
Bivector & $\la\psibar | j \hat{\gamma}_{\mu\nu} \ket{\psi}$ &
$(\gamdm \wdg \gamdn) \dt (\psi \isk \psirev)$ &
$\psi\isk\psirev = S$ \\
Pseudovector & $\la\psibar | \hgamdm\hat{\gamma}_{5}\ket{\psi}$ &
$\gamdm \dt (\psi\gk\psirev)$ &
$\psi\gk\psirev = s$ \\
Pseudoscalar & $\la\psibar|j \hat{\gamma}_{5}\ket{\psi}$ &
$\la\psi\psirev i\ra$ & $-\rho\sin\!\beta$ \\
\hline
\end{tabular}
\end{center}
\caption[dummy1]{\sl Bilinear covariants in the Dirac theory.}
\label{3tab-bicov}
\end{table}

Double-sided application of $R$ on a vector $a$ produces a Lorentz
transformation~\cite{DGL93-notreal}.  The full Dirac spinor $\psi$
therefore contains an instruction to rotate the fixed $\{\gamdm\}$ frame
into the frame of observables.  The analogy with rigid-body dynamics
first encountered in Section~\ref{3-Pau} with Pauli spinors therefore
extends to the relativistic theory.  In particular, the unit vector
$v$~\eqref{3Diracv} is both future-pointing and timelike and has been
interpreted as defining an electron velocity~\cite{hes71,hes-interp}
(see also the critical discussion in~\cite{DGL93-paths}).  The
`$\beta$-factor' appearing in the decomposition of $\psi$~\eqref{3beta}
has also been the subject of much
discussion~\cite{DGL93-paths,hes-interp,bou-dual,kru-sols} since, for
free-particle states, $\beta$ determines the ratio of particle to
anti-particle solutions.  It remains unclear whether this idea extends
usefully to interacting systems.

In Section~\ref{3-Pau} we argued that, for Pauli spinors, any
dynamical consequences derived from the algebraic properties of the
Pauli matrices were questionable, since the algebra of the Pauli
matrices merely expresses the geometrical relations between a set of
orthonormal vectors in space.  Precisely the same is true of any
consequences inferred from the properties of the Dirac matrices.  This
observation has the happy consequence of removing one particularly
prevalent piece of nonsense --- that the observed velocity of an
electron must be the speed of light~\cite{fey-qed,sak-aqm}.  The
`proof' of this result is based on the idea that the velocity operator
in the $k$-direction is the $\hgam_k\hgam_0$ matrix.  Since the square
of this matrix is 1, its eigenvalues must be $\pm1$.  But in the STA
the fact that the square of the $\{\gamdm\}$ matrices is $\pm 1$
merely expresses the fact that they form an orthonormal basis.  This
cannot possibly have any observational consequences.  To the extent
that one can talk about a velocity in the Dirac theory, the relevant
observable must be defined in terms of the current $J$.  The
$\{\gamdm\}$ vectors play no other role than to pick out the
components of this current in a particular frame.  The shift from
viewing the $\{\gamdm\}$ as operators to viewing them as an arbitrary,
fixed frame is seen clearly in the definition of the current
$J$~\eqref{3DiracJ}.  In this expression it is now the $\psi$ that
`operates' to align the $\go$ vector with the observable current.
Since $\psi$ transforms single-sidedly under rotations, the fixed
initial $\go$-vector is never affected by the rotor and its presence
does not violate Lorentz invariance~\cite{DGL93-states}.

We end this subsection by briefly listing how the $C$, $P$ and $T$
symmetries are handled in the STA.  Following the conventions of
Bjorken~\& Drell~\cite{bjo-rel1} we find that
\begin{equation}
\begin{array}{rcl}
\hat{P} \ket{\psi}  & \trans & \go \psi(\bar{x})\go   \\
\hat{C} \ket{\psi}  & \trans & \psi \si   \\
\hat{T} \ket{\psi}  & \trans & i \go \psi(-\bar{x}) \gi,
\label{3dcpt}
\end{array}
\end{equation}
where $\bar{x}=\go x \go$ is (minus) a reflection of $x$ in the
timelike $\go$ axis.  The combined $CPT$ symmetry corresponds to
\begin{equation}
\psi \mapsto - i \psi(-x)
\end{equation}
so that $CPT$ symmetry does not require singling out a preferred
timelike vector.  A more complete discussion of the symmetries and
conserved quantities of the Dirac theory from the STA viewpoint is
given in~\cite{DGL93-lft}.  There the `multivector derivative' was
advocated as a valuable tool for extracting conserved quantities from
Lagrangians.

\subsubsection*{Plane-Wave States}

In most applications of the Dirac theory, the external fields applied
to the electron define a rest-frame, which is taken to be the
$\go$-frame.  The rotor $R$ then decomposes relative to the $\go$
vector into a boost $L$ and a rotation $\Phi$,
\begin{equation}
R = L \Phi,
\end{equation}
where
\begin{align}
L^\dagger &= L \\
\Phi^\dagger &= \tilde{\Phi}
\end{align}
and $L\Lrev=\Phi\Phirev=1$.  A positive-energy plane-wave state is
defined by
\begin{equation}
\psi = \psi_0 \et{-i\sk p\dt x}
\end{equation}
where $\psi_0$ is a constant spinor.  From the Dirac
equation~\eqref{3Dirac} with $A=0$, it follows that $\psi_0$ satisfies
\begin{equation}
p \psi_0 = m \psi_0 \go.
\end{equation}
Postmultiplying by $\psirev_0$ we see that 
\begin{equation}
p \psi\psirev = m J
\end{equation}
from which it follows that $\exp(i\beta)=\pm 1$.  Since $p$ has
positive energy we must take the positive solution ($\beta=0$).  It
follows that $\psi_0$ is just a rotor with a normalisation constant.
The boost $L$ determines the momentum by
\begin{equation}
p = m L \go \Lrev = m L^2 \go,
\end{equation}
which is solved by
\begin{equation}
L = \sqrt{p\go/m} = \frac{E+m +\bp}{\sqrt{2m(E+m)}},
\label{3Dwaves1}
\end{equation}
where
\begin{equation}
p\go = E+\bp.
\end{equation}
The Pauli rotor $\Phi$ determines the `comoving' spin bivector $\Phi
i\sk\Phirev$.  This is boosted by $L$ to give the spin $S$ as seen in
the laboratory frame.  A $\Phi\sk\Phirev$ gives the relative spin in
the rest-frame of the particle, we refer to this as the `rest-spin'.
In Section~\ref{Ss-spinprec} we show that the rest-spin is equivalent
to the `polarisation' vector defined in the traditional matrix
formulation.

Negative energy solutions are constructed in a similar manner, but
with an additional factor of $i$ or $\sk$ on the right (the choice of
which to use is simply a choice of phase).  The usual positive- and
negative-energy basis states employed in scattering theory are
(following the conventions of Itzykson~\& Zuber~\cite[Section
2-2]{itz-quant})
\begin{align}
\mbox{positive energy} & \quad  \psi^{(+)}(x) = u_r(p) e^{-i\sk p\cdot x} \\  
\mbox{negative energy} & \quad \psi^{(-)}(x) = v_r(p) e^{i\sk p\cdot x}
\end{align}
with
\begin{align}
u_r(p) &= L(p) \chi_r \\ 
v_r(p) &= L(p) \chi_r \sk.
\end{align}
Here $L(p)$ is given by equation~\eqref{3Dwaves1} and $\chi_r=\{1,-i\sj\}$
are spin basis states.  The decomposition into a boost and a rotor
turns out to be very useful in scattering theory, as is demonstrated
in Section~\ref{S-scatt}.

The main results for Dirac operators and spinors are summarised in
Table~\ref{3tab-Dirac}.

\begin{table}
\renewcommand{\arraystretch}{1.2}
\begin{center}
\begin{tabular}{lll}
\hline \hline
\\
&  \begin{tabular}{l}
Dirac \\
Matrices
\end{tabular} &
\begin{minipage}[c]{8.4cm}
\fbox{
\( 
\hat{\gamma}_0 = \begin{pmatrix}
        I & 0 \\
        0 & -I
      \end{pmatrix} 
\hs{0.3}
\hat{\gamma}_k = \begin{pmatrix}
        0 & -\hat{\sigma}_k \\
        \hat{\sigma}_k & 0
      \end{pmatrix} 
\hs{0.3}
\hat{\gamma}_5 = \begin{pmatrix}
        0 & I \\
        I & 0
      \end{pmatrix} 
\) }
\end{minipage} \\
\\
& \begin{tabular}{l}
Spinor \\
Equivalence
\end{tabular} &
\begin{minipage}[c]{8cm}
\fbox{ 
\(\ket{\psi} = \begin{pmatrix}
a^0 + ja^3  \\
-a^2 + ja^1  \\
b^0 + jb^3  \\
-b^2 + jb^1
\end{pmatrix} \trans \psi = 
\begin{array}{l} a^0 + a^k i\sigma_k + \\
 (b^0 + b^ki\sigma_k) \sk \end{array}
\) }
\end{minipage} \\
\\
& \begin{tabular}{l}
Operator \\
Equivalences
\end{tabular} &
\begin{minipage}[c]{4.6cm}
\fbox{ 
\begin{tabular}{rcl}
$\hgamdm \ket{\psi} $ & $\lra$ & $\gamma_{\mu} \psi \gamma_0$ \\
$ j \ket{\psi} $ & $\lra$ & $ \psi i\sk $ \\
$\hat{\gamma}_5 \ket{\psi}$ & $\lra$ & $\psi \sk$ \\
$ \la \psibar | \psi' \ra $  & $\lra$ & $ \la \psirev \psi' \ra_S$
\end{tabular} }
\end{minipage} \\
\\
& \begin{tabular}{l}
Dirac \\
Equation
\end{tabular} &
\fbox{ 
\( \grad \psi i \sk -e A\psi = m\psi \go \) } \\
\\
& \begin{tabular}{l}
Observables 
\end{tabular} &
\begin{minipage}[c]{5cm}
\fbox{ 
\begin{tabular}{cc}
\( \rho \et{i \beta} =  \psi \psirev \) & 
\(J = \psi \go \psirev \) \\
\(S = \psi i\sk \psirev \) & 
\(s = \psi \gk \psirev \)
\end{tabular} }
\end{minipage} \\
\\
& \begin{tabular}{l}
Plane-Wave \\
States
\end{tabular} &
\begin{minipage}[c]{7.5cm}
\fbox{ 
\begin{tabular}{l}
$\psi^{(+)}(x) = L(p)\Phi e^{-i\sk p\cdot x}$ \\
$\psi^{(-)}(x) = L(p)\Phi \sk e^{i\sk p\cdot x}$ \\
$L(p) = (p\go +m)/\sqrt{2m(E+m)}$
\end{tabular} }
\end{minipage} \\
\\
\hline \hline
\end{tabular}
\end{center}
\caption[dummy1]{\sl Summary of the main results for the STA
representation of Dirac spinors.  The matrices and spinor equivalence
are for the Dirac--Pauli representation.  The spinor equivalences for
other representations are constructed via the method outlined in the
text.}
\label{3tab-Dirac}
\end{table}
\section{Operators, Monogenics and the Hydrogen \\ Atom}
\label{S-opers}

So far, we have seen how the STA enables us to formulate the Dirac
equation entirely in the real geometric algebra of spacetime.  In so
doing, one might worry that contact has been lost with traditional,
operator-based techniques, but in fact this is not the case.  Operator
techniques are easily handled within the STA, and the use of a
coordinate-free language greatly simplifies manipulations.  The STA
furthermore provides a sharper distinction between the roles of scalar
and vector operators.

This section begins by constructing a Hamiltonian form of the Dirac
equation.  The standard split into even and odd operators then enables
a smooth transition to the non-relativistic Pauli theory.  We next
study central fields and construct angular-momentum operators that
commute with the Hamiltonian.  These lead naturally to the
construction of the spherical \textit{monogenics}, a basis set of
orthogonal eigenfunctions of the angular-momentum operators.  We
finally apply these techniques to two problems --- the Hydrogen atom,
and the Dirac `oscillator'.

\subsection{Hamiltonian Form and the Non-Relativistic Reduction} 
\label{Ss-Ham-Pauli}

The problem of how to best formulate operator techniques within the
STA is really little more than a question of finding a good notation.
We could of course borrow the traditional Dirac `bra-ket' notation,
but we have already seen that the bilinear covariants are better
handled without it.  It is easier instead to just juxtapose the
operator and the wavefunction on which it acts.  But we saw in
Section~\ref{S-spinors} that the STA operators often act
double-sidedly on the spinor $\psi$.  This is not a problem, as the
only permitted right-sided operations are multiplication by $\go$ or
$\isk$, and these operations commute.  Our notation can therefore
safely suppress these right-sided multiplications and lump all
operations on the left.  The overhat notation is useful to achieve
this and we define
\begin{equation}
\hgamdm \psi = \gamdm \psi \go.
\end{equation}
It should be borne in mind that all operations are now defined in the
STA, so the $\hgamdm$ are not intended to be matrix operators, as they
were in Section~\ref{3-Dirac}.

It is also useful to have a symbol for the operation of right-sided
multiplication by $\isk$.  The symbol $j$ carries the correct
connotations of an operator that commutes with all others and squares
to $-1$, and we define
\begin{equation}
j \psi = \psi \isk.
\end{equation}
The Dirac equation~\eqref{3Dirac} can now be written in the `operator' form
\begin{equation}
j \hat{\grad} \psi - e \hat{A} \psi = m \psi.
\label{4opform}
\end{equation}
where 
\begin{equation}
\hat{\grad} \psi = \grad \psi \go,
\hs{0.4} \mathrm{and} \hs{0.4} \hat{A} \psi = A \psi \go.
\end{equation}
Writing the Dirac equation in the form~\eqref{4opform} does not add
anything new, but does confirm that we have an efficient notation for
handling operators in the STA.

In many applications we require a Hamiltonian form of the Dirac
equation.  To express the Dirac equation~\eqref{3Dirac} in Hamiltonian
form we simply multiply from the left by $\go$.  The resulting
equation, with the dimensional constants temporarily put back in, is
\begin{equation}
j \hbar \dift \psi = c \bpht \psi +eV \psi -ce \bA \psi + m
c^2 \psibar
\label{4DHam}
\end{equation}
where
\begin{align}
\bpht \psi &= -j \hbar \bgrad \psi \\ 
\psibar &= \go \psi \go \\
\textrm{and} \hs{0.8} \go A &= V -c\bA.
\end{align}
Choosing a Hamiltonian is a non-covariant operation, since it picks
out a preferred timelike direction.  The Hamiltonian relative to the
$\go$ direction is the operator on the right-hand side of
equation~\eqref{4DHam}.  We write this operator with the symbol $\clh$.

\subsubsection*{The Pauli Equation}

As a first application, we consider the non-relativistic reduction of
the Dirac equation.  In most modern texts, the non-relativistic
approximation is carried out via the Foldy-Wouthuysen
transformation~\cite{bjo-rel1,itz-quant}.  Whilst the theoretical
motivation for this transformation is clear, it has the defect that
the wavefunction is transformed by a unitary operator which is very
hard to calculate in all but the simplest cases.  A simpler approach,
dating back to Feynman~\cite{fey-qed}, is to separate out the
fast-oscillating component of the waves and then split into separate
equations for the Pauli-even and Pauli-odd components of $\psi$.  Thus
we write (with $\hbar=1$ and the factors of $c$ kept in)
\begin{equation}
\psi = (\phi + \eta) \et{-i\sk m c^2 t}
\end{equation}
where $\phibar=\phi$ and $\bar{\eta} = -\eta$.  The Dirac
equation~\eqref{4DHam} now splits into the two equations
\begin{align}
\cle \phi - c \clo \eta &= 0 \label{4NR1} \\
(\cle + 2mc^2) \eta - c \clo \phi &= 0, \label{4NR2}
\end{align}
where
\begin{align}
\cle \phi &= (j \dift - eV) \phi \label{4NRopp1}\\
\clo \phi &= (\bpht - e \bA) \phi. \label{4NRopp2}
\end{align}
The formal solution to the second equation~\eqref{4NR2} is
\begin{equation}
\eta = \frac{1}{2mc} \left( 1 + \frac{\cle}{2mc^2} \right)^{-1} \clo
\phi,
\end{equation}
where the inverse on the right-hand side is understood to denote a
power series.  The power series is well-defined in the
non-relativistic limit as the $\cle$ operator is of the order of the
non-relativistic energy.  The remaining equation for $\phi$ is
\begin{equation}
\cle \phi - \frac{\clo}{2m} \left( 1 - \frac{\cle}{2mc^2} + \cdots
\right) \clo \phi = 0,
\label{4NRexact}
\end{equation}
which can be expanded out to the desired order of magnitude.  There is
little point in going beyond the first relativistic correction, so we
approximate~\eqref{4NRexact} by
\begin{equation}
\cle \phi + \frac{\clo \cle \clo}{4m^2c^2} \phi = \frac{\clo^2}{2m}
\phi. 
\label{4NRapprox}
\end{equation}

We seek an equation of the form $\cle\phi =\clh\phi$, where $\clh$
is the non-relativistic Hamiltonian.  We therefore need to replace the
$\clo\cle\clo$ term in equation~\eqref{4NRapprox} by a term that does not
involve $\cle$.  To do so we would like to utilise the approximate
result that
\begin{equation}
\cle \phi \approx \frac{\clo^2}{2m} \phi,
\label{4pau_res1}
\end{equation}
but we cannot use this result directly in the $\clo\cle\clo$ term
since the $\cle$ does not operate directly on $\phi$.  Instead we
employ the operator rearrangement
\begin{equation}
2 \clo\cle\clo = [\clo, [\cle, \clo]] + \cle \clo^2 + \clo^2 \cle
\end{equation}
to write equation~\eqref{4NRapprox} in the form
\begin{equation}
\cle \phi = \frac{\clo^2}{2m} \phi - \frac{\cle \clo^2 + \clo^2
\cle}{8m^2c^2} \phi - \frac{1}{8m^2c^2} [\clo, [\cle, \clo]] \phi.
\end{equation}
We can now make use of~\eqref{4pau_res1} to write
\begin{equation}
\cle \clo^2 \phi \approx \clo^2 \cle \phi \approx \frac{\clo^4}{2m} +
\mathrm{O}(c^{-2})  
\end{equation}
and so approximate~\eqref{4NRapprox} by
\begin{equation}
\cle \phi =  \frac{\clo^2}{2m} \phi - \frac{1}{8m^2c^2} [\clo, [\cle,
\clo]] \phi - \frac{\clo^4}{8m^3c^2} \phi,
\label{4Pauli1}
\end{equation}
which is valid to order $c^{-2}$.  The commutators are easily
evaluated, for example
\begin{equation}
[\cle,\clo] = -je(\dift \bA +\bgrad V) = je \bE .
\end{equation}
There are no time derivatives left in this commutator, so we do
achieve a sensible non-relativistic Hamiltonian.  The full commutator
required in equation~\eqref{4Pauli1} is
\begin{align}
[\clo, [\cle, \clo]] &= [-j\bgrad -e\bA, je\bE] \nn \\
&= (e\bgrad \bE) -2e \bE \wdg \bgrad -2j e^2 \bA \wdg \bE 
\label{4NRcomm}
\end{align}
in which the STA formulation ensures that we are manipulating spatial
vectors, rather than performing abstract matrix manipulations.

The various operators~\eqref{4NRopp1}, \eqref{4NRopp2} and~\eqref{4NRcomm} can
now be fed into equation~\eqref{4Pauli1} to yield the STA form of the
Pauli equation
\begin{align}
\dift \phi \isk &= \frac{1}{2m}(\bpht - e\bA)^2 \phi + eV \phi -
\frac{\bpht^4}{8m^3c^2} \phi \nn \\
& -\frac{1}{8m^2c^2}[e(\bgrad \bE -2 \bE \wdg \bgrad) \phi -2 e^2 \bA
\wdg \bE \phi \isk],
\label{4Peqn}
\end{align}
which is valid to O($c^{-2}$).  (We have assumed that $|\bA|\sim
c^{-1}$ to replace the $\clo^4$ term by $\bpht^4$.)  Using the
translation scheme of Table~\ref{3tab-Pauli} it is straightforward to
check that equation~\eqref{4Peqn} is the same as that found in standard
texts~\cite{bjo-rel1}.  In the standard approach, the geometric
product in the $\bgrad\bE$ term~\eqref{4Peqn} is split into a `spin-orbit'
term $\bgrad\wdg\bE$ and the `Darwin' term $\bgrad\dt\bE$.  The STA
approach reveals that these terms arise from a single source.

A similar approximation scheme can be adopted for the observables of
the Dirac theory.  For example the current, $\psi\go\psirev$, has a
three-vector part
\begin{equation}
\bJ = (\psi \go \psirev) \wdg \go = \phi \eta^\dagger + \eta \phidag,
\end{equation}
which is approximated to first-order by
\begin{equation}
\bJ \approx - \frac{1}{m} (\la \bgrad \phi \isk \phidag \ra_v - \bA
\phi \phidag ).
\label{4Paucurr}
\end{equation}
Not all applications of the Pauli theory correctly
identify~\eqref{4Paucurr} as the conserved current in the Pauli theory ---
an inconsistency first noted by Hestenes and Gurtler~\cite{hes74} (see
also the discussion in~\cite{DGL93-paths}).

\subsection{Angular Eigenstates and Monogenic Functions}
\label{4S-angmonos}
 
Returning to the Hamiltonian of equation~\eqref{4DHam}, let us now
consider the problem of a central potential $V=V(r)$, $\bA=0$, where
$r=|\bx|$.  We seek a set of angular-momentum operators which commute
with this Hamiltonian.  Starting with the scalar operator
$B\dt(\bx\wdg\bgrad)$, where $B$ is a spatial bivector, we find that
\begin{align}
[B\dt(\bx\wdg\bgrad), \clh] 
&= [B\dt(\bx\wdg\bgrad), -j\bgrad] \nn \\
&= j \dot{\bgrad} B \dt (\dot{\bx} \wdg \bgrad) \nn \\
&= -j B\dt\bgrad.
\end{align}
But, since $B\dt\bgrad=[B,\bgrad]/2$ and $B$ commutes with the rest of
$\clh$, we can rearrange the commutator into
\begin{equation}
[B\dt(\bx\wdg\bgrad) - \half B, \clh] = 0,
\label{4ang-comm}
\end{equation}
which gives us the required operator.  Since $B\dt(\bx\wdg\bgrad)-B/2$
is an anti-Hermitian operator, we define a set of Hermitian operators
as
\begin{equation}
J_B = j (B\dt(\bx\wdg\bgrad) - \half B).
\label{4defJ_B}
\end{equation}
The extra term of $\half B$ is the term that is conventionally viewed as
defining `spin-1/2'.  However, the geometric algebra derivation shows
that the result rests solely on the commutation properties of the
$B\dt(\bx\wdg\bgrad)$ and $\bgrad$ operators.  Furthermore, the factor
of one-half required in the $J_B$ operators would be present in a space
of any dimension.  It follows that the factor of one-half in~\eqref{4defJ_B}
cannot have anything to do with representations of the 3-D rotation
group.

From the STA point of view, $J_B$ is an operator-valued function of the
bivector~$B$.  In conventional quantum theory, however, we would view
the angular-momentum operator as a vector with components
\begin{equation}
\hat{J}_i = \hat{L}_i + \half \hat{\Sigma}_i
\end{equation}
where $\hat{\Sigma}_i= (j/2)\eps_{ijk}\hgam_j\hgam_k$.  The standard
notation takes what should be viewed as the sum of a scalar operator
and a bivector, and forces it to look like the sum of two vector
operators!  As well as being conceptually clearer, the STA approach is
easier to compute with.  For example, it is a simple matter to
establish the commutation relation
\begin{equation}
[J_{B_1}, J_{B_2}] = - j J_{B_1 \crs B_2} ,
\end{equation}
which forms the starting point for the representation theory of the
angular-momentum operators.

\subsubsection*{The Spherical Monogenics}

The key ingredients in the solution of the Dirac equation for
problems with radial symmetry are the spherical monogenics.  These are
Pauli spinors (even elements of the Pauli algebra~\eqref{1pau-bas}) which
satisfy the eigenvalue equation
\begin{equation}
- \bx \wdg \bgrad \psi = l \psi.
\label{4angmono}
\end{equation}
Such functions are called spherical monogenics because they are obtained
from the `monogenic equation'
\begin{equation}
\bgrad \Psi = 0
\label{4mono}
\end{equation}
by separating $\Psi$ into $r^l \psi(\theta,\phi)$.  Equation~\eqref{4mono}
generalises the concept of an analytic function to higher
dimensions~\cite{hes-gc,hes-unified}.

To analyse the properties of equation~\eqref{4angmono} we first note that
\begin{equation}
[ J_B, \bx \wdg \bgrad ] = 0,
\end{equation}
which is proved in the same manner as equation~\eqref{4ang-comm}.  It
follows that $\psi$ can simultaneously be an eigenstate of the
$\bx\wdg\bgrad$ operator and one of the $J_B$ operators.  To simplify
the notation we now define
\begin{equation}
J_k \psi = J_{i\sig_k} \psi = (i \sig_k \dt( \bx \wdg \bgrad) -
\half i\sig_k) \psi \isk. 
\end{equation}
We choose $\psi$ to be an eigenstate of $J_3$, and provisionally
write
\begin{equation}
- \bx \wdg \bgrad \psi = l \psi, \hs{0.8} J_3 \psi = \mu \psi.
\end{equation}

Before proceeding, we must introduce some notation for a
spherical-polar coordinate system.  We define the $\{r,\theta,\phi\}$
coordinates via
\begin{equation}
r = \sqrt{\bx^2}, \hs{0.6} \cos\!\theta = \sk\dt\bx /r,
\hs{0.6} \tan\!\phi = \sj\dt\bx / \si\dt\bx.
\end{equation}
The associated coordinate frame is
\begin{align}
\blde_r &=  \sin\!\theta (\cos\!\phi\, \si + \sin\!\phi\, \sj) +
\cos\!\theta\, \sk \nn \\ 
\blde_\theta &= r \cos\!\theta (\cos\!\phi\, \si + \sin\!\phi\, \sj)
- r \sin\!\theta \, \sk \\
\blde_\phi &= r \sin\!\theta (-\sin\!\phi\, \si + \cos\!\phi\,
\sj). \nn
\end{align}
From these we define the orthonormal vectors
$\{\sig_r,\sig_\theta,\sig_\phi\}$ by
\begin{align}
\sig_r &= \blde_r \nn \\
\sig_\theta &= \blde_\theta / r \\
\sig_\phi &= \blde_\phi / (r\sin\!\theta). \nn
\end{align}
The $\{\sig_r,\sig_\theta,\sig_\phi\}$ form a right-handed set, since
\begin{equation}
\sig_r\sig_\theta\sig_\phi = i.
\end{equation}

The vector $\sig_r$ satisfies
\begin{equation}
\bx \wdg \bgrad \sr = 2 \sr.
\end{equation}
It follows that
\begin{equation}
- \bx \wdg \bgrad (\sr\psi\sk) = -(l+2) \sr\psi\sk
\label{4negl}
\end{equation}
so, without loss of generality, we can choose $l$ to be positive and
recover the negative-$l$ states through multiplying by $\sr$.  In
addition, since
\begin{equation}
\bx\wdg\bgrad (\bx\wdg\bgrad \psi) = l^2 \psi 
\end{equation}
we find that
\begin{equation}
\frac{1}{\sin\!\theta} \deriv{}{\theta} \left(\sin\! \theta
\deriv{\psi}{\theta} \right) + \frac{1}{\sin^2\!\theta}
\frac{\partial^2\psi}{\partial \phi^2} = -l(l+1)\psi.
\end{equation}
Hence, with respect to a constant basis for the STA, the components of
$\psi$ are spherical harmonics and $l$ must be an integer for any
physical solution.

The next step is to introduce ladder operators to move between
different $J_3$ eigenstates.  The required analysis is standard, and
has been relegated to Appendix~\ref{App-monos}.  The conclusions are
that, for each value of $l$, the allowed values of the eigenvalues of
$J_3$ range from $(l+1/2)$ to $-(l+1/2)$.  The total degeneracy is
therefore $2(l+1)$.  The states can therefore be labeled by two
\textit{integers} $l$ and $m$ such that
\begin{align}
-\bx\wdg\bgrad \psi_l^m = l \psi_l^m  \qquad & l \geq 0 \\
J_3 \psi_l^m = (m+\half) \psi_l^m  \qquad & -1-l \leq m \leq l.
\end{align}
Labelling the states in this manner is unconventional, but provides
for many simplifications in describing the properties of the
$\psi^m_l$.

To find an explicit expression for the $\psi_l^m$ we start from the
highest-$m$ eigenstate, which is given by
\begin{equation}
\psi_l^l = \sin^l\!\theta\, \et{l\phi\isk},
\end{equation}
and act on this with the lowering operator $J_-$.  This procedure is
described in detail in Appendix~\ref{App-monos}.  The result is the
following, remarkably compact formula:
\begin{equation}
\psi_l^m = [ (l+m+1)P_l^m(\cos\!\theta) -
P_l^{m+1}(\cos\!\theta)i\sig_\phi ] \et{m\phi\isk},
\label{4monogens}
\end{equation}
where the associated Legendre polynomials follow the conventions of
Gradshteyn~\& Ryzhik~\cite{gr-tables}.  The expression~\eqref{4monogens}
offers a considerable improvement over formulae found elsewhere in
terms of both compactness and ease of use.  The formula~\eqref{4monogens}
is valid for non-negative $l$ and both signs of $m$.  The positive and
negative $m$-states are related by
\begin{equation}
\psi_l^m (-i\sj) = (-1)^m \frac{(l+m+1)!}{(l-m)!} \psi_l^{-(m+1)}.
\end{equation}

\begin{table}[t!]
\begin{center}
\begin{tabular}{c|c|c}
\hline
   $l$  & Eigenvalues of $J_3$  & Degeneracy \\  
\hline
\vdots  &     \vdots        & \vdots \\
  2     & $5/2 \cdots -5/2$ &  6  \\
  1     & $3/2 \cdots -3/2$ &  4  \\
  0     & $1/2 \cdots -1/2$ &  2  \\
($-1$)  &        ?          &  ?  \\
 $-2$   & $1/2 \cdots -1/2$ &  2  \\
\vdots  &     \vdots        & \vdots \\
\hline
\end{tabular}
\end{center}
\caption[dummy1]{\sl Eigenvalues and degeneracies for the $\psi_l^m$
monogenics.} 
\label{4tab-evals}
\end{table}

The negative-$l$ states are constructed using~\eqref{4negl} and the $J_3$
eigenvalues are unchanged by this construction.  The possible
eigenvalues and degeneracies are summarised in Table~\ref{4tab-evals}.
One curious feature of this table is that we appear to be missing a
line for the eigenvalue $l=-1$.  In fact solutions for this case do
exist, but they contain singularities which render them
unnormalisable.  For example, the functions
\begin{equation}
\frac{i\sig_\phi}{\sin\!\theta}, \hs{0.6} \mathrm{and} \hs{0.6}
\frac{\et{-i\sk\phi}}{\sin\!\theta} 
\end{equation}
have $l=-1$ and $J_3$ eigenvalues $+1/2$ and $-1/2$ respectively.
Both solutions are singular along the $z$-axis, however, so are of
limited physical interest.

\subsection{Applications}
\label{Ss-Op-apps}

Having established the properties of the spherical monogenics, we can
proceed quickly to the solution of various problems.  We have chosen
to consider two --- the standard case of the Hydrogen atom, and the
`Dirac Oscillator'~\cite{mos89}.

\subsubsection*{The Coulomb Problem}

The Hamiltonian for this problem is
\begin{equation}
\clh \psi = \bpht \psi - \frac{Z\alp}{r} \psi + m \psibar,
\label{4Hatom1}
\end{equation}
where $\alp=e^2/4\pi$ is the fine-structure constant and $Z$ is the
atomic charge.  Since the $J_B$ operators commute with $\clh$, $\psi$
can be placed in an eigenstate of $J_3$.  The operator $J_iJ_i$ must
also commute with $\clh$, but $\bx\wdg\bgrad$ does not, so both the
$\psi_l^m$ and $\sr\psi_l^m\sk$ monogenics are needed in the
solution.  

Though $\bx\wdg\bgrad$ does not commute with $\clh$, the operator
\begin{equation}
K = \hgo (1-\bx\wdg\bgrad) 
\end{equation}
does, as follows from
\begin{align}
[\hgo (1-\bx\wdg\bgrad), \bgrad] 
&= 2\hgo\bgrad - \hgo \dot{\bgrad}\dot{\bx}\wdg\bgrad \nn \\
&= 0.
\end{align}
We can therefore work with eigenstates of the $K$ operator, which
means that the spatial part of $\psi$ goes either as
\begin{equation}
\psi(\bx,l+1) = \psi_l^m u(r) + \sr \psi_l^m v(r) \isk 
\label{4Hatom3}
\end{equation}
or as
\begin{equation}
\psi(\bx,-(l+1)) = \sr \psi_l^m \sk u(r) +  \psi_l^m i v(r). 
\label{4Hatom4}
\end{equation}
In both cases the second label in $\psi(\bx,l+1)$ specifies the
eigenvalue of $K$.  The functions $u(r)$ and $v(r)$ are initially
`complex' superpositions of a scalar and an $\isk$ term.  It turns
out, however, that the scalar and $\isk$ equations decouple, and it is
sufficient to treat $u(r)$ and $v(r)$ as scalars.

We now insert the trial functions~\eqref{4Hatom3} and~\eqref{4Hatom4} into the
Hamiltonian of equation~\eqref{4Hatom1}.  Using the results that 
\begin{equation}
-\bgrad \psi_l^m = l/r \, \sr \psi_l^m, \hs{0.5} -\bgrad \sr \psi_l^m =
-(l+2)/r \, \psi_l^m,
\end{equation}
and looking for stationary-state solutions of energy $E$, the radial
equations reduce to
\begin{equation}
\begin{pmatrix}
u' \\ v' 
\end{pmatrix}
=
\begin{pmatrix}
	(\kappa -1)/r   & - (E + Z\alp/r +m) \\
	E + Z\alp/r -m  & (-\kappa -1)/r    
\end{pmatrix}
\begin{pmatrix}
u \\ v 
\end{pmatrix},
\end{equation}
where $\kappa$ is the eigenvalue of $K$.  ($\kappa$ is a non-zero
positive or negative integer.)  The solution of these radial equations
can be found in many textbooks (see, for example,
\cite{bjo-rel1,itz-quant,gran-rqm}).  The solutions can be
given in terms of confluent hypergeometric functions, and the energy
spectrum is obtained from the equation
\begin{equation}
E^2 = m^2 \left[ 1- \frac{(Z\alp)^2}{n^2 +2n\nu +(l+1)^2} \right],
\end{equation}
where $n$ is a positive integer and
\begin{equation}
\nu = [(l+1)^2 + (Z\alp)^2]^{1/2}.
\end{equation}

Whilst this analysis does not offer any new results, it should
demonstrate how easy it is to manipulate expressions involving the
spherical monogenics.

\subsubsection*{The Dirac `Oscillator'}

The equation describing a Dirac `oscillator' was introduced as by
Moshinsky and Szczepaniak as recently as 1989~\cite{mos89}.  The
equation is one for a chargeless particle with an anomalous magnetic
moment~\cite{bjo-rel1} which, in the STA, takes the form
\begin{equation}
\grad \psi i\sk -i\mu F \psi \gk = m\psi\go.
\end{equation}
This equation will be met again in Section~\ref{S-spinmeas}, where it is
used to analyse the effects of a Stern-Gerlach apparatus.  The
situation describing the Dirac oscillator is one where the $F$ field
exerts a linear, confining force described by
\begin{equation}
F = \frac{m\om}{\mu} \bx.
\end{equation}
The Hamiltonian in this case is
\begin{equation}
\clh \psi = \bpht \psi -jm\om \bx \psibar + m \psibar.
\end{equation}
It is a simple matter to verify that this Hamiltonian commutes with
both the $J_B$ and $K$ operators defined above, so we can again take
the wavefunction to be of the form of equations~\eqref{4Hatom3}
and~\eqref{4Hatom4}.  The resulting equations in this case are
\begin{equation}
\begin{pmatrix}
u' \\ v' 
\end{pmatrix}
=
\begin{pmatrix}
	(\kappa -1)/r -m\om r   & - (E + m) \\
	E - m  & (-\kappa -1)/r +m\om r   
\end{pmatrix}
\begin{pmatrix}
u \\ v 
\end{pmatrix} .
\end{equation}
The equations are simplified by transforming to the dimensionless
variable $\rho$,
\begin{equation}
\rho = (m\om)^{1/2}r,
\end{equation}
and removing the asymptotic behaviour via
\begin{align}
u &= \rho^l \et{-\rho^2/2} u_1 \\
v &= \rho^l \et{-\rho^2/2} u_2.
\end{align}
The analysis is now slightly different for the positive- and
negative-$\kappa$ equations, which we consider in turn.

\textbf{Positive}-{\boldmath $\kappa$}.  The equations reduce to
\begin{equation}
\frac{d}{d\rho} \begin{pmatrix}
u_1 \\
u_2
\end{pmatrix} =
\begin{pmatrix}
0   & -(E+m)/\sqrt{m\om} \\
(E-m)/\sqrt{m\om} & -2(l+1)/\rho + 2\rho
\end{pmatrix}
\begin{pmatrix}
u_1 \\
u_2
\end{pmatrix},
\end{equation}
which are solved with the power series
\begin{align}
u_1 &= \sum_{n=0} A_n \rho^{2n} \\
u_2 &= \sum_{n=0} B_n \rho^{2n+1}.
\end{align}
The recursion relations are
\begin{align}
2n A_n &= - \frac{E+m}{\sqrt{m\om}}B_{n-1} \\
(2n + 2l + 3) B_n &=  \frac{E-m}{\sqrt{m\om}} A_n + 2B_{n-1},
\end{align}
and the requirement that the series terminate produces the eigenvalue
spectrum
\begin{equation}
E^2 - m^2 = 4n m\om \hs{1} n=1,2 \ldots
\end{equation}
Remarkably, the energy levels do not depend on $l$, so are infinitely
degenerate!

\textbf{Negative}-{\boldmath $\kappa$}.  In this case the equations
reduce to
\begin{equation}
\frac{d}{d\rho} \begin{pmatrix}
u_1 \\
u_2
\end{pmatrix} =
\begin{pmatrix}
-2(l+1)/\rho       & -(E+m)/\sqrt{m\om} \\
(E-m)/\sqrt{m\om}  & 2\rho
\end{pmatrix}
\begin{pmatrix}
u_1 \\
u_2
\end{pmatrix},
\end{equation}
and are solved with the power series
\begin{align}
u_1 &= \sum_{n=0} A_n \rho^{2n+1} \\
u_2 &= \sum_{n=0} B_n \rho^{2n}.
\end{align}
The recursion relations become
\begin{align}
(2n+2l+3) A_n &= - \frac{E+m}{\sqrt{m\om}}B_n \\
2n B_n &=  \frac{E-m}{\sqrt{m\om}} A_{n-1} + 2 B_{n-1},
\end{align}
and this time the eigenvalues are given by the formula
\begin{equation}
E^2 - m^2 = 2 (2n+2l+1) m\om \hs{1} n=1,2 \ldots.
\end{equation}

The energy spectrum only contains $E$ through equations for $E^2$.  It
follows that both positive and negative energies are allowed.  The
lowest positive-energy state (the ground state) has $E^2=m^2+4m\om$,
leading to a non-relativistic energy of $\sim2\hbar\om$.  The
groundstate is infinitely degenerate, whereas the first excited state
has a degeneracy of two.  The energy spectrum is clearly quite bizarre
and does not correspond to any sensible physical system.  In
particular, this system does not reduce to a simple harmonic
oscillator in the non-relativistic limit.  The simple, if bizarre,
nature of the energy spectrum is obscured in other
approaches~\cite{mos89,mart92}, which choose a less clear labeling
system for the eigenstates.

\section{Propagators and Scattering Theory}
\label{S-scatt}

In this section we give a brief review of how problems requiring
propagators are formulated and solved in the STA.  The STA permits a
first-order form of both Maxwell and Dirac theories involving the same
differential operator --- the vector derivative $\grad$.  The key
problem is to find Green's functions for the vector derivative that
allow us to propagate initial data off some surface.  We therefore
start by studying the characteristic surfaces of the vector
derivative.  We then turn to the use of spinor potentials, which were
dealt with in greater detail in~\cite{DGL93-paths}.  The section
concludes with a look at single-particle scattering theory.  Using the
mappings established in Sections~\ref{S-spinors} and~\ref{S-opers} it
is a simple matter to reformulate in the STA the standard matrix
approach to scattering problems as described
in~\cite{bjo-rel1,itz-quant}.  The STA approach allows for a number of
improvements, however, particularly in the treatment of spin.  This is
a subject which was first addressed by Hestenes~\cite{hes-geom82}, and
our presentation closely follows his work.

\subsection{Propagation and Characteristic Surfaces}

One of the simplest demonstrations of the insights provided by the STA
formulation of both Maxwell and Dirac theories is in the treatment of
characteristic surfaces.  In the second-order theory, characteristic
surfaces are usually found by algebraic methods.  Here we show how the
same results can be found using a simple geometric argument applied to
first-order equations.  Suppose, initially, that we have a generic
equation of the type
\begin{equation}
\grad \psi = f(\psi,x),
\label{5eq1}
\end{equation}
where $\psi(x)$ is any multivector field (not necessarily a spinor
field) and $f(\psi,x)$ is some arbitrary, known function.  If we are
given initial data over some 3-D surface are there any obstructions
to us propagating this information off the surface?  If so, the
surface is a characteristic surface.  We start at a point on the
surface and pick three independent vectors $\{a,b,c\}$ tangent to the
surface at the chosen point.  Knowledge of $\psi$ on the surface
enables us to calculate
\begin{equation}
a \dt \grad \psi, \quad b \dt \grad \psi \quad \mbox{and} \quad c \dt
\grad \psi.
\end{equation}
We next form the trivector $a \wdg b \wdg c$ and dualise to define
\begin{equation}
n = i a \wdg b \wdg c.
\end{equation}
We can now multiply  equation~\eqref{5eq1} by $n$ and use 
\begin{align}
n \grad \psi 
&= n \dt \grad \psi + n \wdg \grad \psi \nn \\
&= n \dt \grad \psi + i (a \wdg b \wdg c) \dt \grad \psi \\
&= n \dt \grad \psi + i (a \wdg b \,  c \dt \grad \psi - a \wdg c\, b \dt
\grad \psi + b \wdg c \, a \dt \grad \psi),
\end{align}
to obtain
\begin{equation}
n \dt \grad \psi = nf(\psi,x) - i (a \wdg b \,  c\dt\grad\psi - a\wdg
c \, b\dt\grad\psi + b \wdg c \, a \dt \grad \psi).
\label{5eq2}
\end{equation}
All of the terms on the right-hand side of equation~\eqref{5eq2} are
known, so we can find $n\dt\grad\psi$ and use this to propagate $\psi$
in the $n$ direction ({\em i.e.} off the surface).  The only situation
in which we fail to propagate, therefore, is when $n$ remains in the
surface.  This occurs when
\begin{align}
n \wdg (a \wdg b \wdg c) &= 0 \nn \\
\implies \hs{.5} n \wdg (n i) &= 0 \nn \\
\implies \hs{.5} n \dt n &= 0.
\end{align}
Hence we only fail to propagate when $n^2=0$, and it follows
immediately that the characteristic surfaces of equation~\eqref{5eq1} are
\textit{null} surfaces.  This result applies to any first-order equation based
on the vector derivative $\grad$, including the Maxwell and Dirac
equations.  The fundamental significance of null directions in these
theories is transparent in their STA form.  Furthermore, the technique
extends immediately to a gravitational background, as described
in~\cite{DGL-grav}.

\subsection{Spinor Potentials and Propagators}

A simple method to generate propagators for the Dirac theory is to
introduce a spinor potential satisfying a scalar second-order
equation.  Suppose that $\psi$ satisfies the Dirac equation
\begin{equation}
\grad \psi \isk -m\psi\go = 0.
\end{equation}
$\psi$ can be generated from the (odd multivector) potential $\phi$
via
\begin{equation}
\psi = \grad \phi \isk + m \phi \go
\end{equation}
provided that
\begin{equation}
(\grad^2 + m^2) \phi = 0.
\end{equation}
The standard second-order theory can then be applied to $\phi$, and
then used to recover $\psi$.  In~\cite{DGL93-paths} this technique was
applied to constant-energy waves
\begin{equation}
\psi=\psi(\bx) \et{-i\sk Et}.
\end{equation}
The Dirac equation then becomes
\begin{equation}
\bgrad \psi i\sk + E\psi -m\psibar = 0
\end{equation}
which is solved by
\begin{equation}
\psi = -\bgrad \phi i\sk + E\phi + m\phibar
\end{equation}
where
\begin{equation}
\phi(\bx) = - \frac{1}{4 \pi} \oint |dS'| \, \bn' \psi(\bx')
\frac{\et{\isk pr}}{r}.
\end{equation}
In this integral the initial data $\psi(\bx')$ is given over some
closed spatial surface with normal $\bn'=\bn(\bx')$, and $p$ and $r$
are defined by
\begin{equation}
p = \sqrt{E^2+m^2} \hs{0.4} \mbox{and} \hs{0.4}
r = |\bx-\bx'|. 
\end{equation}
Similar techniques can be applied to the propagation of
electromagnetic waves (see~\cite{DGL93-paths} for details).

\subsection{Scattering Theory}
\label{Ss-sct}

We finish this short section with a brief look at how the matrix
approach to scattering theory is handled in the STA, closely following
the work of Hestenes~\cite{hes-geom82}.  We continue to employ the
symbol $j$ for $i\sk$ in places where it simplifies the notation.  In
particular, we employ the $j$ symbol in the exponential terms
introduced by Fourier transforming to momentum space.  Where the $j$'s
play a more significant geometric role they are left in the $\isk$
form.

We start by replacing the Dirac equation~\eqref{3Dirac} with the integral
equation
\begin{equation}
\psi(x) = \psi_i(x) + e \int d^4x' \, S_F(x-x') A(x') \psi(x') 
\label{5sct1}
\end{equation}
where $\psi_i$ is the asymptotic in-state which solves the free-particle
equation, and $S_F(x-x')$ is the STA form of the Feynman propagator.
Substituting~\eqref{5sct1} into the Dirac equation, we find that
$S_F(x-x')$ must satisfy
\begin{equation}
\grad_x S_F(x-x') M(x') \isk - m S_F(x-x') M(x') \go = \delta(x-x') M(x') 
\end{equation}
for an arbitrary multivector $M(x')$.  The solution to this equation
is
\begin{equation}
S_F(x-x') M(x') = \int \frac{d^4p}{(2\pi)^4} \, \frac{ pM(x')+mM(x')
\go}{p^2-m^2} \et{-jp\dt(x-x')}
\end{equation}
where, for causal propagation, the $dE$ integral must arrange that
positive-frequency waves propagate into the future ($t>t'$) and
negative-frequency waves propagate into the past ($t'>t$).  The result
of performing the $dE$ integral is 
\begin{align}
S_F(x-x') M &= - \theta(t-t') \int \frac{d^3\bp}{(2\pi)^3}
\frac{1}{2E} (pM + m M \go) \isk \et{-j p \dt (x-x')} \nn \\
& + \theta(t'-t) \int \frac{d^3\bp}{(2\pi)^3}
\frac{1}{2E} (pM - m M \go) \isk \et{j p \dt (x-x')}
\end{align}
where $E=+\sqrt{\bp^2+m^2}$ and $M=M(x')$.

With $\psi_{\mathrm{diff}}(x)$ defined by 
\begin{equation}
\psi_{\mathrm{diff}}(x) = \psi(x) - \psi_i(x)
\end{equation}
we find that, as $t$ tends to $+\infty$, $\psi_{\mathrm{diff}}(x)$ is
given by
\begin{equation}
\psi_{\mathrm{diff}}(x) = - e \int \!\! d^4 x'  \int \frac{d^3
  \bp}{(2\pi)^3} 
\frac{1}{2E} [p A(x') \psi(x') + m A(x') \psi(x') \go ] i\sk \et{-j p
\dt (x-x')} .
\end{equation}
We therefore define a set of final states $\psi_f(x)$ by
\begin{equation}
\psi_f(x) = -e \int  \frac{d^4 x'}{2E_f} \, [p_f A(x') \psi(x') + m A(x')
\psi(x') \go] i\sk \et{-j p_f \dt (x-x')},
\label{5sctpsif}
\end{equation}
which are plane-wave solutions to the free-field equations with
momentum $p_f$.  $\psi_{\mathrm{diff}}(x)$ can now be expressed as a
superposition of these plane-wave states,
\begin{equation}
\psi_{\mathrm{diff}}(x) = \int \frac{d^3 \bp_f}{(2\pi)^3} \, 
\psi_f(x).
\end{equation}

\subsubsection*{The Born Approximation and Coulomb Scattering}

In order to find $\psi_f(x)$ we must evaluate the
integral~\eqref{5sctpsif}.  In the Born approximation, we simplify the
problem by approximating $\psi(x')$ by $\psi_i(x')$.  In this case,
since
\begin{equation}
\psi_i(x') = \psi_i \et{-jp_i \dt x'}, \hs{0.4} \mbox{and} \hs{0.4} 
m\psi_i\go= p_i \psi_i,
\end{equation}
we can write
\begin{equation}
\psi_f(x) = - e \int \frac{d^4 x'}{2E_f} \, [p_f A(x') + A(x')p_i]
\psi_i i\sk \et{j q \dt x'} \et{-j p_f \dt x}, 
\label{5sct4}
\end{equation}
where
\begin{equation}
q = p_f - p_i.
\end{equation}
The integral in~\eqref{5sct4} can now be evaluated for any given
$A$-field.

As a simple application consider Coulomb scattering, for which $A(x')$
is given by 
\begin{equation}
A(x') = \frac{Ze}{4\pi |\bx'|} \go.
\end{equation}
Inserting this in~\eqref{5sct4} and carrying out the integrals, we obtain
\begin{equation}
\psi_f(x) = - S_{fi} \psi_i \isk \frac{(2\pi)^2}{E_f} \delta(E_f-E_i) 
\label{5sct6}
\end{equation}
where
\begin{align}
S_{fi} &= \frac{Z\alp}{4\pi} [p_f \go + \go p_i] \int d^3x \,
\frac{\et{ -j\bq\dt\br}}{r} \nn \\
&= \frac{Z\alp}{\bq^2} (2E + \bq).
\label{5sct7}
\end{align}
Here $E=E_f=E_i$ and $\alp=e^2/(4\pi)$ is the fine-structure constant.
The quantity $S_{fi}$ contains all the information about the
scattering process.  Its magnitude determines the cross-section
via~\cite{hes-geom82}
\begin{equation}
\frac{d\sigma}{d\Om_f} = S_{fi} \Srev_{fi}
\label{5sct8}
\end{equation}
and the remainder of $S_{fi}$ determines the change of momentum and
spin vectors.  This is clear from~\eqref{5sct6}, which shows that $S_{fi}$
must contain the rotor $R_f \Rrev_i$, where $R_i$ and $R_f$ are the
rotors for the initial and final plane-wave states.

Substituting~\eqref{5sct7} into~\eqref{5sct8} we immediately recover the Mott
scattering cross-section
\begin{equation}
\frac{d\sigma}{d\Om_f} = \frac{Z^2\alp^2}{\bq^4} (4E^2 - \bq^2) = 
\frac{Z^2\alp^2}{4 \bp^2 \bet^2 \sin^4(\theta/2)}
\left(1-\bet^2\sin^2(\theta/2) \right),
\end{equation}
where 
\begin{equation}
\bq^2 = (\bp_f-\bp_i)^2 = 2\bp^2 (1- \cos\!\theta) 
\hs{0.4} \mbox{and} \hs{0.4} \beta = |\bp|/E.
\end{equation}
The notable feature of this derivation is that no spin sums are
required.  Instead, all the spin dependence is contained in the
directional information in $S_{fi}$.  As well as being computationally
more efficient, the STA method for organising cross-section
calculations offers deeper insights into the structure of the theory.
For example, for Mott-scattering the directional information is
contained entirely in the quantity~\cite{hes-geom82}
\begin{equation}
S_{fi}' = \frac{1}{m} [p_f \go + \go p_i] = L_f^2 + \Lrev_i^2
\end{equation}
where $L_f$ and $L_i$ are the boosts contained in $R_f$ and $R_i$
respectively.  The algebraic structure 
\begin{equation}
S_{fi} = p_fM + Mp_i,
\end{equation}
where $M$ is some odd multivector, is common to many scattering
problems.

Since $S_{fi}$ contains the quantity $R_f\Rrev_i$, we obtain a spatial
rotor by removing the two boost terms.  We therefore define the
(unnormalised) rotor
\begin{equation}
U'_{fi} = \Lrev_f (L_f^2 + \Lrev_i^2) L_i = L_f L_i + \Lrev_f
\Lrev_i,
\end{equation}
so that $U_{fi} =U_{fi}' /|U_{fi}|$ determines the rotation from
initial to final rest-spins.  A simple calculation gives
\begin{equation}
U'_{fi} = 2[(E+m)^2 + \bp_f \bp_i],
\end{equation}
hence the rest-spin vector precesses in the $\bp_f\wdg\bp_i$ plane
through an angle $\del$, where
\begin{equation}
\tan(\del/2) = \frac{\sin\!\theta}{(E+m)/(E-m)\, + \cos\!\theta}.
\label{5prec_ang}
\end{equation}

Whilst the derivations of the Mott scattering formula and the
polarisation precession angle are only presented in outline here
(further details are contained in~\cite{hes-geom82}) it should be
clear that they offer many advantages over the usual
derivations~\cite{bjo-rel1,itz-quant}.  All the features of the
scattering are contained in the single multivector $S_{fi}$, algebraic
form of which is very simple.  Much work remains, however, if these
techniques are to be extended to the whole of QED.

\section{Plane Waves at Potential Steps}
\label{S-steps}

We now turn to a discussion of the matching of Dirac plane waves at a
potential step.  The case of perpendicular incidence is a standard
problem and is treated in most
texts~\cite{bjo-rel1,itz-quant,sak-aqm}.  In order to demonstrate the
power of the STA approach we treat the more general case of oblique
incidence, adapting an approach used in electrical engineering to
analyse the propagation of electromagnetic waves.  A number of
applications are given as illustrations, including the tunnelling of
monochromatic waves and spin precession on total reflection at a
barrier.  We conclude the section with a discussion of the Klein
paradox.

\begin{figure}[t!]
\begin{center}
\begin{picture}(250,250)(200,225)
\put(200,225){\hbox{\psfig{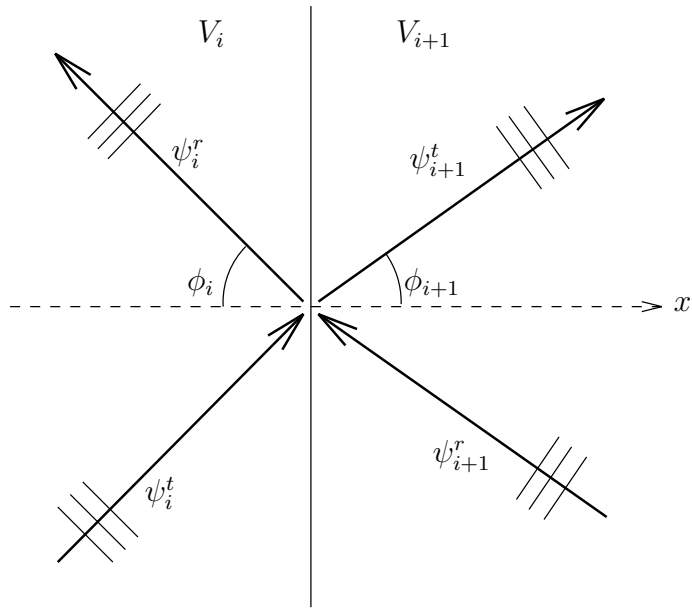}}}
\put(250,266){\large $\psi_i^t$}
\put(359,281){\large $\psi_{i+1}^r$}
\put(260,395){\large $\psi_i^r$}
\put(350,392){\large $\psi_{i+1}^t$}
\put(266,346){\large $\phi_i$}
\put(348,346){\large $\phi_{i+1}$}
\put(450,338){\large $x$}
\put(270,440){\large $V_i$}
\put(345,440){\large $V_{i+1}$}
\end{picture}
\end{center}
\caption[dummy1]{\sl Plane waves at a potential step.  The spatial
component of the wavevector lies in the $x$--$y$ plane and the step
lies in the $y$--$z$ plane.}
\label{6fig-barr1}
\end{figure}

The problem of interest is that of plane waves incident on a
succession of potential steps.  The steps are taken as lying along the
$x$ direction, with infinite extent in the $y$ and $z$ directions.
Since the spatial components of the incoming and outgoing wavevectors
lie in a single plane, the matching problem can be reduced to one in
two dimensions.  The analysis is simplified further if the wavevectors
are taken to lie in the $\isk$ plane.  (Other configurations can
always be obtained by applying a rotation.)  The arrangement is
illustrated in Figure~\ref{6fig-barr1}.  The waves all oscillate at a
single frequency $E$, and the Dirac equation in the $i$th region is
\begin{equation}
(E-eV_i) \psi = -\bgrad \psi\isk + m \go \psi \go.
\end{equation}
By continuity of $\psi$ at each boundary, the $y$ component of the
wavevector, $p_y$, must be the same in all regions.  For the $i$th
region we define
\begin{equation}
E_i' = E-eV_i
\end{equation}
and, depending on the magnitude of $V_i$, the waves in the this region
will be either travelling or evanescent.  For travelling waves we
define (dropping the subscripts)
\begin{equation}
p_x^2 = {E'}^2 - p_y^2 - m^2 \hs{1}  |E-eV| > \sqrt{p_y^2 +m^2}.
\end{equation}
In terms of the angle of incidence $\phi$ we also have
\begin{equation}
p_x = p \cos\!\phi, \qquad p_y = p \sin\!\phi, \qquad {E'}^2 = p^2
+ m^2. 
\end{equation}
For evanescent waves we write
\begin{equation}
\kap^2 = - {E'}^2 + p_y^2 + m^2 \hs{1}  |E-eV| < \sqrt{p_y^2
+m^2}.
\label{6defkap}
\end{equation}

In all the cases that we study, the incoming waves are assumed to be
positive-energy travelling waves in a region where $eV < E-\sqrt{p_y^2
+m^2}$.  Recalling the plane-wave solutions found in
Section~\ref{Ss3-Deqn}, the travelling waves are given by
\begin{align}
\psi^t = [\cosh(u/2) + \sinh(u/2) (\cos\!\phi\, \si + \sin\!\phi \,
\sj)] \Phi \et{-\isk(Et - p_xx -p_yy)} T  \\
\psi^r = [\cosh(u/2) + \sinh(u/2) (-\cos\!\phi\, \si + \sin\!\phi \,
\sj)] \Phi \et{-\isk(Et + p_xx -p_yy)} R, 
\label{6waves}
\end{align}
where
\begin{equation}
\tanh(u/2)= p/(E'+m)
\end{equation}
so that
\begin{equation}
\sinh\!u = p/m, \hs{1} \cosh\!u = E'/m.
\end{equation}
The transmission and reflection coefficients $T$ and $R$ are scalar
$+\isk$ combinations, always appearing on the right-hand side of
the spinor.  The fact that $p_y$ is the same in all regions gives the
electron equivalent of Snell's law,
\begin{equation}
\sinh\!u \, \sin\!\phi = \mbox{constant}.
\end{equation}
The Pauli spinor $\Phi$ describes the rest-spin of the particle, with
$\Phi=1$ giving spin-up and $\Phi=-i\sj$ spin down.  Other situations
are, of course, built from superpositions of these basis states.  For
these two spin basis states the spin vector is $\pm\sk$, which lies in
the plane of the barrier and is perpendicular to the plane of motion.
Choosing the states so that the spin is aligned in this manner
simplifies the analysis, as the two spin states completely decouple.
Many treatments (including one published by some of the present
authors~\cite{DGL93-paths}) miss this simplification.

There are three matching situations to consider, depending on whether
the transmitted waves are travelling, evanescent or in the Klein
region ($eV > E+\sqrt{p_y^2 +m^2}$).  We consider each of these in
turn.

\subsection{Matching Conditions for Travelling Waves}

The situation of interest here is when there are waves of
type~\eqref{6waves} in both regions.  The matching condition in all of
these problems is simply that $\psi$ is continuous at the boundary.
The work involved is therefore, in principle, less than for the
equivalent non-relativistic problem.  The matching is slightly
different for the two spins, so we consider each in turn.

\subsubsection*{Spin-Up ($\Phi=1$)}

We simplify the problem initially by taking the boundary at $x=0$.
Steps at other values of $x$ are then dealt with by inserting suitable
phase factors.  The matching condition at $x=0$ reduces to
\begin{multline}
[\cosh(u_i/2) + \sinh(u_i/2) \si \et{\phi_i \isk}] \Tup_i 
+  [\cosh(u_i/2) - \sinh(u_i/2) \si \et{-\phi_i \isk}] \Rup_i 
\\
\begin{array}{l}
 = [\cosh(u_{i+1}/2) + \sinh(u_{i+1}/2) \si \et{\phi_{i+1}
\isk}]\Tup_{i+1} \\
\quad + [\cosh(u_{i+1}/2) - \sinh(u_{i+1}/2) \si \et{-\phi_{i+1}
\isk}] \Rup_{i+1}. 
\end{array}
\label{6match1}
\end{multline}
Since the equations for the reflection and transmission coefficients
involve only scalar and $\isk$ terms, it is again convenient to
replace the $\isk$ bivector with the symbol $j$.  If we now define the
$2\times 2$ matrix
\begin{equation}
\bA_i =  \begin{pmatrix}
\cosh(u_i/2)       &\cosh(u_i/2) \\
\sinh(u_i/2) \et{j \phi_i}  & - \sinh(u_i/2) \et{-j \phi_i}
\end{pmatrix}
\label{6def-bA}
\end{equation}
we find that equation~\eqref{6match1} can be written concisely as
\begin{equation}
\bA_i \begin{pmatrix}
\Tup_i \\ \Rup_i   \end{pmatrix}
= \bA_{i+1} \begin{pmatrix}
\Tup_{i+1} \\ \Rup_{i+1}   \end{pmatrix}.
\label{6match2}
\end{equation}

The $\bA_i$ matrix has a straightforward inverse, so
equation~\eqref{6match2} can be easily manipulated to describe various
physical situations.  For example, consider plane waves incident on a
single step.  The equation describing this configuration is simply
\begin{equation}
\bA_1  \begin{pmatrix}
\Tup_1 \\ \Rup_1   \end{pmatrix}
= \bA_2  \begin{pmatrix}
\Tup_2 \\ 0   \end{pmatrix}
\end{equation}
so that
\begin{multline}
\begin{pmatrix}
\Tup_1 \\ \Rup_1   
\end{pmatrix}
= 
\frac{\Tup_2}{\sinh\!u_1 \cos\!\phi_1} \cdot \\
\begin{pmatrix}
\sinh(u_1/2) \cosh(u_2/2) \et{-j \phi_1} + \cosh(u_1/2) \sinh(u_2/2)
\et{j \phi_2} \\
\sinh(u_1/2) \cosh(u_2/2) \et{j \phi_1} - \cosh(u_1/2) \sinh(u_2/2)
\et{j \phi_2} 
\end{pmatrix}
\end{multline}
from which the reflection and transmission coefficients can be read
off.  The case of perpendicular incidence is particularly simple as
equation~\eqref{6match2} can be replaced by
\begin{equation}
\begin{pmatrix}
T_{i+1} \\ R_{i+1}
\end{pmatrix}
=
\frac{1}{\sinh\!u_{i+1}} 
\begin{pmatrix}
\sinh\! \half(u_{i+1}+u_i)   &\sinh\! \half(u_{i+1}-u_i) \\
\sinh\! \half (u_{i+1}-u_i)  & \sinh\! \half (u_{i+1}+u_i)
\end{pmatrix}
\begin{pmatrix}
T_i \\ R_i
\end{pmatrix},
\end{equation}
which is valid for all spin orientations.  So, for perpendicular
incidence, the reflection coefficient $r=R_1/T_1$ and transmission
coefficient $t=T_2/T_1$ at a single step are
\begin{equation}
r = \frac{\sinh\!\half(u_1-u_2)}{\sinh\!\half(u_1+u_2)}, \hs{1}
t = \frac{\sinh\! u_1}{\sinh\! \half(u_1+u_2)},
\end{equation}
which agree with the results given in standard texts (and
also~\cite{gul-steps}).

\subsubsection*{Spin-Down ($\Phi=-i\sj$)}

The matching equations for the case of opposite spin are
\begin{multline}
\begin{array}[b]{l}
[\cosh(u_i/2) + \sinh(u_i/2) \si \et{\phi_i \isk}](-\isj) \Tdn_i \\
+  [\cosh(u_i/2) - \sinh(u_i/2) \si \et{-\phi_i \isk}](-\isj) \Rdn_i 
\end{array} 
\\ 
\begin{array}[b]{l}
 =  [\cosh(u_{i+1}/2) + \sinh(u_{i+1}/2) \si \et{\phi_{i+1}
\isk}](-\isj) \Tdn_{i+1} \hs{1} \\
\quad + \, [\cosh(u_{i+1}/2) - \sinh(u_{i+1}/2) \si \et{-\phi_{i+1} 
\isk}] (-\isj) \Rdn_{i+1}.
\end{array}  
\end{multline}
Pulling the $\isj$ out on the right-hand side just has the effect of
complex-conjugating the reflection and transmission coefficients, so
the matrix equation~\eqref{6match1} is unchanged except that it now
relates the complex conjugates of the reflection and transmission
coefficients.  The analog of equation~\eqref{6match2} is therefore
\begin{equation}
\bA_i^* \begin{pmatrix}
\Tdn_i \\ \Rdn_i   \end{pmatrix}
= \bA_{i+1}^* \begin{pmatrix}
\Tdn_{i+1} \\ \Rdn_{i+1}   \end{pmatrix}.
\label{6match3}
\end{equation}
As mentioned earlier, the choice of alignment of spin basis states
ensures that there is no coupling between them.

\begin{figure}[t!]
\begin{center}
\begin{picture}(335,160)(196,325)
\put(196,325){\hbox{\psfig{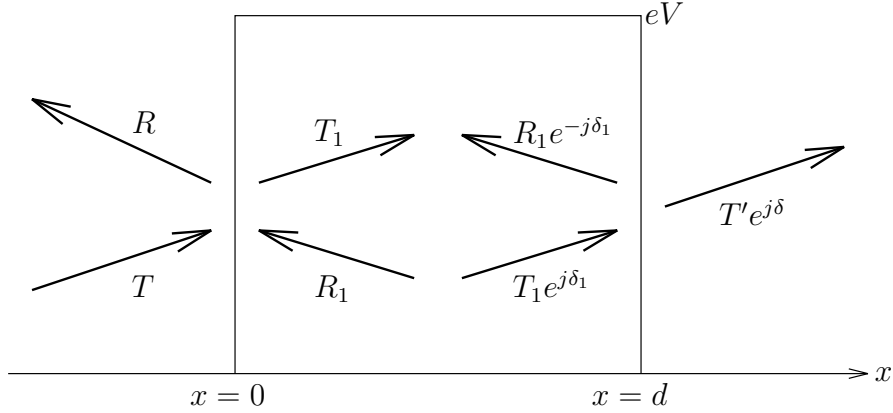}}}
\put(264,314){\large $x=0$}
\put(416,314){\large $x=d$}
\put(523,323){\large $x$}
\put(436,458){\large $eV$}
\put(242,355){\large $T$}
\put(242,417){\large $R$}
\put(311,355){\large $R_1$}
\put(311,414){\large $T_1$}
\put(386,355){\large $T_1e^{j\delta_1}$}
\put(386,413){\large $R_1e^{-j\delta_1}$}
\put(464,382){\large $T'e^{j\delta}$}
\end{picture}
\end{center}
\caption[dummy1]{\sl Plane waves scattering from a barrier.  The
barrier has height $eV$ and width~$d$.  Quantities inside the barrier
are labeled with a subscript 1, and the free quantities have no
subscripts.  The phases are given by
$\delta_1=md\sinh\!u_1\cos\!\phi_1$ and $\delta=p_x d$.}
\label{6fig-barr-waves}
\end{figure}

One can string together series of barriers by including suitable
`propagation' matrices.  For example, consider the set-up described in
Figure~\ref{6fig-barr-waves}.  The matching equations for spin-up are,
at the first barrier,
\begin{equation}
\bA \begin{pmatrix}
\Tup \\ \Rup   \end{pmatrix}
= \bA_1  \begin{pmatrix}
\Tup_1 \\ \Rup_1   \end{pmatrix}
\end{equation}
and at the second barrier,
\begin{equation}
\bA_1 \begin{pmatrix}
\et{j\delta_1} & 0 \\
0 & \et{-j\delta_1}
\end{pmatrix}
\begin{pmatrix}
\Tup_1 \\ \Rup_1   \end{pmatrix}
= \bA \begin{pmatrix}
\et{j\delta} & 0 \\
0 & \et{-j\delta}
\end{pmatrix}
\begin{pmatrix}
{\Tup}' \\ 0 \end{pmatrix}.
\label{6match4}
\end{equation}
Equation~\eqref{6match4} demonstrates neatly how matrices of the type
\begin{equation}
\mathbf{P} = 
\begin{pmatrix}
\et{j\delta} & 0 \\
0 & \et{-j\delta}
\end{pmatrix},
\end{equation}
where $\delta=p_x d$ and $d$ is the distance between steps, can be used
to propagate from one step to the next.  In this case the problem is
reduced to the equation
\begin{equation}
\begin{pmatrix}
\Tup \\ \Rup   \end{pmatrix}
=
\left[
\cos\!\delta_1 \, \bI - j \sin\!\delta_1\, \bA^{-1} \bA_1 
\begin{pmatrix}
1 & 0 \\
0 & -1 \end{pmatrix}
\bA^{-1}_1 \bA  \right] 
\begin{pmatrix}
\et{j\delta} {\Tup}' \\ 0 \end{pmatrix},
\end{equation}
which quickly yields the reflection and transmission coefficients.

\subsection{Matching onto Evanescent Waves}

Before studying matching onto evanescent waves, we must first solve
the Dirac equation in the evanescent region.  Again, the two spin
orientations behave differently and are treated separately.  Taking
spin-up first and looking at the transmitted (decaying) wave in the
evanescent region, the solution takes the form
\begin{equation}
\psi^t = [\cosh(u/2) + \sinh(u/2)\sj] \et{-\kap x} \et{-\isk(Et
-p_yy)} T.
\end{equation}
Substituting this into the Dirac equation yields
\begin{equation}
(E' \go + p_y\gj) \et{u\sj/2} =  \et{u\sj/2} (m \go -\kap \gj)
\label{6ev1}
\end{equation}
which is consistent with the definition of $\kap$~\eqref{6defkap}.  From
equation~\eqref{6ev1} we find that
\begin{equation}
\tanh(u/2) = \frac{E'-m}{p_y-\kap} = \frac{p_y+\kap}{E'+m},
\end{equation}
which completes the solution.  For the incoming (growing) wave we flip
the sign of $\kap$.  We therefore define $u^\pm$ via
\begin{equation}
\tanh(u^\pm/2) = \frac{p_y \pm \kap}{E'+m}
\end{equation}
and write the outgoing and incoming spin-up waves in the evanescent
region as
\begin{align}
& \psi^t = [\cosh(u^+/2) + \sinh(u^+/2)\sj] \et{-\kap x} \et{-\isk(Et 
-p_yy)} \Tup & \\ 
& \psi^r = [\cosh(u^-/2) + \sinh(u^-/2)\sj] \et{\kap x} \et{-\isk(Et 
-p_yy)} \Rup. & 
\label{6ev-up}
\end{align}

If we now consider matching at $x=0$, the continuity equation becomes
\begin{multline}
\begin{array}{l} 
[\cosh(u_i/2) + \sinh(u_i/2) \si \et{\phi_i \isk}]\Tup_i \\
+ \, [\cosh(u_i/2) - \sinh(u_i/2) \si \et{-\phi_i \isk}]\Rup_i 
\end{array} \\
\begin{array}{l} 
= \, [\cosh(u^+_{i+1}/2) + \sinh(u^+_{i+1}/2)\sj] \Tup_{i+1} \\
\quad + \, [\cosh(u^-_{i+1}/2) + \sinh(u^-_{i+1}/2)\sj)]
\Rup_{i+1}.
\end{array}
\label{6ev2}
\end{multline}
On defining the matrix
\begin{equation}
\bB_i^+ = \begin{pmatrix}
\cosh(u^+_i/2)    & \cosh(u^-_i/2) \\
j\sinh(u^+_i/2)    & +j\sinh(u^-_i/2)
\end{pmatrix},
\label{6def-bB}
\end{equation}
we can write equation~\eqref{6ev2} compactly as
\begin{equation}
\bA_i \begin{pmatrix}
\Tup_i \\ \Rup_i   \end{pmatrix}
= \bB_{i+1}^+ \begin{pmatrix}
\Tup_{i+1} \\ \Rup_{i+1}   \end{pmatrix}.
\label{6ev3}
\end{equation}
Again, either of the matrices can be inverted to analyse various
physical situations.  For example, the case of total reflection by a
step is handled by
\begin{equation}
\bA_1  \begin{pmatrix}
\Tup_1 \\ \Rup_1   \end{pmatrix}
= \bB_2^+  \begin{pmatrix}
\Tup_2 \\ 0   \end{pmatrix},
\end{equation}
from which one finds the reflection coefficient 
\begin{equation}
\rup = - \frac{\tanh(u^+/2) + \tanh(u/2) j
\et{j\phi}}{\tanh(u^+/2) - \tanh(u/2) j \et{-j\phi}} 
\label{6prec1}
\end{equation}
which has $|r^\uparrow|=1$, as expected.  The subscripts on $u_1$,
$u^\pm_2$ and $\phi_1$ are all obvious, and have been dropped.

The case of spin-down requires some sign changes.  The spinors in the
evanescent region are now given by
\begin{align}
& \psi^t = [\cosh(u^-/2) + \sinh(u^-/2)\sj] (-\isj) \et{-\kap x}
\et{-\isk(Et-p_yy)} \Tdn & \\
& \psi^r = [\cosh(u^+/2) + \sinh(u^+/2)\sj] (-\isj) \et{\kap x}
\et{-\isk(Et-p_yy)} \Rdn &
\label{6ev-down}
\end{align}
and, on defining
\begin{equation}
\bB_i^- = \begin{pmatrix}
\cosh(u^-_i/2)    & \cosh(u^+_i/2) \\
j\sinh(u^-_i/2)    & -j\sinh(u^+_i/2)
\end{pmatrix},
\end{equation}
the analog of equation~\eqref{6ev3} is
\begin{equation}
\bA_i^* \begin{pmatrix}
\Tdn_i \\ \Rdn_i   \end{pmatrix}
= \bB_{i+1}^{-^*} \begin{pmatrix}
\Tdn_{i+1} \\ \Rdn_{i+1}   \end{pmatrix}.
\end{equation}
These formulae are now applied to two situations of physical interest.

\subsection{Spin Precession at a Barrier}
\label{Ss-spinprec}

When a monochromatic wave is incident on a single step of sufficient
height that the wave cannot propagate there is total reflection.  In
the preceding section we found that the reflection coefficient for
spin-up is given by equation~\eqref{6prec1}, and the analogous calculation
for spin-down yields
\begin{equation}
\rdn = - \frac{\tanh(u^-/2) - \tanh(u/2) j\et{-j\phi}}
{\tanh(u^-/2) + \tanh(u/2) j \et{j\phi}}.
\label{6prec2}
\end{equation}
Both $\rup$ and $\rdn$ are pure phases, but there is an overall phase
difference between the two.  If the rest-spin vector
$\bs=\Phi\sk\Phirev$ is not perpendicular to the plane of incidence,
then this phase difference produces a precession of the spin vector.
To see how, suppose that the incident wave contains an arbitrary
superposition of spin-up and spin-down states,
\begin{equation}
\Phi = \cos(\theta/2)\et{\isk \phi_1} - \sin(\theta/2) i\sj \et{\isk
\phi_2} = \et{\isk \phi/2} \et{-\isj \theta/2} \et{-\isk \eps/2},
\end{equation}
where 
\begin{equation}
\phi = \phi_1-\phi_2, \hs{1} \eps = \phi_1+\phi_2,
\end{equation}
and the final pure-phase term is irrelevant.  After reflection, suppose
that the separate up and down states receive phase shifts of
$\delta^\uparrow$ and $\delta^\downarrow$ respectively.  The Pauli
spinor in the reflected wave is therefore
\begin{align}
\Phi^r &= \cos(\theta/2)\et{\isk (\phi_1+\delta^\uparrow)} -
\sin(\theta/2) i\sj \et{\isk (\phi_2+ \delta^\downarrow)} \nn \\  
&= \et{\isk\delta/2} \et{\isk\phi/2} \et{-\isj \theta/2} \et{-\isk
\eps'/2}, 
\end{align}
where
\begin{equation}
\delta = \delta^\uparrow - \delta^\downarrow
\end{equation}
and again there is an irrelevant overall phase.  The rest-spin vector
for the reflected wave is therefore
\begin{equation}
\bs^r = \Phi^r \sk \tilde{\Phi}^r = \et{\isk\delta/2} \bs
\et{-\isk\delta/2},
\end{equation}
so the spin-vector precesses in the plane of incidence through an
angle $\delta^\uparrow - \delta^\downarrow$.  If $\delta^\uparrow$ and
$\delta^\downarrow$ are defined for the asymptotic (free) states then
this result for the spin precession is general.

To find the precession angle for the case of a single step we return
to the formulae~\eqref{6prec1} and~\eqref{6prec2} and write
\begin{align}
\et{j\delta} &= r^\uparrow {r^\downarrow}^* \nn \\
&= \frac{(\tanh(u^+/2) + \tanh(u/2) j\et{j\phi})(\tanh(u^-/2) +
\tanh(u/2) j \et{j\phi})} {(\tanh(u^+/2) - \tanh(u/2) j
\et{-j\phi})(\tanh(u^-/2) - \tanh(u/2) j \et{-j\phi})} 
\end{align}
If we now recall that
\begin{equation}
\tanh(u/2)= p/(E+m), \hs{1} \tanh(\frac{u^\pm}{2}) = \frac{p_y \pm
\kappa}{E-eV+m},
\end{equation}
we find that
\begin{equation}
\et{j\delta} = \et{2j\phi} \frac{m \cos\!\phi - jE \sin\!\phi}{m
\cos\!\phi + jE\sin\!\phi}.
\label{6prec5}
\end{equation}
The remarkable feature of this result is that all dependence on the
height of the barrier has vanished, so that the precession angle is
determined solely by the incident energy and direction.  To proceed we write
\begin{equation}
m \cos\!\phi - jE \sin\!\phi = \rho \et{j\alp}
\end{equation}
so that
\begin{equation}
\tan\!\alp = - \cosh\!u \tan\!\phi.
\end{equation}
Equation~\eqref{6prec5} now yields
\begin{equation}
\tan(\delta/2 - \phi) = - \cosh\!u \tan\!\phi,
\end{equation}
from which we obtain the final result that
\begin{equation}
\tan(\delta/2) = - \frac{(\cosh\!u -1) \tan\!\phi}{1+\cosh\!u
\tan^2\!\phi}.
\label{6prec-ang}
\end{equation}
A similar result for the precession angle of the rest-spin vector was
obtained by Fradkin \& Kashuba~\cite{frad74} using standard
techniques.  Readers are invited to compare their derivation with the
present approach.  The formula~\eqref{6prec-ang} agrees with
equation~\eqref{5prec_ang} from Section~\ref{Ss-sct}, since the angle
$\theta$ employed there is related to the angle of incidence $\phi$ by
\begin{equation}
\theta = \pi - 2 \phi.
\end{equation}

Since the decomposition of the plane-wave spinor into a boost term and
a Pauli spinor term is unique to the STA, it is not at all clear how
the conventional approach can formulate the idea of the rest-spin.  In
fact, the rest-spin vector is contained in the standard approach in
the form of the `polarisation operator'~\cite{frad74,ros-ret} which,
in the STA, is given by
\begin{equation}
\hat{O}(\bn) = - \frac{j}{\bpht^2} [i \bpht \bpht \dt \bn + \hgo
i (\bn \wdg \bpht) \dt \bpht]
\end{equation}
where $\bn$ is a unit spatial vector.  This operator is Hermitian,
squares to 1 and commutes with the free-field Hamiltonian.  If we
consider a free-particle plane-wave state, then the expectation value of
the $\hat{O}(\sig_i)$ operator is
\begin{align}
\frac{\la \psidag \hat{O}(\sig_i) \psi \ra}{\la \psidag \psi \ra}
&= \frac{m}{E \bp^2} \la \tilde{\Phi} L(\sig_i \dt \bp \bp L \Phi \sk
+ \sig_i \wdg \bp \bp \Lrev \Phi \sk) \ra \nn \\
&=  \frac{m}{E \bp^2} [ L(\sig_i \dt \bp \bp L + \sig_i \wdg \bp \bp
\Lrev)] \dt \bs
\label{6prec7}
\end{align}
where $L$ is the boost 
\begin{equation}
L(\bp) = \frac{E+m+ \bp}{\sqrt{2m(E+m)}}
\end{equation}
and $\bpht$ is replaced by its eigenvalue $\bp$.  To manipulate
equation~\eqref{6prec7} we use the facts that $L$ commutes with $\bp$ and
satisfies
\begin{equation}
L^2 = (E+\bp)/m
\end{equation}
to construct
\begin{align}
\frac{m}{E \bp^2} [ L(\sig_i \dt \bp \bp L + \sig_i \wdg \bp
\bp \Lrev)] 
&= \frac{m}{E \bp^2} [ \sig_i \dt \bp \frac{(E+\bp)}{m} \bp -
\sig_i\dt\bp \bp  \nn \\
& + \frac{E-m}{2m}(E+m+\bp)\sig_i(E+m-\bp) ].
\end{align}
Since only the relative vector part of this quantity is needed in
equation~\eqref{6prec7} we are left with
\begin{align}
\frac{E-m}{2E \bp^2} (2 \sig_i \dt \bp \bp + (E+m)^2 \sig_i
- \bp \sig_i \bp)
&= \frac{1}{2E(E+m)} (\bp^2 \sig_i + (E+m)^2\sig_i) \nn \\
&= \sig_i.
\end{align}
The expectation value of the `polarisation' operators is therefore
simply 
\begin{equation}
\frac{\la \psidag \hat{O}(\sig_i) \psi \ra}{\la \psidag \psi \ra} =
\sig_i \dt \bs,
\end{equation}
which just picks out the components of the rest-spin vector, as
claimed.  

For the case of the potential step, $\hat{O}(\bn)$ still commutes with
the full Hamiltonian when $\bn$ is perpendicular to the plane of
incidence.  In their paper Fradkin~\& Kashuba decompose the incident
and reflected waves into eigenstates of $\hat{O}(\bn)$, which is
equivalent to aligning the spin in the manner adopted in this section.
As we have stressed, removing the boost and working directly with
$\Phi$ simplifies many of these manipulations, and removes any need
for the polarisation operator.

\subsection{Tunnelling of Plane Waves}
\label{6S-tunn}

\begin{figure}
\begin{center}
\begin{picture}(340,120)(128,365)
\put(128,365){\hbox{\psfig{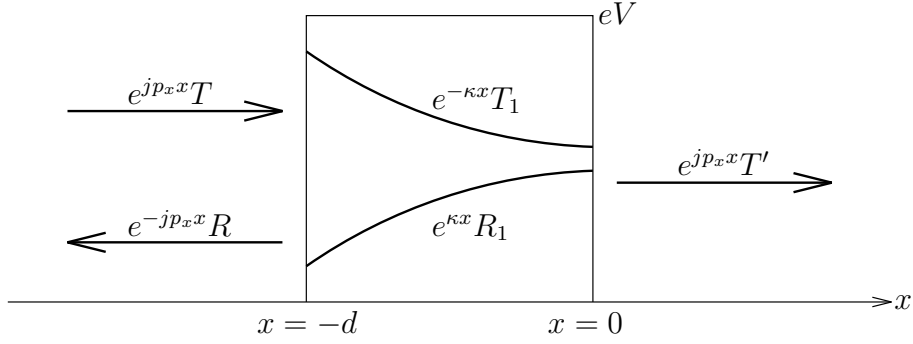}}}
\put(222,354){\large $x=-d$}
\put(332,354){\large $x=0$}
\put(463,363){\large $x$}
\put(173,391){\large $e^{-jp_xx} R$}
\put(173,441){\large $e^{jp_xx}T$}
\put(288,439){\large $e^{-\kappa x}T_1$}
\put(288,391){\large $e^{\kappa x}R_1$}
\put(381,414){\large $e^{jp_x x} T'$}
\put(351,471){\large $eV$}
\end{picture}
\end{center}
\caption[dummy1]{\sl Schematic representation of plane-wave
tunnelling.}  
\label{6fig-plnwv-tunn}
\end{figure}

Suppose now that a continuous beam of plane waves is incident on a
potential barrier of finite width.  We know that,
quantum-mechanically, some fraction of the wave tunnels through to the
other side.  In Section~\ref{S-tunn} we address the question `how long
does the tunnelling process take?'.  To answer this we will need to
combine plane-wave solutions to construct a wavepacket, so here we
give the results for plane waves.  The physical set-up is illustrated
in Figure~\ref{6fig-plnwv-tunn}.

The matching equation at the $x=0$ boundary is, for spin-up, 
\begin{equation}
\bB^+ \begin{pmatrix}
\Tup_1 \\ \Rup_1 \end{pmatrix}
=
\bA \begin{pmatrix}
{\Tup}' \\ 0  \end{pmatrix},
\label{6tunn2}
\end{equation}
where the $\bA$ and $\bB^+$ matrices are as defined in~\eqref{6def-bA}
and~\eqref{6def-bB} respectively.  All subscripts can be dropped again as
$\bA$ always refers to free space and $\bB^+$ to the barrier region.
The matching conditions at $x=-d$ require the inclusion of suitable
propagators and the resulting equation is
\begin{equation}
\bA \begin{pmatrix}
\et{-jdp_x} & 0 \\
0 & \et{jdp_x}
\end{pmatrix}
\begin{pmatrix}
\Tup \\ \Rup   \end{pmatrix}
= \bB^+ \begin{pmatrix}
\et{\kap d} & 0 \\
0 & \et{-\kap d}
\end{pmatrix}
\begin{pmatrix}
\Tup_1 \\ \Rup_1 \end{pmatrix}.
\label{6tunn3}
\end{equation}
Equation~\eqref{6tunn3} shows that the relevant propagator matrix for
evanescent waves is
\begin{equation}
\bP = \begin{pmatrix}
\et{\kap d} & 0 \\
0 & \et{-\kap d}
\end{pmatrix}.
\end{equation}
The problem now reduces to the matrix equation
\begin{equation}
\begin{pmatrix}
\et{-jdp_x} \Tup \\ \et{jdp_x} \Rup
\end{pmatrix}
=
\left[ \cosh(\kappa d) + \sinh(\kappa d) 
\bA^{-1} \bB^+ 
\begin{pmatrix}
1 & 0 \\
0 & -1 \end{pmatrix}
{\bB^+}^{-1} \bA  \right] 
\begin{pmatrix}
{\Tup}' \\ 0 \end{pmatrix},
\label{6tunn4}
\end{equation}
from which the reflection and transmission coefficients are easily
obtained.

The applications to tunnelling discussed in Section~\ref{S-tunn} deal
mainly with perpendicular incidence, so we now specialise to this
situation.  For perpendicular incidence we can set $u^+=-u^-=u'$,
where
\begin{equation}
\tanh\!u' = \frac{\kap}{E'+m}.
\end{equation}
It follows that the equations for spin-up and spin-down are the same,
and we can remove the up-arrows from the preceding equations.
Equation~\eqref{6tunn2} now yields
\begin{equation}
\begin{pmatrix}
T_1 \\ R_1 \end{pmatrix} 
= \frac{T'}{\sinh\!u'} \begin{pmatrix}
\sinh(u'/2)\cosh(u/2) - j \cosh(u'/2)\sinh(u/2) \\
\sinh(u'/2)\cosh(u/2) + j \cosh(u'/2)\sinh(u/2)
\end{pmatrix}
\label{6tunn-coeffs}
\end{equation}
and from $T_1$ and $R_1$ the current in the evanescent region can be
constructed.  The ratio $J_1/J_0$ may be interpreted as defining a
`velocity' inside the barrier.  The consequences of this idea were
discussed in~\cite{DGL93-paths} where it was concluded that the
tunnelling times predicted by this velocity are not related to
tunnelling times measured for individual particles.  The reasons for
this are discussed in Section~\ref{S-tunn}.

Multiplying out the matrices in equation~\eqref{6tunn4} is
straightforward, and yields
\begin{equation}
\begin{pmatrix}
\et{-jdp} T \\ \et{jdp} R
\end{pmatrix}
=
T' \begin{pmatrix}
\cosh(\kap d) -j \sinh(\kap d) (EE' - m^2) / (\kap p)  \\ 
-j \sinh(\kap d) eV m / (\kap p) 
\end{pmatrix},
\end{equation}
which solves the problem.  The transmission coefficient is
\begin{equation}
t = \frac{\kap p \et{-jdp}}{\kap p \cosh(\kap d) -j (p^2- eV E)
\sinh(\kap d) }
\end{equation}
which recovers  the familiar non-relativistic formula in the limit
$E\approx m$.

\subsection{The Klein Paradox}

In the Klein region, $eV-E>\sqrt{p_y^2+m^2}$, travelling wave
solutions exist again.  To find these we observe that plane-wave
solutions must now satisfy
\begin{equation}
(p-eV \go) \psi = m \psi \go
\end{equation}
and, as $p-eV \go$ has a negative time component, $\psi\psirev$ must
now be $-1$.  We could achieve this flip by inserting a
`$\beta$-factor' of the type described in Section~\eqref{Ss3-Deqn}, but
this would mix the rest-spin states.  It is more convenient to work
with solutions given by
\begin{align}
\psi^t &= [\cosh(u/2) + \sinh(u/2) (\cos\!\phi \si - \sin\!\phi \sj)]
\si \Phi \et{-i\sk(Et +p_x x - p_y y)} T \label{6kli1} \\
\psi^r &= [\cosh(u/2) + \sinh(u/2) (-\cos\!\phi \si - \sin\!\phi \sj)]
\si \Phi \et{-i\sk(Et -p_x x - p_y y)} R ,  \label{6kli2}
\end{align}
where the choice of $\si$ or $\sj$ on the right-hand side of the boost
is merely a phase choice.  To verify that $\psi^t$ is a solution we
write the Dirac equation as
\begin{align}
\lefteqn{ [(E-eV)\go -p_x \gi + p_y \gj] \et{(\cos\!\phi \si -
\sin\!\phi \sj)u/2} \si } \hs{5} \nn \\
& & = m  \et{(\cos\!\phi \si - \sin\!\phi \sj)u/2} \si \go
\end{align}
which holds provided that
\begin{equation}
\tanh(u/2) = \frac{p}{m+eV-E}.
\end{equation}
It follows that
\begin{equation}
m \cosh\!u = eV- E, \hs{1} m \sinh\!u = p.
\end{equation}

The current obtained from $\psi^t$ is found to be
\begin{equation}
\psi^t \go \psirev^t = (eV-E)\go +p_x \gi - p_y \gj
\end{equation}
which is future pointing (as it must be) and points in the
positive-$x$ direction.  It is in order to obtain the correct
direction for the current that the sign of $p_x$ is changed
in~\eqref{6kli1} and~\eqref{6kli2}.  As has been pointed out by various
authors~\cite{gul-steps,man88,gre-strong}, some texts on quantum
theory miss this argument and match onto a solution inside the barrier
with an incoming group velocity~\cite{bjo-rel1,itz-quant}.  The result
is a reflection coefficient greater than 1.  This is interpreted as
evidence for pair production, though in fact the effect is due to the
choice of boundary conditions.  To find the correct reflection and
transmission coefficients for an outgoing current, we return to the
matching equation which, for spin-up, gives
\begin{multline}
\begin{array}{l}
[\cosh(u_i/2) + \sinh(u_i/2) \si \et{\phi_i \isk}] \Tup_i \\ 
+ \, [\cosh(u_i/2) - \sinh(u_i/2) \si \et{-\phi_i \isk}] \Rup_i 
\end{array} \\
\begin{array}{l}
= \, [\cosh(u_{i+1}/2) + \sinh(u_{i+1}/2) \si \et{-\phi_{i+1}
\isk}] \si \Tup_{i+1} \\
\quad + \, [\cosh(u_{i+1}/2) - \sinh(u_{i+1}/2) \si \et{\phi_{i+1}
\isk}] \si \Rup_{i+1}. 
\end{array}
\label{6kli4}
\end{multline}
This time we define the matrix
\begin{equation}
\bC_i = 
\begin{pmatrix}
\sinh(u_i/2) \et{j \phi_i}  & - \sinh(u_i/2) \et{-j \phi_i} \\
\cosh(u_i/2)       &   \cosh(u_i/2) 
\end{pmatrix}
\end{equation}
so that equation~\eqref{6kli4} becomes
\begin{equation}
\bA_i \begin{pmatrix}
\Tup_i \\ \Rup_i   \end{pmatrix}
= \bC_{i+1} \begin{pmatrix}
\Tup_{i+1} \\ \Rup_{i+1}   \end{pmatrix}.
\end{equation}
It should be noted that
\begin{equation}
\bC_i = \begin{pmatrix} 
0 & 1 \\
1 & 0 \end{pmatrix} \bA_i.
\end{equation}
The corresponding equation for spin-down is simply
\begin{equation}
\bA_i^* \begin{pmatrix}
\Tdn_i \\ \Rdn_i   \end{pmatrix}
= \bC_{i+1}^* \begin{pmatrix}
\Tdn_{i+1} \\ \Rdn_{i+1}   \end{pmatrix}.
\end{equation}

The Klein `paradox' occurs at a single step, for which the matching
equation is
\begin{equation}
\bA_1  \begin{pmatrix}
\Tup_1 \\ \Rup_1   \end{pmatrix}
= \bC_2  \begin{pmatrix}
\Tup_2 \\ 0   \end{pmatrix}.
\end{equation}
Inverting the $\bA_1$ matrix yields
\begin{equation}
\begin{pmatrix}
\Tup_1 \\ \Rup_1   \end{pmatrix}
= \frac{\Tup_2}{\sinh\!u \cos\!\phi} 
\begin{pmatrix} 
\cosh(u+u')/2 \\ 
\sinh(u/2) \sinh(u'/2) \et{2j\phi} - \cosh(u/2) \cosh(u'/2) 
\end{pmatrix}
\end{equation}
from which the reflection and transmission coefficients can be read
off.  (The primed quantities relate to the barrier region, as usual.)
In particular, for perpendicular incidence, we recover
\begin{equation}
r = -\frac{\cosh(u-u')/2}{\cosh(u+u')/2}, \hs{1} t =
\frac{\sinh\!u}{\cosh(u+u')/2}
\end{equation}
as found in~\cite{gul-steps}.  The reflection coefficient is always
$\leq 1$, as it must be from current conservation with these boundary
conditions.  But, although a reflection coefficient $\leq 1$ appears
to ease the paradox, some difficulties remain.  In particular, the
momentum vector inside the barrier points in an opposite direction to
the current.

A more complete understanding of the Klein barrier requires quantum
field theory since, as the barrier height is $>2m$, we expect pair
creation to occur.  An indication that this must be the case comes
from an analysis of boson modes based on the Klein-Gordon equation.
There one finds that superradiance ($r>1$) does occur, which has to be
interpreted in terms of particle production.  For the fermion case the
resulting picture is that electron-positron pairs are created and
split apart, with the electrons travelling back out to the left and
the positrons moving into the barrier region.  If a single electron is
incident on such a step then it is reflected and, according to the
Pauli principle, the corresponding pair-production mode is suppressed.

A complete analysis of the Klein barrier has been given by
Manogue~\cite{man88} to which readers are referred for further
details.  Manogue concludes that the fermion pair-production rate is
given by
\begin{equation}
\Gamma = \int \! \frac{d^2\bk}{(2\pi)^2} \int \! \frac{d\om}{2\pi} \,
\frac{\sinh\!u'}{\sinh\!u} \, \sum_i |T^i|^2 
\label{6kli7}
\end{equation}
where the integrals run over the available modes in the Klein region,
and the sum runs over the two spin states.  This formula gives a
production rate per unit time, per unit area, and applies to any shape
of barrier.  The integrals in~\eqref{6kli7} are not easy to evaluate, but
a useful expression can be obtained by assuming that the barrier
height is only slightly greater than $2m$,
\begin{equation}
eV = 2m(1 + \eps).
\end{equation}
Then, for the case of a single step, we obtain a pair-production rate of
\begin{equation}
\Gamma = \frac{\pi m^3 \eps^3}{32},
\end{equation}
to leading order in $\eps$.  The dimensional term is $m^3$ which, for
electrons, corresponds to a rate of $10^{48}$ particles per second,
per square meter.  Such an enormous rate would clearly be difficult to
sustain in any physically-realistic situation!

The results obtained in this section are summarised in
Table~\ref{6tab-summ}.

\begin{table}
\renewcommand{\arraystretch}{1.2}
\begin{tabular}{c}
\hline \hline
\\
Travelling Waves \\
\begin{minipage}[c]{10.5cm}
\fbox{
\begin{tabular}{c}
$\psi^t = [\cosh(u/2) + \sinh(u/2)\si \et{\phi \isk} ] \Phi
\et{-\isk(Et - p_xx -p_yy)}$ \\
$\psi^r = [\cosh(u/2) - \sinh(u/2)\si \et{-\phi \isk} ] \Phi
\et{-\isk(Et + p_xx -p_yy)}$ \\
$\tanh(u/2)= p/(E-eV+m)$
\end{tabular} } 
\end{minipage} \\
\\
Evanescent Waves \\
\begin{minipage}[c]{10cm}
\fbox{ 
\begin{tabular}{c}
$ \psi^t = [\cosh(u^\pm/2) + \sinh(u^\pm/2)\sj] \et{-\kap x} \et{-\isk(Et
-p_yy)} T$ \\
$ \psi^r = [\cosh(u^\mp/2) + \sinh(u^\mp/2)\sj] \et{\kap x} \et{-\isk(Et
-p_yy)} R $ \\
(upper/lower signs $=$ spin up/down) \\
$\tanh(u^\pm/2)= (p_y \pm \kap)/(E-eV+m)$
\end{tabular} }
\end{minipage} \\
\\
Klein Waves \\
\begin{minipage}[c]{11cm}
\fbox{ 
\begin{tabular}{c}
$\psi^t = [\cosh(u/2) + \sinh(u/2)\si \et{-\phi \isk} ] \si \Phi
\et{-\isk(Et + p_xx -p_yy)}$ \\
$\psi^r = [\cosh(u/2) - \sinh(u/2)\si \et{\phi \isk} ] \si \Phi
\et{-\isk(Et - p_xx -p_yy)}$ \\
$\tanh(u/2)= p/(m +eV-E)$
\end{tabular} }
\end{minipage} \\
\\
Matching Matrices \\
\begin{minipage}[c]{7.5cm}
\fbox{ 
\begin{tabular}{c}
\( \bA =  \begin{pmatrix}
\cosh(u/2)       &\cosh(u/2) \\
\sinh(u/2) \et{j \phi}  & - \sinh(u/2) \et{-j \phi}
\end{pmatrix} \) \\
\( \bB^+ = \begin{pmatrix}
\cosh(u^+/2)    & \cosh(u^-/2) \\
j\sinh(u^+/2)    & +j\sinh(u^-/2)
\end{pmatrix} \) \\
\( \bC = \begin{pmatrix}
\sinh(u/2) \et{j \phi}  & - \sinh(u/2) \et{-j \phi} \\
\cosh(u/2)       &   \cosh(u/2) 
\end{pmatrix} \) \\
$\bA^*$, $\bB^{-^*}$, $\bC^*$ for spin down.
\end{tabular}
 }
\end{minipage} \\
\\
Propagators \\
\begin{minipage}[c]{6.5cm}
\fbox{ \(
\begin{pmatrix}
\et{jdp_x} & 0 \\
0 & \et{-jdp_x}
\end{pmatrix}, \hs{0.5} 
\begin{pmatrix}
\et{\kap d} & 0 \\
0 & \et{-\kap d}
\end{pmatrix} 
\) }
\end{minipage}\\
\\
\hline \hline
\end{tabular}
\caption[dummy1]{\sl Summary of results for plane waves incident on a
potential step.  The waves travel in the $x-y$ plane and the steps lie
in the $y-z$ plane.   The matching matrices relate $T$ and $R$ on either
side of a step.}
\label{6tab-summ}
\end{table}
\section{Tunnelling Times}
\label{S-tunn}

In this Section we study tunnelling phenomena.  We do so by setting up
a wavepacket and examining its evolution as it impinges on a potential
barrier.  The packet splits into reflected and transmitted parts, and
the streamlines of the conserved current show which parts of the
initial packet end up being transmitted.  The analysis can be used to
obtain a distribution of arrival times at some fixed point on the far
side of the barrier, which can be compared directly with experiment.
The bulk of this section is concerned with packets in one spatial
dimension, and compares our approach to other studies of the
tunnelling-time problem.  The section ends with a discussion of the
complications introduced in attempting 2- or 3-dimensional
simulations.

The study of tunnelling neatly combines the solutions found in
Section~\ref{S-steps} with the views on operators and the
interpretation of quantum mechanics expressed in
Section~\ref{S-spinors}.  Tunnelling also provides a good illustration
of how simple it is study electron physics via the Dirac theory once
the STA is available.

\subsection{Wavepacket Tunnelling}

In Section~\ref{6S-tunn} we studied tunnelling of a continuous plane
wave through a potential barrier.  It was found that the growing and
decaying waves in the barrier region are given by equations~\eqref{6ev-up}
for spin-up and~\eqref{6ev-down} for spin-down.  Restricting to the case
of perpendicular incidence, the amplitudes of the reflected and
transmitted waves are given by equation~\eqref{6tunn-coeffs}.  It follows
that, for arbitrary spin, the wavefunction in the barrier region is
\begin{align}
\psi_1 &= [\cosh(u'/2) \Phi + \sinh(u'/2)\sj\sk\Phi\sk] \et{-\kap x}
\et{-\isk Et} \alp + \nn \\
& [\cosh(u'/2) \Phi - \sinh(u'/2)\sj\sk\Phi\sk] \et{\kap x}
\et{-\isk Et} \alp^*
\end{align}
where
\begin{equation}
\alp = \frac{T'}{\sinh\!u'} [\sinh(u'/2)\cosh(u/2) - \isk
\cosh(u'/2)\sinh(u/2) ]
\end{equation}
and
\begin{equation}
\tanh(u'/2) = \frac{\kap}{E'+m} = \frac{\kap}{E-eV+m}, \hs{1}
\kap^2=m^2-{E'}^2.
\end{equation}

\begin{figure}[t!]
\centerline{
\hbox{\psfig{figure=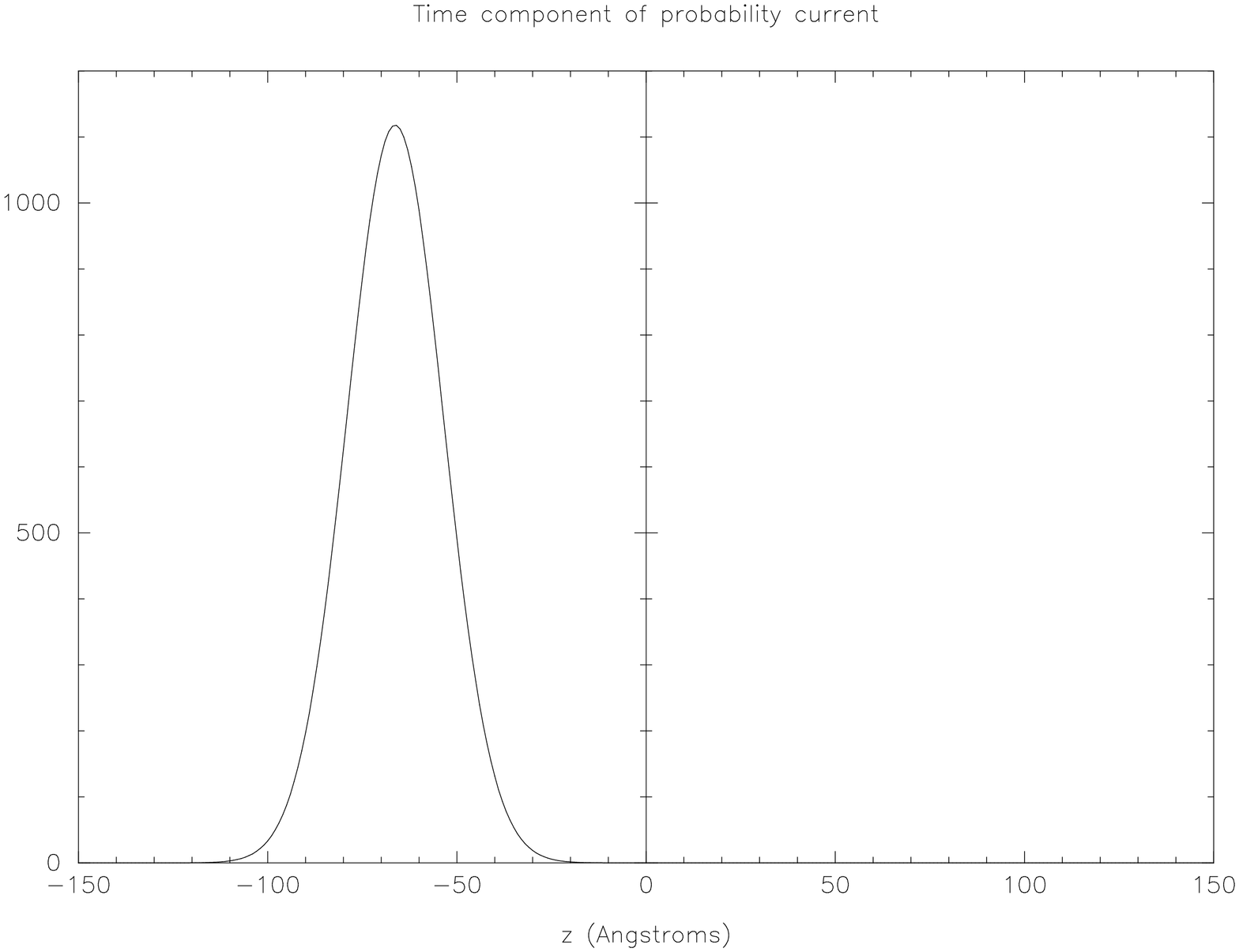,height=5.8cm,angle=-90}}
\hs{-0.2}  
\hbox{\psfig{figure=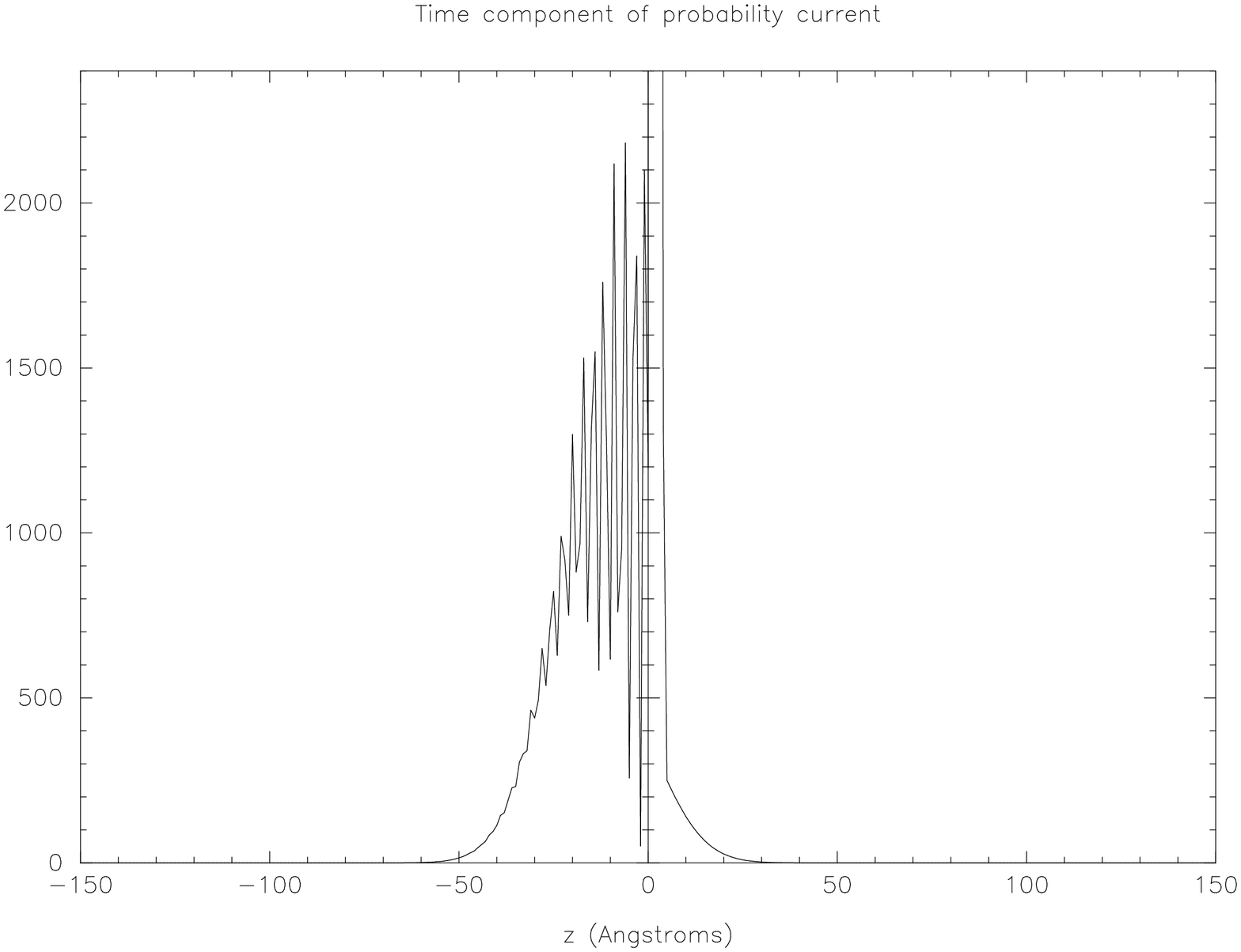,height=5.8cm,angle=-90}} }
\centerline{
\hbox{\psfig{figure=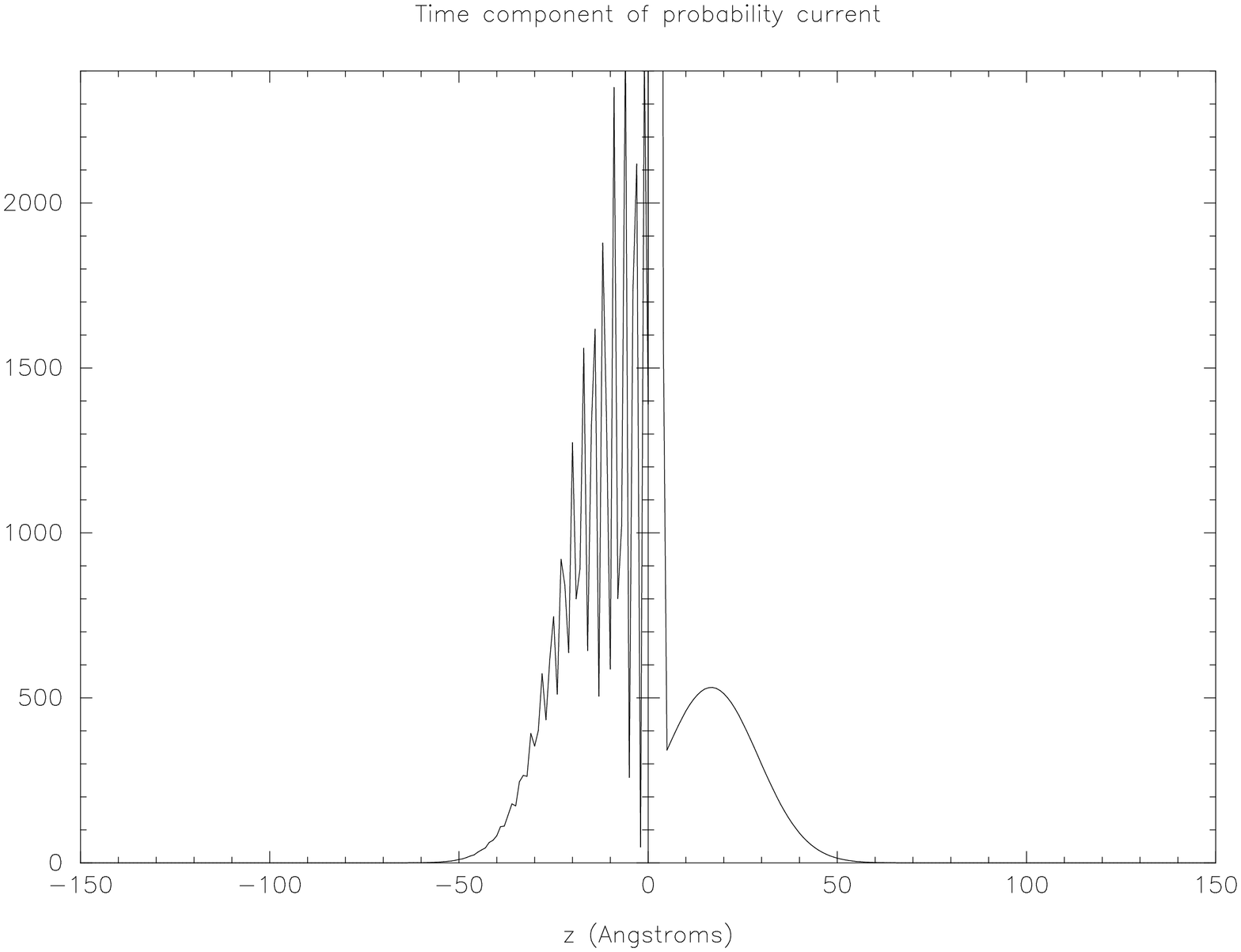,height=5.8cm,angle=-90}}
\hs{-0.2}  
\hbox{\psfig{figure=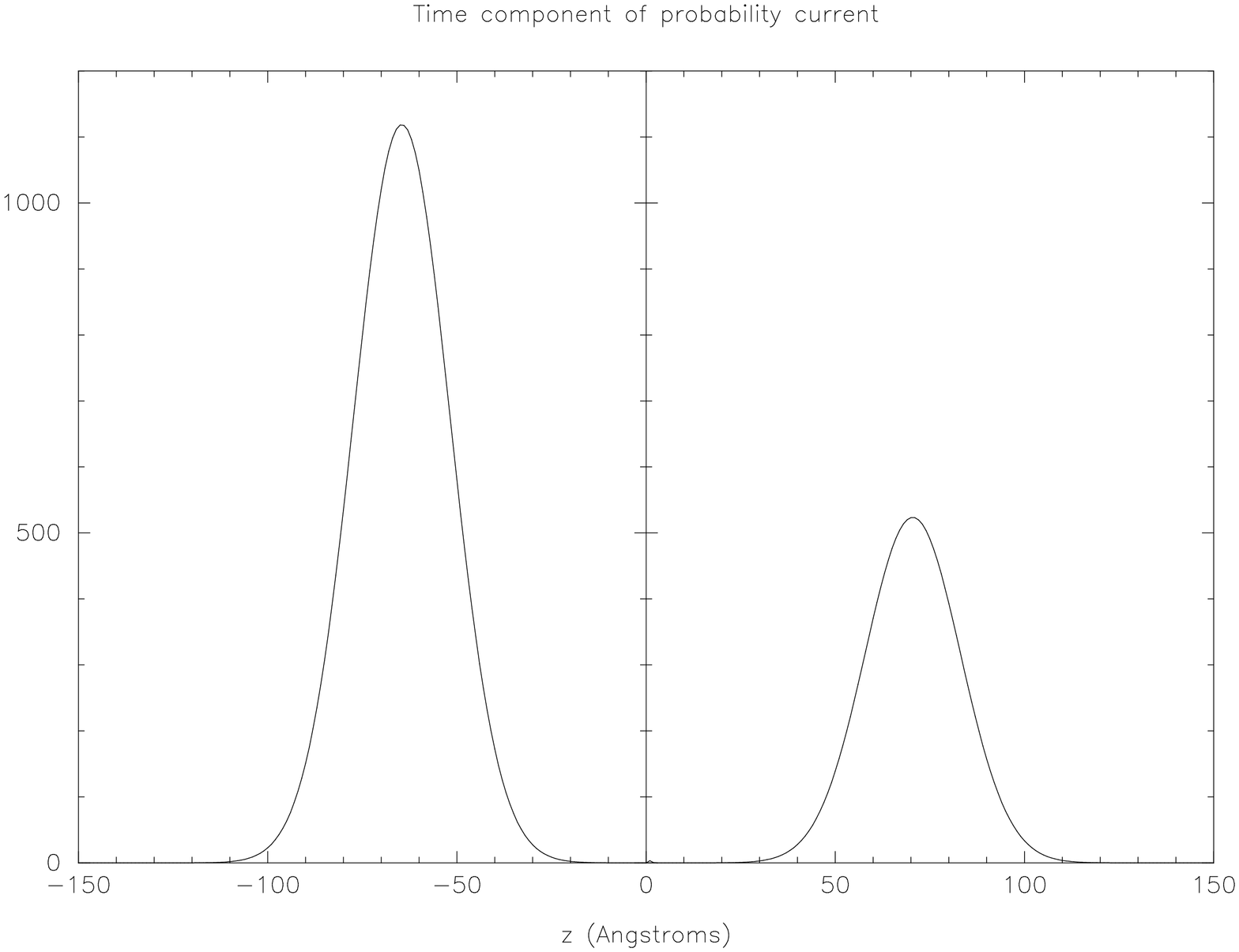,height=5.8cm,angle=-90}} }
\caption[dummy1]{\sl Evolution of the density $J^0$ as a function of
time.  The initial packet is a Gaussian of width $\Delta
k=0.04$\AA$^{-1}$ and energy 5eV.  The barrier starts at the origin
and has width 5{\AA} and height 10eV.  The top line shows the
density profile at times $-0.5\times 10^{-14}$s and $-0.1\times
10^{-14}$s, and the bottom line shows times $0.1\times 10^{-14}$s and
$0.5\times 10^{-14}$s.  In all plots the vertical scale to the right
of the barrier is multiplied by $10^4$ to enhance the features of the
small, transmitted packet.}
\label{7fig-tunn1}
\end{figure}

The current in the barrier region is
\begin{align}
\psi_1\go\psirev_1 &= \frac{|T'|^2}{m\kap^2} [ m^2 eV \cosh(2\kap x)
+E'(p^2-Eev) \nn \\
& + p\kap^2 \si -m\kap eV \sinh(2\kap x) (i\si)\dt\bs ] \go
\label{7curr}
\end{align}
from which we can define a `velocity'
\begin{equation}
\frac{dx}{dt} = \frac{J\dt\gam^1}{J\dt\go} = \frac{p\kap^2}{ m^2
eV \cosh(2\kap x) +E'(p^2-Eev) }.
\label{7vel}
\end{equation}
In fact, the velocity~\eqref{7vel} does not lead to a sensible definition
of a tunnelling time for an individual particle~\cite{DGL93-paths}.
As we shall see shortly, an additional phenomenon underlies wavepacket
tunnelling, leading to much shorter times than those predicted
from~\eqref{7vel}.  To study wavepacket tunnelling it is useful,
initially, to simplify to a one-dimensional problem.  To achieve this
we must eliminate the transverse current in~\eqref{7curr} by setting
$\bs=\pm\si$.  This is equivalent to aligning the spin vector to point
in the direction of motion.  (In this case there is no distinction
between the laboratory and comoving spin.)  With $\Phi$ chosen so that
$\bs=\si$ it is a now a simple matter to superpose solutions at $t=0$
to construct a wavepacket centred to the left of the barrier and
moving towards the barrier.  The wavepacket at later times is then
reassembled from the plane-wave states, whose time evolution is
known.  The density $J^0=\go\dt J$ can then be plotted as a function
of time and the result of such a simulation is illustrated in
Figure~\ref{7fig-tunn1}.

\begin{figure}[t!]
\centerline{
\hbox{\psfig{figure=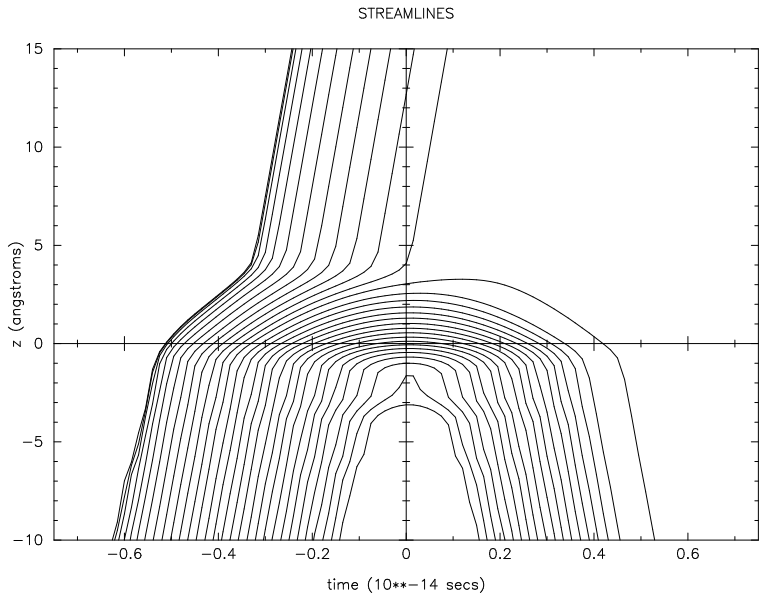,height=11cm,angle=-90}} }
\caption[dummy1]{\sl Particle streamlines for the packet evolution
shown in Figure~\ref{7fig-tunn1}.  Only the streamlines from the very
front of the packet cross the barrier, with the individual streamlines
slowing down as they pass through.}
\label{7fig-streamlines}
\end{figure}

The Dirac current $J=\psi\go\psirev$ is conserved even in the presence
of an electromagnetic field.  It follows that $J$ defines a set of
streamlines which never end or cross.  Furthermore, the time-component
of the current is positive-definite so the tangents to the streamlines
are always future-pointing timelike vectors.  According to the
standard interpretation of quantum mechanics, $J^0(\bx,t)$ gives the
probability density of locating a particle at position $\bx$ at time
$t$.  But, considering a flux tube defined by adjacent streamlines, we
find that 
\begin{equation}
\rho(t_0,x_0) dx_0 = \rho(t_1,x_1) dx_1
\end{equation}
where $(t_0,x_0)$ and $(t_1,x_1)$ are connected by a streamline.  It
follows that the density $J^0$ flows along the streamlines without
`leaking' between them.  So, in order to study the tunnelling process,
we should follow the streamlines from the initial wavepacket through
spacetime.  A sample set of these streamlines is shown in
Figure~\ref{7fig-streamlines}.  A significant feature of this plot is
that a continuously-distributed set of initial input conditions has
given rise to a disjoint set of outcomes (whether or not a streamline
passes through the barrier).  Hence the deterministic evolution of the
wavepacket alone is able to explain the discrete results expected in a
quantum measurement, and all notions of wavefunction collapse are
avoided.  This is of fundamental significance to the interpretation of
quantum mechanics.  Some consequences of this view for other areas of
quantum measurement have been explored by Dewdney \etal~\cite{dew88}
and Vigier \etal~\cite{vig-causal}, though their work was founded in
the Bohmian interpretation of non-relativistic quantum mechanics.  The
results presented here are, of course, independent of any
interpretation --- we do not need the apparatus of Bohm/de Broglie
theory in order to accept the validity of predictions obtained from
the current streamlines.

The second key feature of the streamline plot in
Figure~\ref{7fig-streamlines} is that it is only the streamlines
starting near the front of the initial wavepacket that pass through
the barrier.  Relative to the centre of the packet, they therefore
have a `head start' in their arrival time at some chosen point on the
far side of the barrier.  Over the front part of the barrier, however,
the streamlines slow down considerably, as can be seen by the change
in their slope.  These two effects, of picking out the front end of
the packet and then slowing it down, compete against each other and it
is not immediately obvious which dominates.  To establish this, we
return to Figure~\ref{7fig-tunn1} and look at the positions of the
wavepacket peaks.  At $t=0.5\times 10^{-14}$s, the peak of the
transmitted packet lies at $x=70${\AA}, whereas the peak of the
initial packet would have been at $x=66${\AA} had the barrier not been
present.  In this case, therefore, the peak of the transmitted packet
is slightly advanced, a phenomenon often interpreted as showing that
tunnelling particles speed up, sometimes to velocities greater than
$c$~\cite{chi93}.  The plots presented here show that such an
interpretation is completely mistaken.  There is no speeding up, as
all that happens is that it is only the streamlines from the front of
the wavepacket that cross the barrier (slowing down in the process)
and these reassemble to form a localised packet on the far side.  The
reason that tunnelling particles may be transmitted faster than free
particles is due entirely to the spread of the initial wavepacket.

There is considerable interest in the theoretical description of
tunnelling processes because it is now possible to obtain measurements
of the times involved.  The clearest experiments conducted to date
have concerned photon tunnelling~\cite{ste93}, where an ingenious
2-photon interference technique is used to compare photons that pass
through a barrier with photons that follow an unobstructed path.  The
discussion of the results of photon-tunnelling experiments usually
emphasise packet reshaping, but miss the arguments about the
streamlines.  Thus many articles concentrate on a comparison of the
peaks of the incident and transmitted wavepackets and discuss whether
the experiments show particles travelling at speeds
$>c$~\cite{chi93,lan93}.  As we have seen, a full relativistic study
of the streamlines followed by the electron probability density show
clearly that no superluminal velocities are present.  The same result
is true for photons, as we will discuss elsewhere.

Ever since the possibility of tunnelling was revealed by quantum
theory, people have attempted to define how long the process takes.
Reviews of the various different approaches to this problem have been
given by Hauge~\& St{\o}vneng \cite{hau89} and, more recently, by
Landauer~\& Martin~\cite{lan94}.  Most approaches attempt to
define a single tunnelling time for the process, rather than a
distribution of possible outcomes as is the case here.  Quite why one
should believe that it is possible to define a single time in a
probabilistic process such as tunnelling is unclear, but the view is
still regularly expressed in the modern literature.  A further flaw in
many other approaches is that they attempt to define how long the
particle spent in the barrier region, with answers ranging from the
implausible (zero time) to the utterly bizarre (imaginary time).  From
the streamline plot presented here, it is clearly possible to obtain a
distribution of the times spent in the barrier for the tunnelling
particles, and the answers will be relatively long as the particles
slow down in the barrier.  But such a distribution neglects the fact
that the front of the packet is preferentially selected, and anyway
does not appear to be accessible to direct experimental measurement.
As the recent experiments show~\cite{ste93}, it is the arrival time at
a point on the far side of the barrier that is measurable, and not the
time spent in the barrier.

\subsection{2-Dimensional Simulations}

\begin{figure}[t!]
\begin{center}
\begin{picture}(345,250)
\put(-3,0){\hbox{\psfig{figure=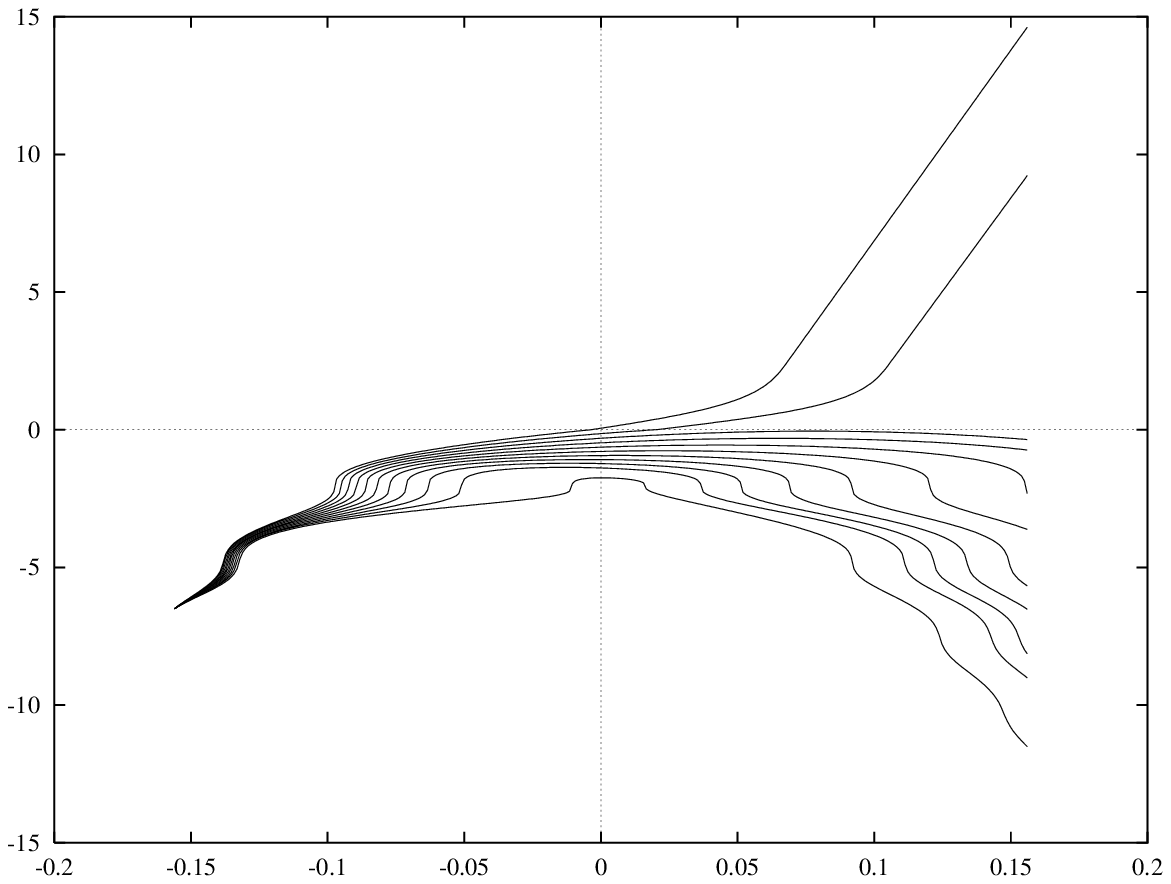,height=4cm}}}
\put(0,59){\scriptsize $x$}
\put(80,-8){\scriptsize $t$}
\put(177,0){\hbox{\psfig{figure=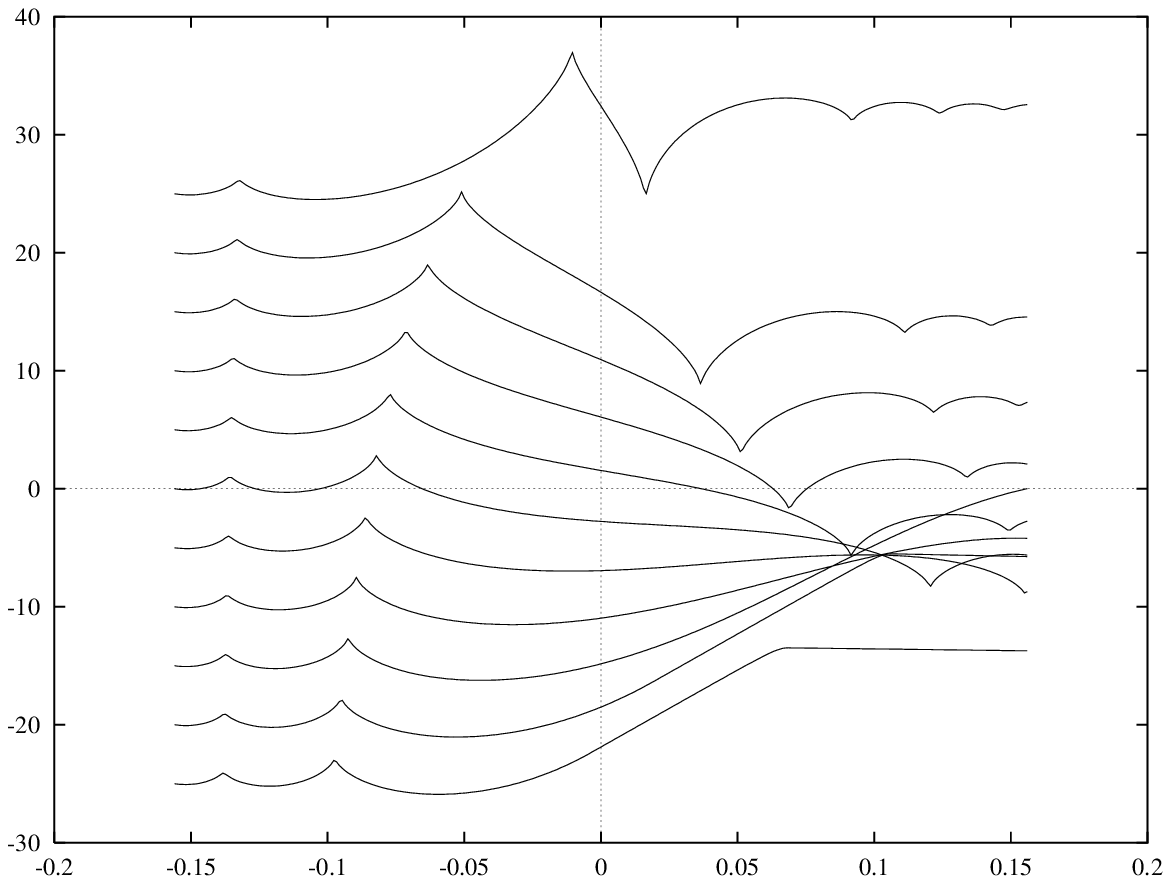,height=4cm}}}
\put(181,59){\scriptsize $y$}
\put(260,-8){\scriptsize $t$}
\put(-6,129){\hbox{\psfig{figure=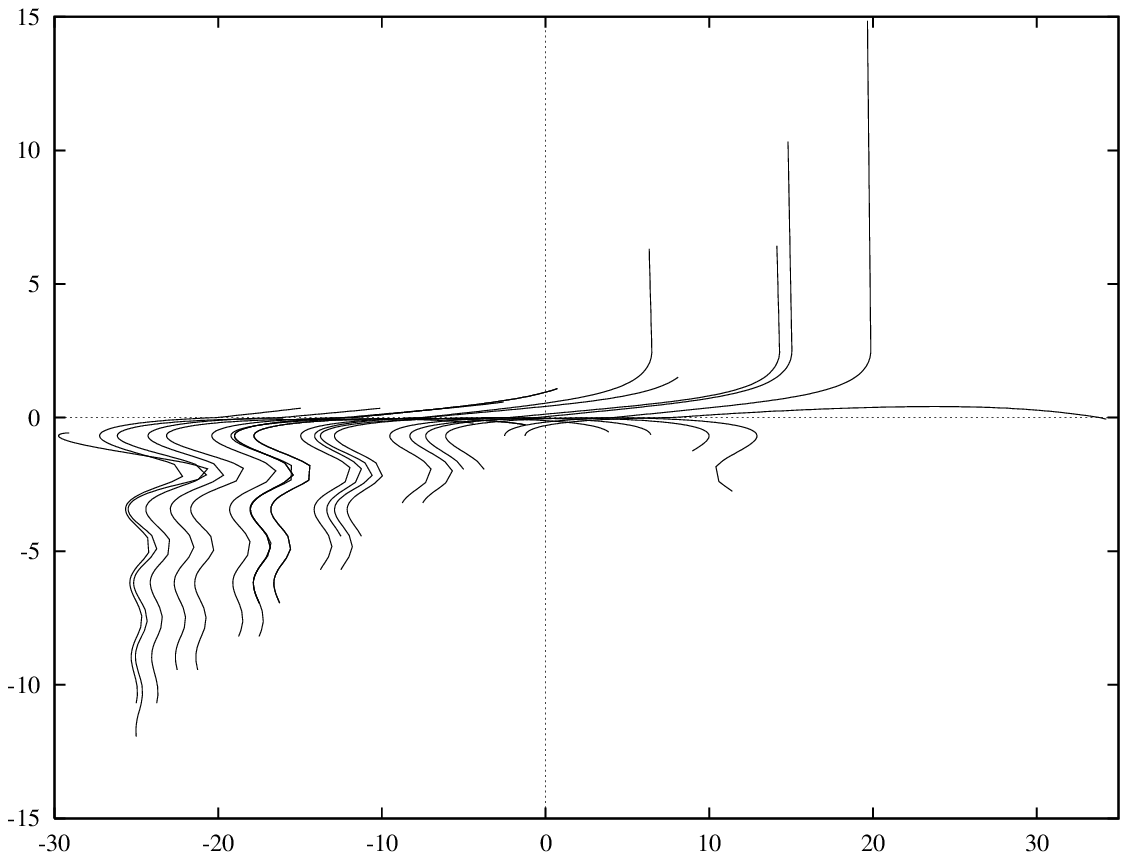,height=4cm}}}
\put(0,201){\scriptsize $x$}
\put(74,127){\scriptsize $y$}
\put(177,131){\hbox{\psfig{figure=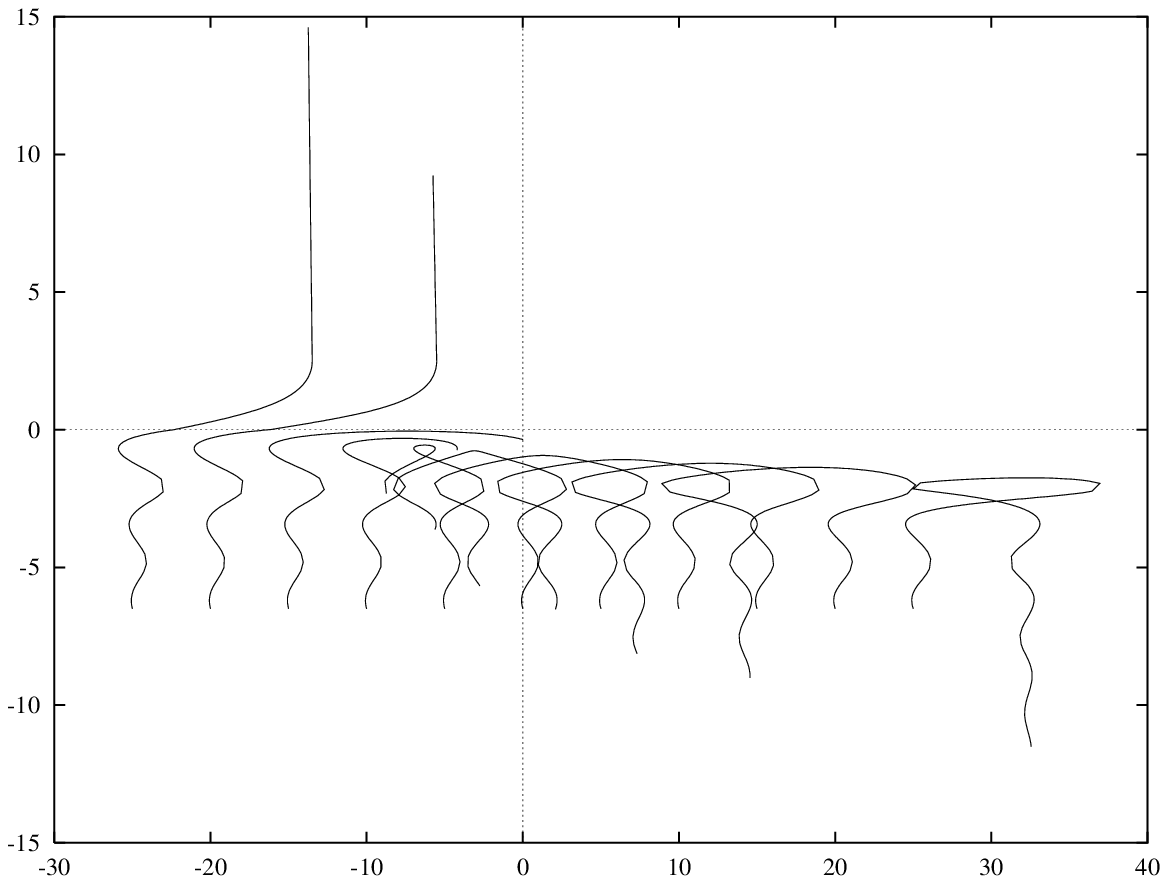,height=4cm}}}
\put(181,201){\scriptsize $x$}
\put(260,127){\scriptsize $y$}
\end{picture}
\end{center}
\caption[dummy1]{\em A 2-Dimensional simulation.  The simulation uses a
2-D wavepacket in the $x$--$y$ plane of width $\Delta
k=0.04$\AA$^{-1}$ and energy 5eV.  The packet is incident
perpendicular to the barrier, and the spin vector lies in the $+$$z$
direction.  The barrier is at $x=0$ and has width $2.5${\AA} and
height 10eV.  The top two plots show streamlines in the $x$--$y$
plane.  The top left plot shows streamlines about the bifurcation
line, illustrating that the left side of the packet, which `spins'
into the barrier, is preferentially transmitted.  The top right plot
shows streamlines for a set of points near the front of the packet
with the same $x$ value.  Again, it is the left side of the packet
that is transmitted.  The reflected trajectories show quite complex
behaviour, and in both plots the effect of the transverse current in
the barrier is clear.  The $t$-dependence of the streamlines in the
top right plot is shown in the bottom two plots.  The left-hand plot
shows the streamlines in $t$--$x$ space.  Since the streamlines were
started from the same $x$ position and different $y$ positions, the
streamlines start from the same point and then spread out as the
individual lines evolve differently.  Again, it is possible to infer
that the streamlines slow down as they pass through the barrier.  The
right-hand plot shows the $t$--$y$ evolution of the same streamlines.
In all plots distance is measured in {\AA} and time in units of
$10^{-14}$s.}
\label{7fig-2dsim}
\end{figure}

In the preceding section we simplified the problem in two ways: by
assuming perpendicular incidence, and by aligning the spin-vector in
the direction of motion.  For other configurations more complicated
two-dimensional or three-dimensional simulations are required.  As
well as the obvious numerical complications introduced there are some
further difficulties.  For the 1-D plots just shown, there was no
difficulty in deciding which part of the wavepacket was transmitted
and which was reflected, since the bifurcation point occurred at some
fixed value of $x$.  For 2-D or 3-D simulations, however, the split of
the initial packet into transmitted and reflected parts occurs along a
line or over a 2-D surface.  Furthermore, this split is spin-dependent
--- if one constructs a moving wavepacket, one finds that the
streamlines circulate around the spin axis~\cite{hes79}. (A similar
circulation phenomena is found in the ground state of the hydrogen
atom~\cite{DGL93-paths}.)  The full picture of how the packet behaves
is therefore quite complicated, though qualitatively it is still the
front portion of the packet that is transmitted.  The results of a 2-D
simulation are shown in Figure~\ref{7fig-2dsim}, and show a number of
interesting features.  For example, it is the part of the wavepacket
that `spins' \textit{into} the barrier that is predominantly
responsible for the transmitted wavepacket.  The significance of the
spin in the barrier region was clear from equation~\eqref{7curr}, which
showed that the spin vector generates a transverse current in the
barrier region.  These transverse currents are clearly displayed in
Figure~\ref{7fig-2dsim}.  The motion near the barrier is highly
complex, with the appearance of current loops suggesting the formation
of vortices.  Similar effects have been described by Hirschfelder
\etal~\cite{hir74} in the context of the Schr\"{o}dinger theory.  The
streamline plots again show a slowing down in the barrier, which
offsets the fact that it is the front of the packet that crosses the
barrier.

\section{Spin Measurements}
\label{S-spinmeas}

We now turn to a second application of the local observables approach
to quantum theory, namely to determine what happens to a wavepacket
when a spin measurement is made.  The first attempts to answer this
question were made by Dewdney \textit{et~al.}~\cite{dew86,dew88}, who
used the Pauli equation for a particle with zero charge and an
anomalous magnetic moment to provide a model for a spin-1/2 particle
in a Stern-Gerlach apparatus.  Written in the STA, the relevant
equation is
\begin{equation}
\dift \Phi \isk = -\frac{1}{2m} \bgrad^2 \Phi - \mu \bB \Phi \sk
\label{9Peqn}
\end{equation}
and the current employed by Dewdney \textit{et al}. is 
\begin{equation}
\bJ = - \frac{1}{m} \dot{\bgrad} \la \Phidot \isk \Phidag \ra .
\label{9dewcurr}
\end{equation}
Dewdney~\textit{et al}. parameterise the Pauli spinor $\Phi$ in terms
of a density and three `Euler angles'.  In the STA, this
parameterisation takes the transparent form
\begin{equation}
\Phi = \rho^{1/2} \et{\isk \phi/2} \et{\isi \theta/2} \et{\isk \psi
/2},
\end{equation}
where the rotor term is precisely that needed to parameterise a
rotation in terms of the Euler angles.  With this parameterisation, it
is a simple matter to show that the current becomes
\begin{equation}
\bJ = \frac{\rho}{2m} (\bgrad \psi + \cos\!\theta \bgrad \phi).  
\end{equation}
But, as was noted in Section~\ref{Ss-Ham-Pauli}, the current defined
by equation~\eqref{9dewcurr} is not consistent with that obtained from the
Dirac theory through a non-relativistic reduction.  In fact, the two
currents differ by a term in the curl of the spin
vector~\cite{DGL93-paths,hes751}.

To obtain a fuller understanding of the spin measurement process, an
analysis based on the Dirac theory is required.  Such an analysis is
presented here.  As well as dealing with a well-defined current,
basing the analysis in the Dirac theory is important if one intends to
proceed to study correlated spin measurements performed over spacelike
intervals (\textit{i.e.} to model an EPR-type experiment).  To study
such systems it is surely essential that one employs relativistic
equations so that causality and the structure of spacetime are
correctly built in.

\subsection{A Relativistic Model of a Spin Measurement}

As is shown in Section~\ref{Ss-Op-apps}, the modified Dirac equation
for a neutral particle with an anomalous magnetic moment $\mu$ is
\begin{equation}
\grad\psi\isk -i\mu F \psi \gk = m\psi\go.
\label{9Deqn}
\end{equation}
This is the equation we use to study the effects of a spin
measurement, and it is not hard to show that equation~\eqref{9Deqn}
reduces to~\eqref{9Peqn} in the non-relativistic limit.  Following Dewdney
\etal~\cite{dew86} we model the effect of a spin measurement by
applying an impulsive magnetic field gradient,
\begin{equation}
F = B z \delta(t) \isk.
\end{equation}
The other components of $\bB$ are ignored, as we are only modelling
the behaviour of the packet in the $z$-direction.  Around $t=0$
equation~\eqref{9Deqn} is approximated by
\begin{equation}
\dift \psi \isk = \Delta p \, z \del(t)\gk \psi \gk,
\label{9r1}
\end{equation}
where
\begin{equation}
\Delta p = \mu B.
\end{equation}
To solve~\eqref{9r1} we decompose the initial spinor $\psi_0$ into
\begin{equation}
\psiup = \half(\psi_0 - \gk \psi_0 \gk ), \hs{1} \psidn =
\half (\psi_0 + \gk \psi_0 \gk ).
\end{equation}
Equation~\eqref{9r1} now becomes, for $\psiup$
\begin{equation}
\dift \psiup = \Delta p \, z \del(t) \psiup \isk 
\end{equation}
with the opposite sign for $\psidn$.  The solution is now
straightforward, as the impulse just serves to insert a phase factor
into each of $\psiup$ and $\psidn$:
\begin{equation}
\psiup \rightarrow \psiup \et{i \sk \Delta p \, z}, \hs{1} \psidn
\rightarrow \psidn \et{-i\sk \Delta p\, z} .
\end{equation}

If we now suppose that the initial $\psi$ consists of a
positive-energy plane-wave
\begin{equation}
\psi_0 = L(\bp) \Phi \et{\isk(\bp\dt\bx - Et)}
\end{equation}
then, immediately after the shock, $\psi$ is given by
\begin{equation}
\psi = \psiup \et{i\sk ( \bp \dt \bx + \Delta p\, z)} + \psidn
\et{i\sk(\bp\dt\bx - \Delta p\, z)} .
\end{equation}
The spatial dependence of $\psi$ is now appropriate to two different
values of the 3-momentum, $\bp^\up$ and $\bp^\down$, where
\begin{equation}
\bp^\up = \bp + \Delta p \, \sk, \hs{1} \bp^\dn = \bp -
\Delta p \, \sk. 
\end{equation}
The boost term $L(\bp)$ corresponds to a different momentum, however,
so both positive and negative frequency waves are required for the
future evolution.  After the shock, the wavefunction therefore
propagates as
\begin{equation}
\psi = \psiup_+ \et{-i\sk p^\up \dt x} + \psiup_- \et{i\sk\bar{p}^\up
\dt x} + \psidn_+ \et{-i \sk p^\dn \dt x} + \psidn_- \et{i \sk
\bar{p}^\dn \dt x}
\label{9newpsi}
\end{equation}
where
\begin{align}
p^\up \go &= E^\up + \bp^\up, \\
\bar{p}^\up \go &= E^\up - \bp^\up,
\end{align}
and
\begin{equation}
E^\up = (m^2 + {\bp^\up}^2)^{1/2}.
\end{equation}
Both $p^\dn$ and $E^\dn$ are defined similarly.

Each term in~\eqref{9newpsi} must separately satisfy the free-particle
Dirac equation, so it follows that
\begin{align}
p^\up \psi^\up_+ &= m \psi^\up_+ \go , \\
-\bar{p}^\up \psi^\up_- &= m \psi^\up_- \go,
\end{align}
which are satisfied together with
\begin{equation}
\psiup = \psiup_+ + \psiup_-.
\end{equation}
The same set of equations hold for $\psidn$.  Dropping the arrows,
we find that
\begin{align}
\psi_+ &= \frac{1}{2E}(p \go \psi + m \psibar ) \\
\psi_- &= \frac{1}{2E}(\bar{p} \go \psi - m \psibar ) 
\end{align}
which hold for both $\psiup$ and $\psidn$.

The effect of the magnetic shock on a monochromatic wave is to split
the wave into four components, each with a distinct momentum.  The
positive frequency waves are transmitted by the device and split into
two waves, whereas the negative frequency states are reflected.  The
appearance of the antiparticle states must ultimately be attributed to
pair production, and only becomes significant for large $\bB$-fields.
We examine this effect after looking at more physical situations.

\subsection{Wavepacket Simulations}

For computational simplicity we take the incident particle to be 
localised along the field direction only, with no momentum components
transverse to the field. This reduces the dimensionality of the
problem to one spatial coordinate and the time coordinate.  This was
the set-up considered by Dewdney \etal~\cite{dew86} and is sufficient
to demonstrate the salient features of the measurement process.  The
most obvious difference between this model and a real experiment where
the electron is moving is that, in our model, all four packets have
group velocities along the field direction.

\begin{figure}[t!]
\begin{center}
\psfig{figure=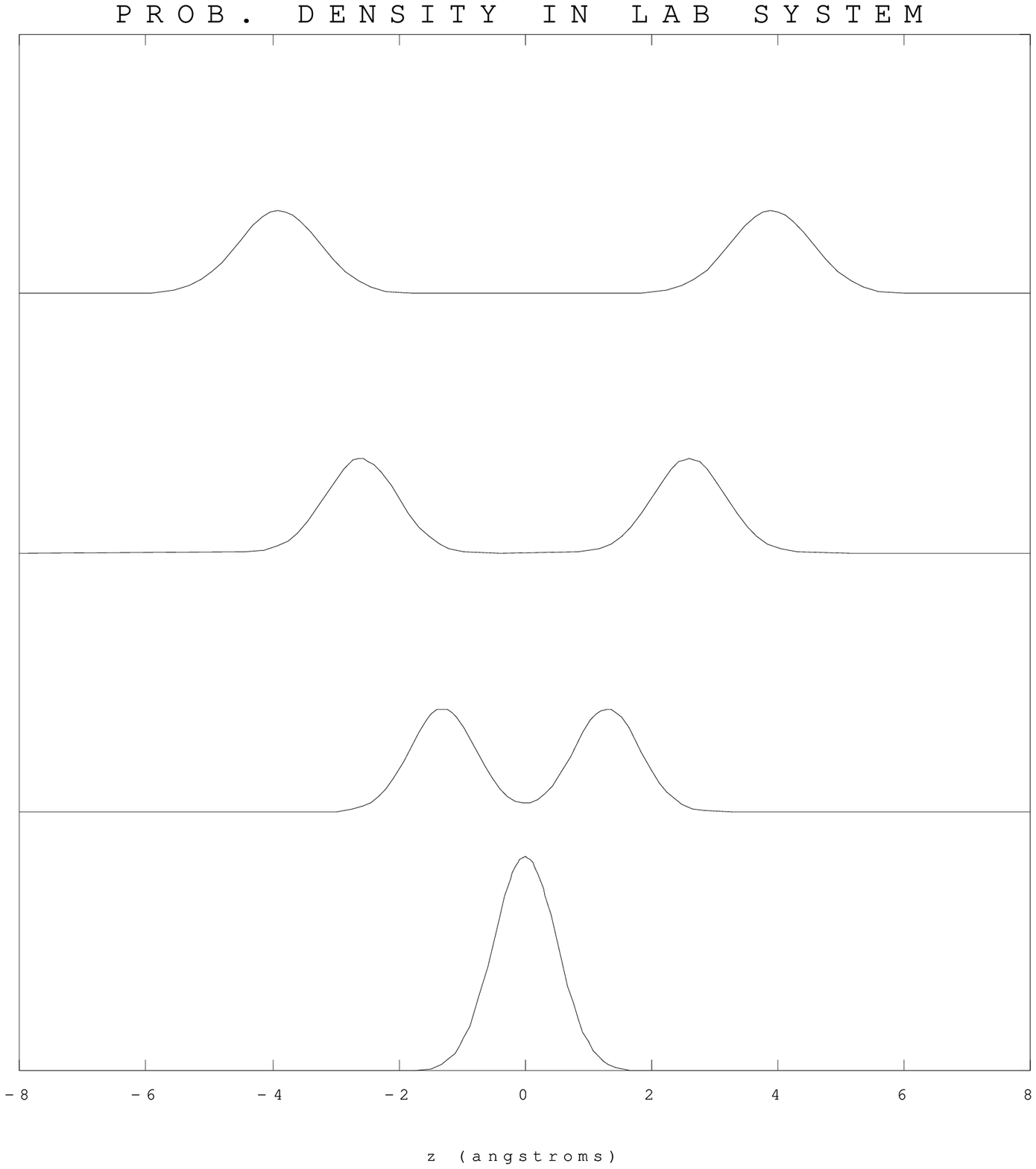,height=7cm,angle=-90} \\
\vspace{1cm} 
\psfig{figure=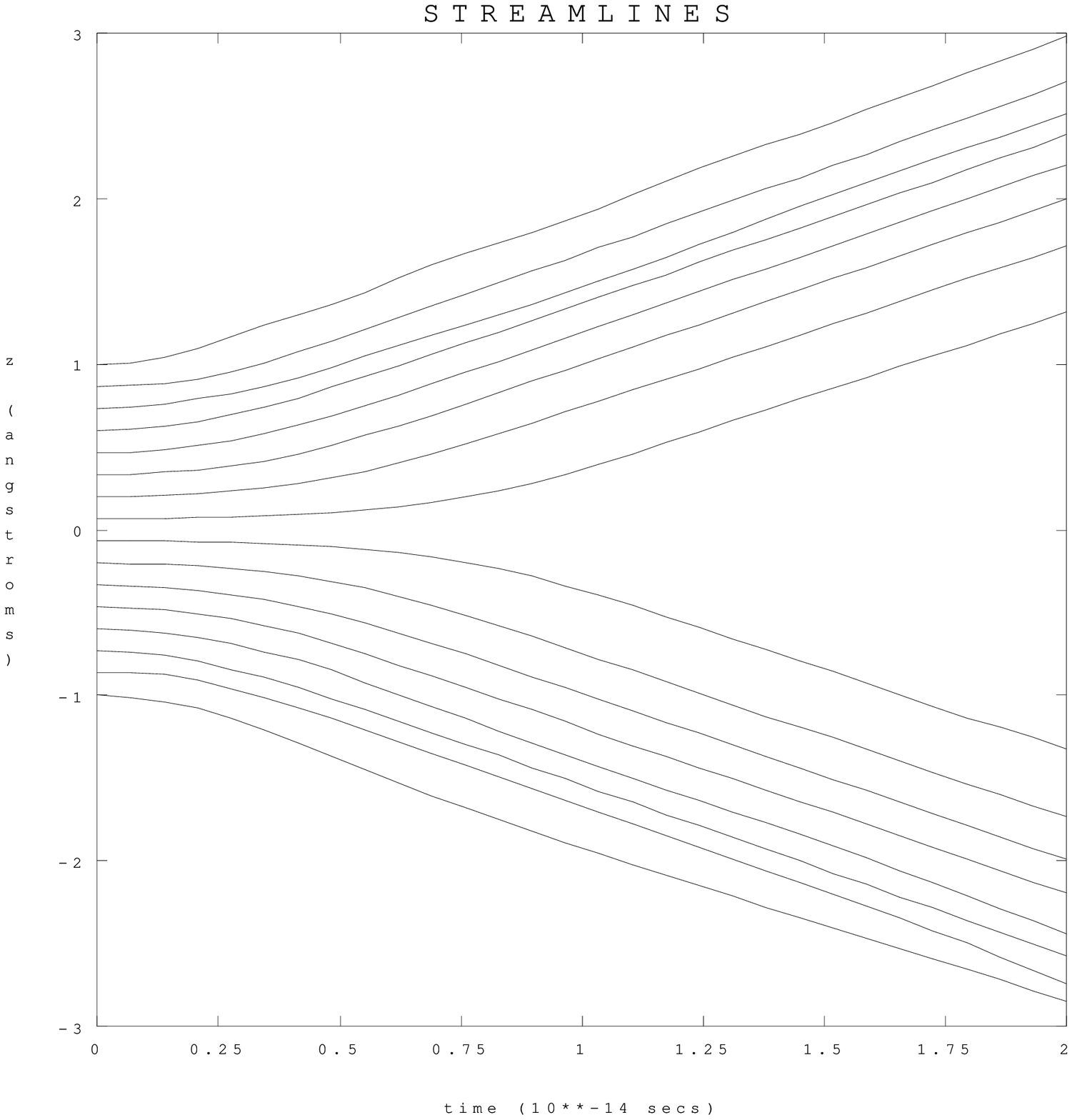,height=7cm,angle=-90}
\end{center} 
\caption[dummy1]{\em Splitting of a wavepacket caused by an impulsive
$\bB$-field.  The initial packet has a width of $1\times 10^{-24}
kg\,m\,s^{-1}$ in momentum space, and receives an impulse of $\Delta p
= 1\times 10^{-23}kg\,m\,s^{-1}$.  The top figure shows the
probability density $J_0$ at $t=0,1.3,2.6,3.9 \times 10^{-14}$s, with
$t$ increasing up the figure.  The bottom figure shows streamlines in
the ($t,z$) plane.}
\label{9fig1}
\end{figure}

The initial packet is built up from plane-wave solutions of the form
\begin{equation}
\psi = \et{u\sk/2} \Phi \et{\isk(pz-Et)}
\end{equation}
which are superposed numerically to form a Gaussian packet.  After the
impulse, the future evolution is found from equation~\eqref{9newpsi} and
the behaviour of the spin vector and the streamlines can be found for
various initial values of $\Phi$.  The results of these simulations
are plotted on the next few pages. 

\begin{figure}[t!]
\centerline{
\hbox{\psfig{figure=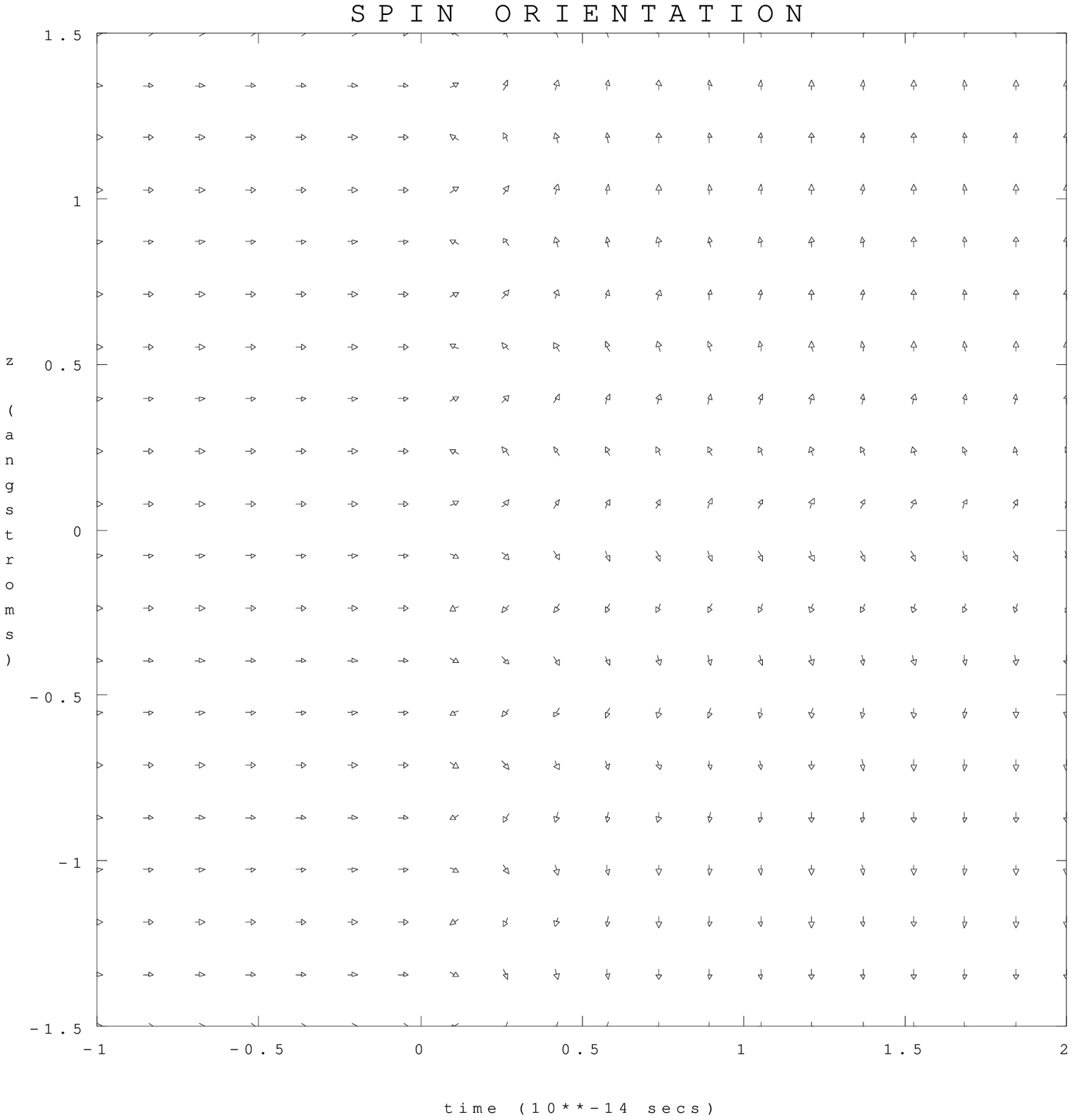,height=11cm,angle=-90}}}
\vspace{0.5cm}
\caption[dummy1]{\em Evolution of the relative spin-vector $s\wdg\go$,
in projection in the ($x,z$) plane.  Immediately after the shock the
spin-vectors point in all directions, but after about $2 \times
10^{-14}$s they sort themselves into the two packets, pointing in the
$+z$ and $-z$ directions.}
\label{9fig2}
\end{figure}
\begin{figure}[t!]
\centerline{
\hbox{\psfig{figure=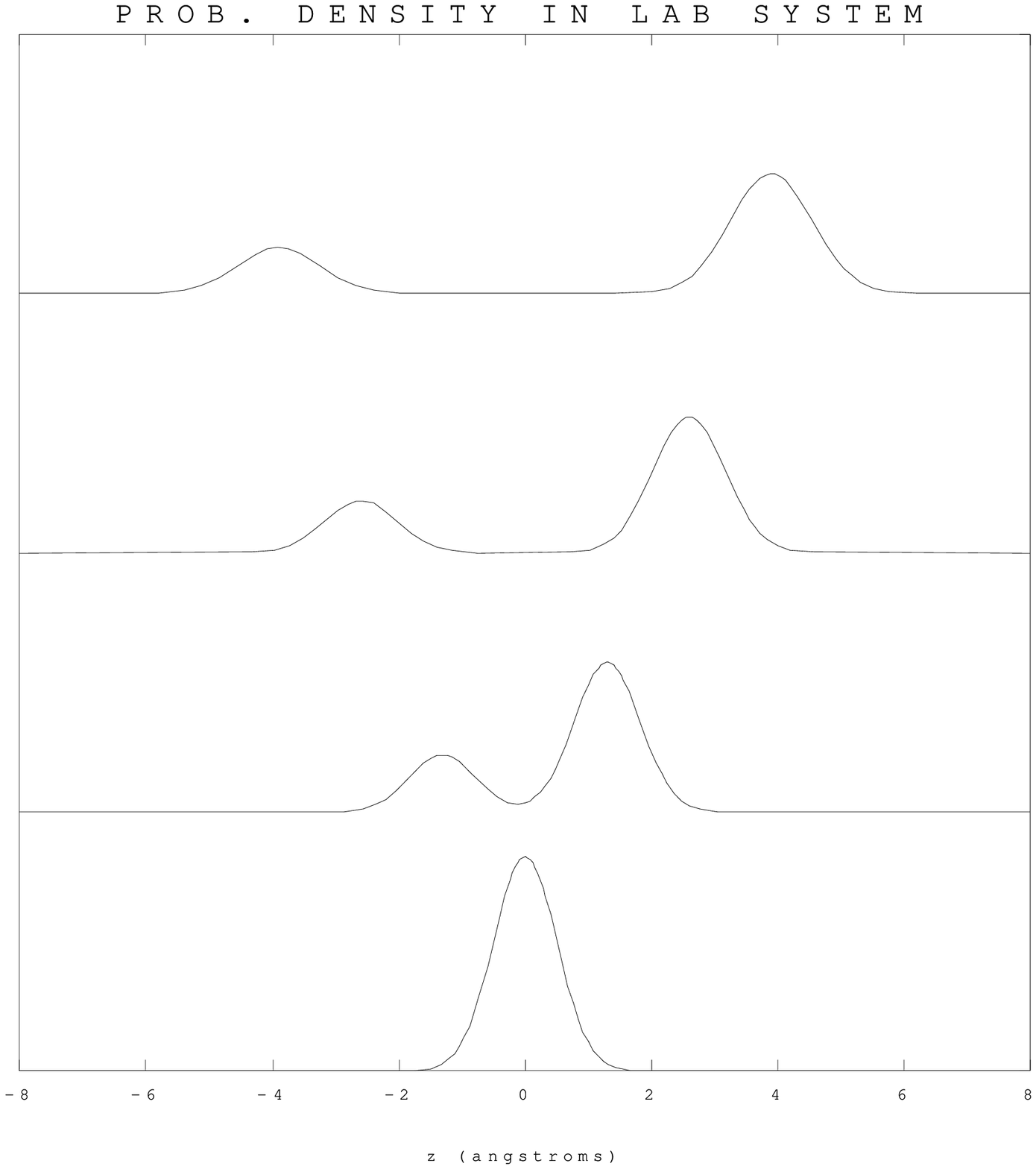,height=6cm,angle=-90}} \hs{0.5} 
\hbox{\psfig{figure=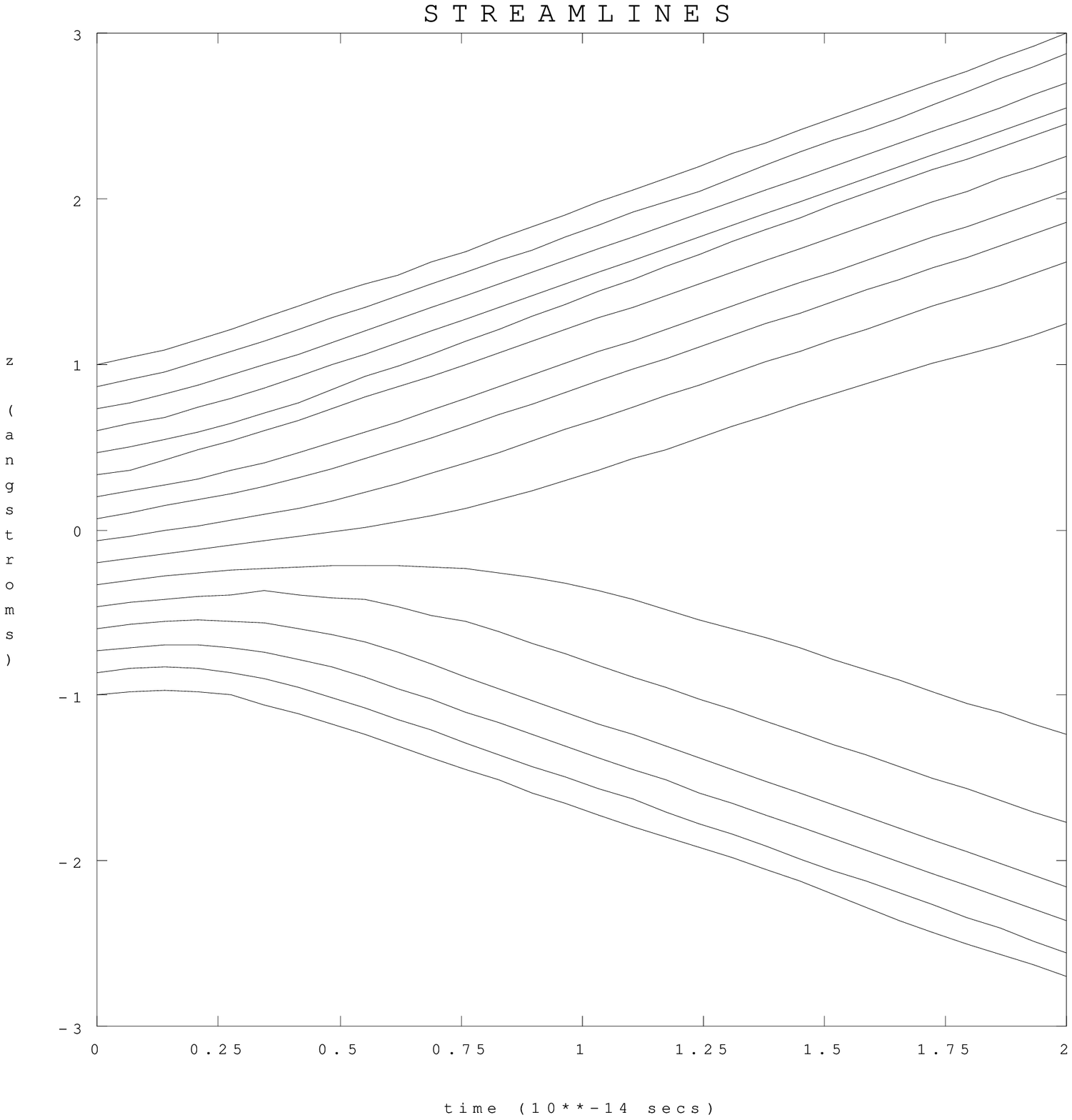,height=6cm,angle=-90}} }
\vspace{1cm}
\centerline{
\hbox{\psfig{figure=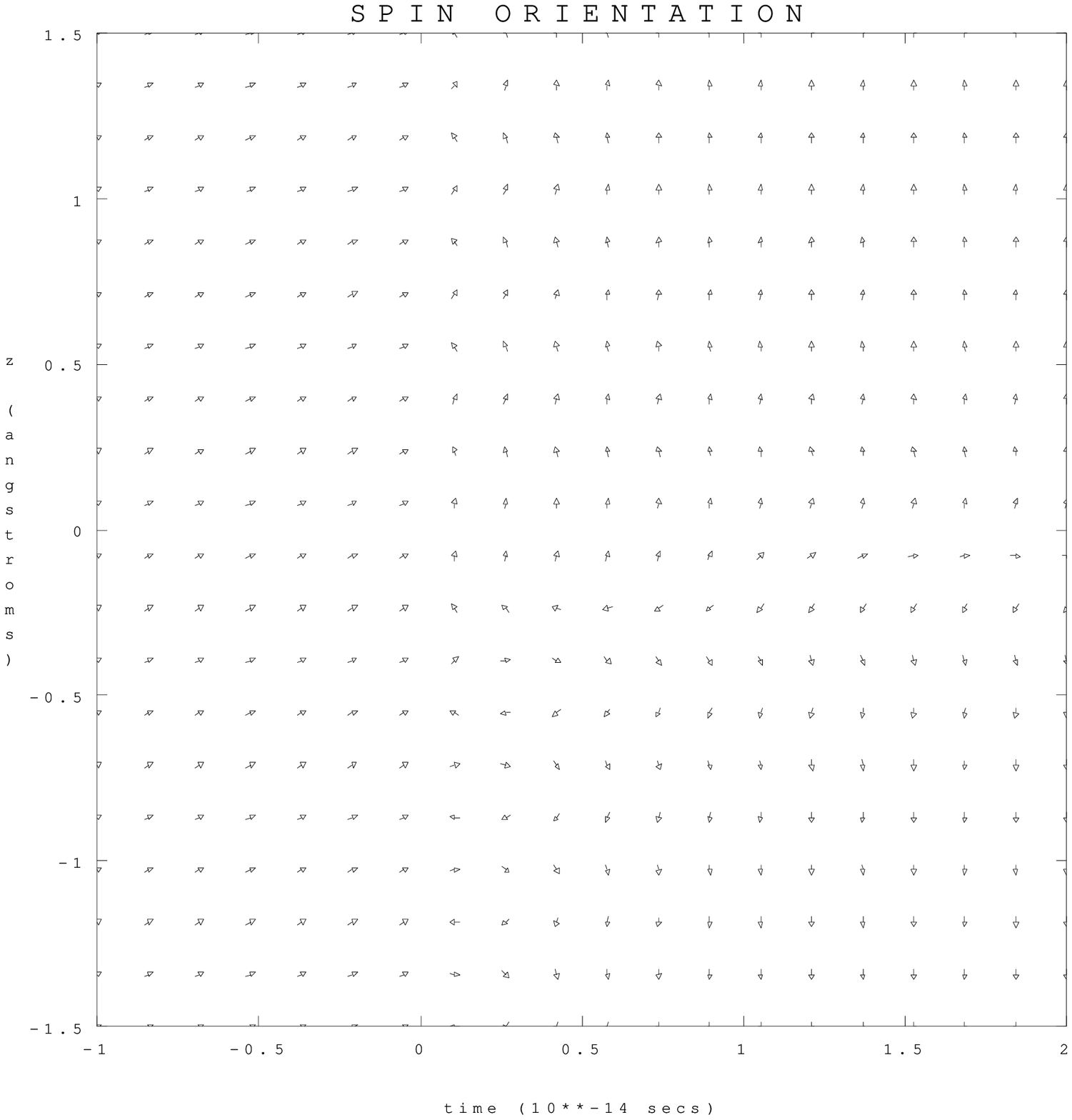,height=10cm,angle=-90}}}
\vspace{0.4cm}
\caption[dummy1]{\em Splitting of a wavepacket with unequal mixtures 
of spin-up and spin-down components.  The initial packet has
$\Phi=1.618 -i\sj$, so more of the streamlines are deflected upwards,
and the bifurcation point lies below the $z=0$ plane.  The evolution
of the spin vector is shown in the bottom plot.}
\label{9fig3}
\end{figure}

In Figures~\ref{9fig1} and~\ref{9fig2} we plot the evolution of a
packet whose initial spin vector points in the $\si$ direction ($\Phi
=\exp\{-i\sj\pi/4\}$).  After the shock, the density splits neatly
into two equal-sized packets, and the streamlines bifurcate at the
origin.  As with the tunnelling simulations, we see that disjoint
quantum outcomes are entirely consistent with the causal wavepacket
evolution defined by the Dirac equation.  The plot of the spin vector
$s\wdg\go$ shows that immediately after the shock the spins are
disordered, but that after a little time they sort themselves into one
of the two packets, with the spin vector pointing in the direction of
motion of the deflected packet.  These plots are in good qualitative
agreement with those obtained by Dewdney \etal~\cite{dew86}, who also
found that the choice of which packet a streamline enters is
determined by its starting position in the incident wavepacket.

Figure~\ref{9fig3} shows the results of a similar simulation, but with
the initial spinor now containing unequal amounts of spin-up and
spin-down components.  This time we observe an asymmetry in the
wavepacket split, with more of the density travelling in the spin-up
packet.  It is a simple matter to compute the ratio of the sizes of
the two packets, and to verify that the ratio agrees with the
prediction of standard quantum theory.

\begin{figure}[t!]
\centerline{
\hbox{\psfig{figure=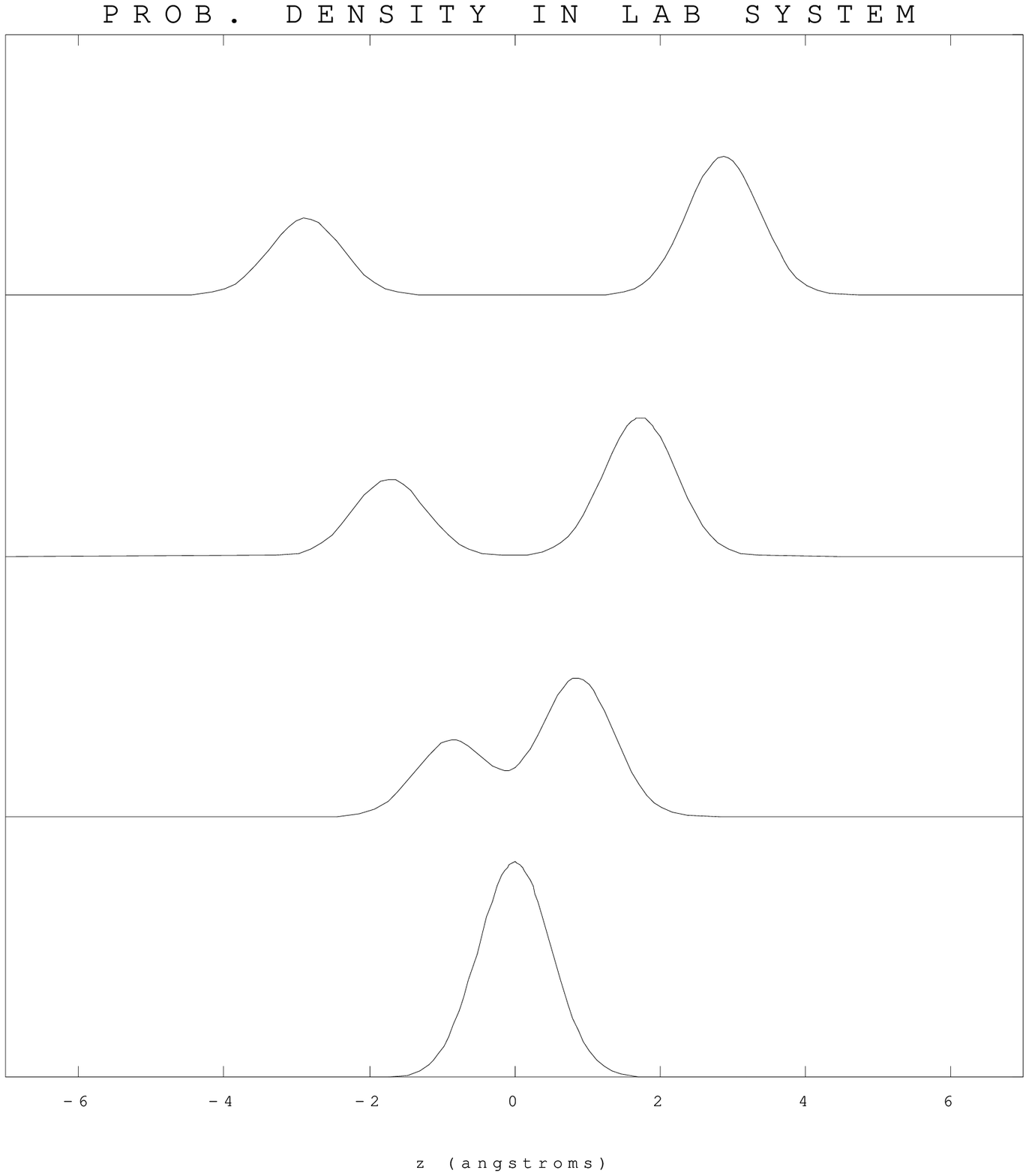,height=5.5cm,angle=-90}} \hs{0.3} 
\hbox{\psfig{figure=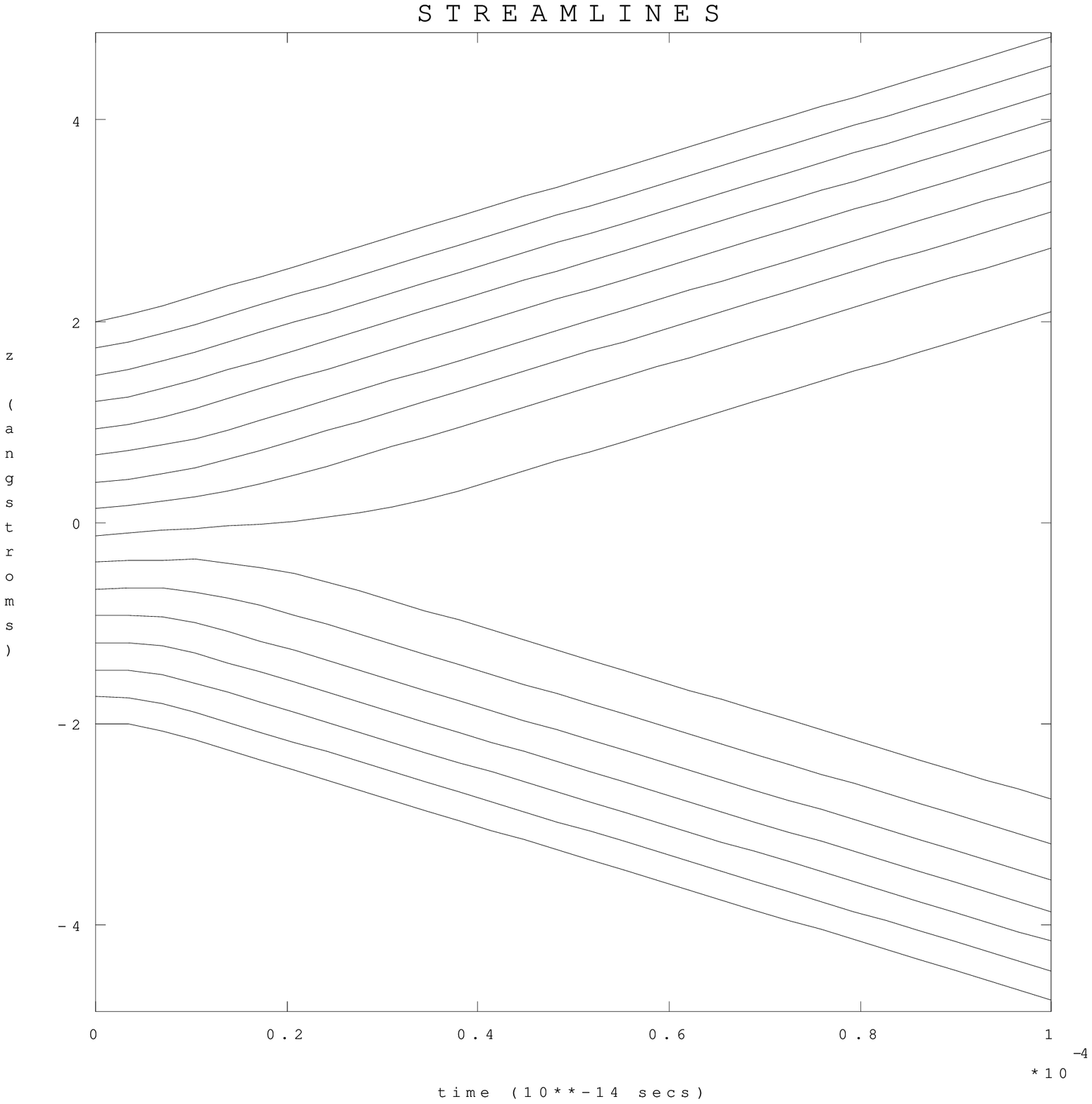,height=5.5cm,angle=-90}} }
\vspace{0.5cm}
\caption[dummy1]{\em Creation of antiparticle states by a strong
magnetic shock.  The impulse used is $\Delta p = 1\times
10^{-18}$kg\,m\,s$^{-1}$ and the initial packet is entirely spin-up
($\Phi=1$).  The packet travelling to the left consists of negative
energy (antiparticle) states.}
\label{9fig5}
\end{figure}

As a final, novel, illustration of our approach, we consider a strong
shock applied to a packet which is already aligned in the spin-up
direction.  For a weak shock the entire packet is deflected but, if
the shock is sufficiently strong that the antiparticle states have
significant amplitude, we find that a second packet is created.  The
significant feature of Figure~\ref{9fig5} is that the antiparticle
states are deflected in the opposite direction, despite the fact that
their spin is still oriented in the $+z$ direction.  The antiparticle
states thus behave as if they have a magnetic moment to mass ratio of
opposite sign.  A more complete understanding of this phenomenon
requires a field-theoretic treatment.  The appearance of antiparticle
states would then be attributed to pair production, with the
antiparticle states having the same magnetic moment, but the opposite
spin.  (One of the crucial effects of the field quantisation of
fermionic systems is to flip the signs of the charges and spins of
antiparticle states.)

The conclusions reached in this section are in broad agreement with
those of Dewdney~{\etal}.  From the viewpoint of the local observables
of the Dirac wavefunction (the current and spin densities), a
Stern-Gerlach apparatus does not fulfil the role of a classical
measuring device.  Instead, it behaves much more like a polariser,
where the ratio of particles polarised up and down is dependent on the
initial wavefunction.  The $\bB$-field dramatically alters the
wavefunction and its observables, though in a causal manner that is
entirely consistent with the predictions of standard quantum theory.
The implications of these observations for the interpretation of
quantum mechanics are profound, though they are only slowly being
absorbed by the wider physics community.  (Some of these issues are
debated in the collection of essays entitled `\textit{Quantum
Implications}'~\cite{quant-imps} and in the recent book by
Holland~\cite{hol-qtm}.)

\section{The Multiparticle STA}
\label{S-multi}

So far we have dealt with the application of the STA to
single-particle quantum theory.  In this section we turn to
multiparticle theory.  The aim here is to develop the STA approach so
that it is capable of encoding multiparticle wavefunctions, and
describing the correlations between them.  Given the advances in
clarity and insight that the STA brings to single-particle quantum
mechanics, we expect similar advances in the multiparticle case.  This
is indeed what we have found, although the field is relatively
unexplored as yet.  Here we highlight some areas where the
multiparticle STA promises a new conceptual approach, rather than
attempting to reproduce the calculational techniques employed in
standard approaches to many-body or many-electron theory.  In
particular, we concentrate on the unique geometric insights that the
multiparticle STA provides --- insights that are lost in the matrix
theory.  A preliminary introduction to the ideas developed here was
given in~\cite{DGL93-states}, though this is the first occasion that a
full relativistic treatment has been presented.

The $n$-particle STA is created simply by taking $n$ sets of basis
vectors $\{\gamdm^i\}$, where the superscript labels the particle
space, and imposing the geometric algebra relations
\begin{equation}
\begin{array}{lcc}
\gamdm^i \gamdn^j + \gamdn^i \gamdm^j = 0, & \hs{0.5} & i\neq j \\
\gamdm^i \gamdn^j + \gamdn^i \gamdm^j = 2 \eta_{\mu\nu} & \hs{0.5} &
i=j.
\end{array}
\end{equation}
These relations are summarised in the single formula
\begin{equation}
\gamdm^i \dt \gamdn^j = \del^{ij} \eta_{\mu\nu}.
\end{equation}
The fact that the basis vectors from distinct particle spaces
anticommute means that we have constructed a basis for the geometric
algebra of a $4n$-dimensional configuration space.  There is nothing
uniquely quantum-mechanical in this idea --- a system of three
classical particles could be described by a set of three trajectories
in a single space, or one path in a nine-dimensional space.  The extra
dimensions serve simply to label the properties of each individual
particle, and should not be thought of as existing in anything other
than a mathematical sense.  This construction enables us, for example,
to define a rotor which rotates one particle whilst leaving all the
others fixed.  The unique feature of the multiparticle STA is that it
implies a separate copy of the time dimension for each particle, as
well as the three spatial dimensions.  To our knowledge, this is the
first attempt to construct a solid conceptual framework for a
multi-time approach to quantum theory.  Clearly, if successful, such
an approach will shed light on issues of locality and causality in
quantum theory.

The $\{\gamdm^i\}$ serve to generate a geometric algebra of enormously
rich structure.  Here we illustrate just a few of the more immediate
features of this algebra.  It is our belief that the multiparticle STA
will prove rich enough to encode all aspects of multiparticle quantum
field theory, including the algebra of the fermionic
creation/annihilation operators.

Throughout, Roman superscripts are employed to label the particle
space in which the object appears.  So, for example, $\psi^1$ and
$\psi^2$ refer to two copies of the same 1-particle object $\psi$, and
not to separate, independent objects.  Separate objects are given
distinct symbols, or subscripts if they represent a quantity such as
the current or spin-vector, which are vectors in configuration space
with different projections into the separate copies of the STA.  The
absence of superscripts denotes that all objects have been collapsed
into a single copy of the STA.  As always, Roman and Greek subscripts
are also used as frame indices, though this does not interfere with the
occasional use of subscripts to determine separate projections.

\subsection{2-Particle Pauli States and the Quantum Correlator}

As an introduction to the properties of the multiparticle STA, we
first consider the 2-particle Pauli algebra and the spin states of
pairs of spin-1/2 particles.  As in the single-particle case, the
2-particle Pauli algebra is just a subset of the full 2-particle STA.
A set of basis vectors is defined by
\begin{align}
\sigi^1 &= \gam_i^1 \go^1 \\
\sigi^2 &= \gam_i^2 \go^2
\end{align}
which satisfy
\begin{equation}
\sigi^1 \sigj^2 =  \gam_i^1 \go^1 \gam_j^2 \go^2 =  \gam_i^1\gam_j^2 \go^2
\go^1 = \gam_j^2 \go^2 \gam_i^1 \go^1 =\sigj^2  \sigi^1.
\end{equation}
So, in constructing multiparticle Pauli states, the basis vectors from
different particle spaces commute rather than anticommute.  Using the
elements $\{1,i\sig_k^1,i\sig_k^2,i\sig_j^1\,i\sig_k^2\}$ as a basis,
we can construct 2-particle states.  Here we have introduced the
abbreviation
\begin{equation}
i\sigma_i^1 = i^1 \sigma_i^1
\end{equation}
since, in most expressions, it is obvious which particle label should
be attached to the $i$.  In cases where there is potential for
confusion, the particle label is put back on the $i$.  The basis set
$\{1,i\sig_k^1,i\sig_k^2,i\sig_j^1\,i\sig_k^2\}$ spans a
16-dimensional space, which is twice the dimension of the direct
product space of two 2-component complex spinors. For example, the
outer-product space of two spin-$1/2$ states can be built from complex
superpositions of the set
\begin{equation}
\begin{array}{rcl}
\begin{pmatrix}
        1 \\
        0
      \end{pmatrix}
\otimes
\begin{pmatrix}
        1 \\
        0
      \end{pmatrix},
\quad
\begin{pmatrix}
        0 \\
        1
      \end{pmatrix}
\otimes
\begin{pmatrix}
        1 \\
        0
      \end{pmatrix},
\quad
\begin{pmatrix}
        1 \\
        0
      \end{pmatrix}
\otimes
\begin{pmatrix}
        0 \\
        1
      \end{pmatrix},
\quad
\begin{pmatrix}
        0 \\
        1
      \end{pmatrix}
\otimes
\begin{pmatrix}
        0 \\
        1
      \end{pmatrix},
\end{array}
\end{equation}
which forms a 4-dimensional complex space (8 real dimensions).  The
dimensionality has doubled because we have not yet taken the complex
structure of the spinors into account.  While the role of $j$ is
played in the two single-particle spaces by right multiplication by
$\iski$ and $\iskj$ respectively, standard quantum mechanics does not
distinguish between these operations.  A projection operator must
therefore be included to ensure that right multiplication by $\iski$
or $\iskj$ reduces to the same operation.  If a 2-particle spin
state is represented by the multivector $\psi$, then $\psi$ must
satisfy
\begin{equation}
\psi \isk^1 = \psi \isk^2 
\end{equation}
from which we find that
\begin{align}
\begin{array}{ll}
& \psi = - \psi  \isk^1 \, \isk^2 \\
\implies & \psi =  \psi \half(1-  \isk^1 \, \isk^2). 
\end{array} 
\end{align}
On defining
\begin{equation}
E = \half(1-  \isk^1 \, \isk^2),
\end{equation}
we find that
\begin{equation}
E^2  = E
\end{equation}
so right multiplication by $E$ is a projection operation.  (The
relation $E^2=E$ means that $E$ is technically referred to as an
`idempotent' element.)  It follows that the 2-particle state $\psi$ must
contain a factor of $E$ on its right-hand side.  We can further define
\begin{equation}
J = E \isk^1 = E \isk^2 = \half (\isk^1 + \isk^2)
\end{equation}
so that 
\begin{equation}
J^2 = -E.
\end{equation}
Right-sided multiplication by $J$ takes on the role of $j$ for
multiparticle states.

The STA representation of a direct-product 2-particle Pauli spinor is
now given by $\psi^1 \phi^2 E$, where $\psi^1$ and $\phi^2$ are
spinors (even multivectors) in their own spaces.  A complete basis for
2-particle spin states is provided by
\begin{equation}
\begin{array}{rcl}
\begin{pmatrix}
        1 \\
        0
      \end{pmatrix} 
\otimes 
\begin{pmatrix}
        1 \\
        0
      \end{pmatrix}
& \trans & E \rule[-.5cm]{0cm}{0.5cm} \\
\begin{pmatrix}
        0 \\
        1
      \end{pmatrix} 
\otimes 
\begin{pmatrix}
        1 \\
        0
      \end{pmatrix}
& \trans & -\isji E \rule[-.5cm]{0cm}{0.5cm} \\
\begin{pmatrix}
        1 \\
        0
      \end{pmatrix} 
\otimes 
\begin{pmatrix}
        0 \\
        1
      \end{pmatrix}
& \trans & - \isjj E \rule[-.5cm]{0cm}{0.5cm} \\
\begin{pmatrix}
        0 \\
        1
      \end{pmatrix} 
\otimes 
\begin{pmatrix}
        0 \\
        1
      \end{pmatrix}
& \trans & \isji \, \isjj E .
\end{array}
\label{92parbas}
\end{equation}

This procedure extends simply to higher multiplicities.   All that is
required is to find the `quantum correlator' $E_n$ satisfying
\begin{equation}
E_n \isk^j = E_n \isk^k = J_n \hs{1} \mbox{for all $j$, $k$}.
\end{equation}
$E_n$ can be constructed by picking out the $j=1$ space, say, and
correlating all the other spaces to this, so that
\begin{equation}
E_n = \prod_{j=2}^{n} \half(1-\isk^1 \, \isk^j).
\end{equation}
The value of $E_n$ is independent of which of the $n$ spaces is singled
out and correlated to.  The complex structure is defined by
\begin{equation}
J_n = E_n \isk^j,
\end{equation}
where $\isk^j$ can be chosen from any of the $n$ spaces.  To
illustrate this consider the case of $n=3$, where
\begin{align}
E_3 &= \qrt(1-\isk^1 \, \isk^2)(1-\isk^1 \, \isk^3) \\
&= \qrt(1-\isk^1 \, \isk^2 - \isk^1 \, \isk^3 - \isk^2 \, \isk^3)
\end{align}
and
\begin{equation}
J_3 = \qrt(\isk^1 + \isk^2 + \isk^3 - \isk^1 \, \isk^2 \, \isk^3).
\end{equation}
Both $E_3$ and $J_3$ are symmetric under permutations of their
indices.

A significant feature of this approach is that all the operations
defined for the single-particle STA extend naturally to the
multiparticle algebra. The reversion operation, for example, still has
precisely the same definition --- it simply reverses the order of
vectors in any given multivector.  The spinor inner
product~\eqref{3pinner} also generalises immediately, to
\begin{equation}
( \psi, \phi )_S = \la E_n \ra^{-1} [\la \psidag \phi \ra - \la
\psidag \phi J_n \ra \isk],
\end{equation}
where the right-hand side is projected onto a single copy of the STA.
The factor of $\la E_n \ra^{-1}$ is included so that the state `1'
always has unit norm, which matches with the inner product used in the
matrix formulation.

\subsubsection*{The Non-Relativistic Singlet State}
\label{Ss-singlet}

As an application of the formalism outlined above, consider the
2-particle singlet state $|\eps \ra$, defined by
\begin{equation}
|\eps \ra = \sqhalf \left\{
\begin{pmatrix}
        1 \\
        0
      \end{pmatrix} \otimes 
\begin{pmatrix}
        0 \\
        1
      \end{pmatrix} -
\begin{pmatrix}
        0 \\
        1
      \end{pmatrix} \otimes 
\begin{pmatrix}
        1 \\
        0
      \end{pmatrix} \right\}.
\end{equation}
This is represented in the 2-particle STA by the multivector 
\begin{equation}
\eps = \sqhalf (\isji - \isjj)\half(1-\iski \, \iskj).
\label{9psinglet}
\end{equation}
The properties of $\eps$ are more easily seen by writing
\begin{equation}
\eps = \half(1+ i\sj^1 \, i\sj^2) \half(1+ i\sk^1 \, i\sk^2) \sqrt{2}
\, i\sj^1, 
\end{equation}
which shows how $\eps$ contains the commuting idempotents $\half(1+
i\sj^1 \, i\sj^2)$ and $\half(1+ i\sk^1 \, i\sk^2)$.  The
normalisation ensures that
\begin{align}
( \eps, \eps)_S &= 2\la \eps^\dagger \eps \ra \nn \\
&= 4 \la \half(1+ i\sj^1i\sj^2) \half(1+ i\sk^1i\sk^2) \ra \nn \\
&= 1.
\end{align}

The identification of the idempotents in $\eps$ leads immediately
to the results that
\begin{equation}
i \sj^1 \eps = \half (i\sj^1-i\sj^2) \half(1+ i\sk^1 \, i\sk^2)
\sqrt{2} i\sj^1  = -i \sj^2 \eps 
\end{equation}
and
\begin{equation}
i \sk^1 \eps = -i \sk^2 \eps, 
\end{equation}
and hence that
\begin{equation}
i \si^1 \eps = \isk^1 \, i\sj^1 \eps = -i\sj^2 \, i\sk^1 \eps  =
i\sj^2 \, i\sk^2 \eps = -i \si^2 \eps .
\end{equation}
If $M^1$ is an arbitrary even element in the Pauli algebra ($M=M^0
+M^ki\sigk^1$), it follows that $\eps$ satisfies
\begin{equation}
M^1\eps = {M^2}^\dagger \eps.
\label{9psinglet1}
\end{equation}
This now provides a novel demonstration of the rotational invariance of
$\eps$.  Under a joint rotation in 2-particle space, a spinor $\psi$
transforms to $R^1 R^2 \psi$, where $R^1$ and $R^2$ are copies of the
same rotor but acting in the two different spaces. From
equation~\eqref{9psinglet1} it follows that, under such a rotation, $\eps$
transforms as
\begin{equation}
\eps \mapsto R^1 R^2 \eps = R^1 {R^1}^\dagger \eps = \eps,
\end{equation}
so that $\eps$ is a genuine 2-particle scalar.

\subsubsection*{Non-Relativistic Multiparticle Observables}

Multiparticle observables are formed in the same way as for
single-particle states.  Some combination of elements from the fixed
$\{\sigk^j\}$ frames is sandwiched between a multiparticle
wavefunction $\psi$ and its spatial reverse $\psidag$.  An important
example of this construction is provided by the multiparticle
spin-vector.  In the matrix formulation, the $k$th component of the
particle-1 spin vector is given by
\begin{equation}
S_{1 k} = \la \psi | \hsig^1_k | \psi \ra 
\end{equation}
which has the STA equivalent
\begin{align} 
S_{1 k} &=  2^{n-1}\left( \la \psidag \sigk^1 \psi \sk^1 \ra - 
\la \psidag i \sigk^1 \psi \ra \isk \right) \nn \\
&= - 2^{n-1} \la i\sigk^1 \psi i\sk^1 \psidag \ra \nn \\
&= - 2^{n-1} (i\sigk^1) \dt (\psi J \psidag).
\end{align}
Clearly, the essential quantity is the bivector part of $\psi
J\psidag$, which neatly generalises the single-particle formula.  If
we denote the result of projecting out from a multivector $M$ the
components contained entirely in the $i$th-particle space by $\la M
\ra^i$, we can then write
\begin{equation}
\bS_a^a = 2^{n-1} \la \psi J \psidag \ra_2^a.
\label{9subsup}
\end{equation}
The various subscripts and superscripts deserve some explanation.  On
both sides of equation~\eqref{9subsup} the superscript~$a$ labels the copy
of the STA of interest.  The subscript on the right-hand side as usual
labels the fact that we are projecting out the grade-2 components of
some multivector.  The subscript $a$ on the left-hand side is
necessary to distinguish the separate projections of $\psi J\psidag$.
Had we not included the subscript, then $\bS^1$ and $\bS^2$ would
refer to two copies of the \textit{same} bivector, whereas $\bS^1_1$
and $\bS^2_2$ are different bivectors with different components.  The
reason for including both the subscript and the superscript on
$\bS_a^a$ is that we often want to copy the individual bivectors from
one space to another, without changing the components.

We can hold all of the individual $\bS_a^a$ bivectors in a single
multiparticle bivector defined by
\begin{equation}
\bS = 2^{n-1} \la \psi J \psidag \ra_2.
\label{9pmultspin}
\end{equation}
Under a joint rotation in $n$-particle space, $\psi$ transforms to
$R_1 \ldots R_n \psi$ and  $\bS$ therefore transforms to
\begin{equation}
R^1 \ldots R^n \bS {R^n}^\dagger \ldots {R^1}^\dagger = R^1
\bS^1_1 {R^1}^\dagger  + \cdots + R^n \bS^n_n {R^n}^\dagger.
\end{equation}
Each of the separate projections of the spin current is therefore
rotated by the same amount, in its own space.  That the
definition~\eqref{9pmultspin} is sensible can be checked with the four
basis states~\eqref{92parbas}.  The form of $\bS$ for each of these is
contained in Table~\ref{9t-spin}.  Multiparticle observables for the
2-particle case are discussed further below.

\begin{table}
\begin{center}
\begin{tabular}{lcc}
\hline \hline
Pauli   & \hspace{.5cm} Multivector \hspace{.5cm}  &  Spin     \\            
State   &                Form                      &  Current  \\
\hline 
$| \up \up \ra     $ & $ E_2              $ & $ \isk^1 + \isk^2  $ \\ 
$| \up \down  \ra  $ & $ -\isj^2 E_2      $ & $ \isk^1 - \isk^2  $ \\ 
$| \down \up \ra   $ & $ -\isj^1 E_2      $ & $ -\isk^1 + \isk^2 $ \\ 
$| \down \down \ra $ & $\isj^1\,\isj^2 E_2$ & $ -\isk^1 - \isk^2 $ \\ 
\hline \hline
\end{tabular}
\end{center}
\caption{\em Spin Currents for 2-Particle Pauli States}
\label{9t-spin}
\end{table}

Other observables can be formed using different fixed multivectors.
For example, a 2-particle invariant is generated by sandwiching a
constant multivector $\Sigma$ between the singlet state $\eps$,
\begin{equation}
M = \eps \Sigma \epsdag.
\end{equation}
Taking $\Sigma=1$ yields
\begin{align}
M = \eps \epsdag &= 2 \half(1+ i\sj^1 \,i\sj^2)
\half(1+i\sk^1\,i\sk^2) \nn \\
&= \half(1 + i\si^1 \, i\si^2 + i\sj^1 \, i\sj^2 + i\sk^1 \, i\sk^2), 
\end{align}
which rearranges to give
\begin{equation}
i\sig_k^1 \, i\sig_k^2 = 2 \eps \epsdag -1.
\end{equation}
This equation contains the essence of the matrix result
\begin{equation}
\hsig^a_{k \,a'} \, \hsig^b_{k \, b'} = 2 \del^a_{b'} \, \del^b_{a'} -
\del^a_{a'} \, \del^b_{b'}  
\end{equation}
where $a$, $b$, $a'$, $b'$ label the matrix components.  This matrix
equation is now seen to express a relationship between 2-particle
invariants.  Further invariants are obtained by taking
$\Sigma=i^1i^2$, yielding
\begin{equation}
M = \eps i^1 i^2 \epsdag = \half(i^1i^2 + \si^1 \, \si^2 + \sj^1 \,
\sj^2 + \sk^1 \, \sk^2) . 
\end{equation}
This shows that both $i\sigk^1\,i\sigk^2$ and $\sigk^1\,\sigk^2$ are
invariants under 2-particle rotations.  In standard quantum
mechanics these invariants would be thought of as arising from the
``inner product'' of the spin vectors $\hsigi^1$ and $\hsigi^2$.
Here, we have seen that the invariants arise in a completely different
way by looking at the full multivector $\eps \epsdag$.

The contents of this section should have demonstrated that the
multiparticle STA approach is capable of reproducing most (if not all)
of standard multiparticle quantum mechanics.  One important result
that follows is that the unit scalar imaginary $j$ can be
completely eliminated from quantum mechanics and replaced by
geometrically meaningful quantities.  This should have significant
implications for the interpretation of quantum mechanics.

\subsection{Comparison with the `Causal' Approach to \\
Non-Relativistic Spin States} 

As an application of the techniques outlined above, we look at the
work of Holland on the `Causal interpretation of a system of two
spin-1/2 particles'~\cite{hol88}.  This work attempts to give a
non-relativistic definition of local observables in the
higher-dimensional space of a 2-particle wavefunction.  As we have
seen, such a construction appears naturally in our approach.
Holland's main application is to a Bell inequality type experiment,
with spin measurements carried out on a system of two correlated
spin-1/2 particles by Stern-Gerlach experiments at spatially separated
positions.  Such an analysis, though interesting, will only be
convincing if carried out in the fully relativistic domain, where
issues of causality and superluminal propagation can be coherently
addressed.  We intend to carry out such an analysis in the future,
using the STA multiparticle methods, and the work below on the
observables of a 2-particle system can be seen as part of this aim.

The aspect of Holland's work that concerns us here (his Section~3 and
Appendix~A) deals with the joint spin-space of a system of two
non-relativistic spin-1/2 particles.  The aim is to show that `all 8
real degrees of freedom in the two body spinor wavefunction may be
interpreted (up to a sign) in terms of the properties of algebraically
interconnected Euclidean tensors'~\cite{hol88}.  Holland's working is
complex and requires a number of index manipulations and algebraic
identities.  Furthermore, the meaning of the expressions derived is
far from transparent.  Using the above techniques, however, the
significant results can be derived more efficiently and in such a way
that their geometric meaning is made much clearer.  Rather than give a
line-by-line translation of Holland's work, we simply state the key
results in our notation and prove them.

Let $\psi=\psi E$ be a 2-particle spinor in the correlated product
space of the one-particle spin-spaces (the even subalgebra of the
Pauli algebra).  The observables of this 2-particle system are formed
from projections of bilinear products of the form
$\psi\Gam\tilde{\psi}$, where $\Gam$ is an element of the 2-particle
Pauli spinor algebra.  For example, the two 3-dimensional
spin-vectors, $\bs^1_1$, $\bs^2_2$, are defined by
\begin{equation}
i\bs^1_1 + i\bs^2_2 = 2\psi J \tilde{\psi} 
\label{9multi-eqn7}
\end{equation}
where the right-hand side can be written in the equivalent forms
\begin{equation}
\psi J \tilde{\psi} = \psi i\sk^1 \tilde{\psi} = \psi i\sk^2
\tilde{\psi}.
\end{equation}
The formula~\eqref{9multi-eqn7} is a special case of
equation~\eqref{9pmultspin} where, as we are working in a 2-particle
system, the projection onto bivector parts is not required.
The vectors $\bs^1_1$ and $\bs^2_2$ correspond to the two spin vectors
defined by Holland, with the explicit correspondence to his $S_{1k}$
and $S_{2k}$ given by
\begin{equation}
S_{1k} = -(i\bs^1_1) \dt (i\sig_k^1), \hs{1} S_{2k} = -(i\bs^2_2)
\dt (i\sig_k^2).
\label{9S12cpts}
\end{equation}

An important relation proved by Holland is that the vectors $\bs^1_1$
and $\bs^2_2$ are of equal magnitude,
\begin{equation}
(\bs^1_1)^2 = (\bs^2_2)^2 = 2\Omega - \rho^2,
\end{equation}
where
\begin{equation}
\rho = 2 \la \psi \psidag \ra
\end{equation}
and an explicit form for $\Omega$ is to be determined.  To prove this
result, we write the formulae for the components of $\bs^1_1$ and
$\bs^2_2$~\eqref{9S12cpts} in the equivalent forms
\begin{equation}
S_{1k} = -2 (\psidag i\sig_k^1 \psi) \dt (i\sk^1), \hs{1} 
S_{2k} = -2 (\psidag i\sig_k^2 \psi) \dt (i\sk^2).
\label{9S12altdef}
\end{equation}
But, in both cases, the term $\psidag i\sig_k^a \psi$ contains a
bivector sandwiched between two idempotents, so is of the form
$E\ldots E$.  This sandwiching projects out the $i\sk^1$ and $i\sk^2$
components of the full bivector, and ensures that that the terms have
equal magnitude.  The inner products in~\eqref{9S12altdef} can therefore
be dropped and we are left with
\begin{equation}
\psidag i\sig_k^a \psi = S_{ak} J
\label{9hol3}
\end{equation}
where the $a=1,2$ labels the two separate spin-vectors.  It follows
immediately from equation~\eqref{9hol3} that
\begin{equation}
(\bs^a_a)^2 = -2 \la \psidag i\sig_k^a \psi \psidag i\sig_k^a \psi \ra, 
\label{9hol4}
\end{equation}
where the $a$'s are not summed.  But the quantity $\psi\psidag$
contains only scalar and 4-vector components, so we find that
\begin{equation}
i\sig_k^1 \psi \psidag i\sig_k^1 = i\sig_k^2 \psi \psidag i\sig_k^2 =  
-3\la\psi\psidag\ra + \la\psi\psidag\ra_4 = \psi\psidag - 2 \rho.
\label{9hol4.1}
\end{equation}
(This result follows immediately from $\sig_k i\ba \sig_k=-i\ba$,
which is valid for any vector $\ba$ in the single-particle Pauli
algebra.)  Inserting equation~\eqref{9hol4.1} back into~\eqref{9hol4} we can
now write
\begin{equation}
(\bs^1_1)^2 = (\bs^2_2)^2 = 2 \rho^2 - 2 \la \psi\psidag\psi\psidag \ra,
\end{equation}
which shows that $\bs^1_1$ and $\bs^2_2$ are indeed of equal magnitude,
and enables us to identify $\Omega$ as
\begin{equation}
\Omega = \half(3\rho^2 - 2 \la \psi\psidag\psi\psidag \ra).
\end{equation}

In addition to $\rho$, $\bs^1_1$ and $\bs^2_2$, Holland defines a tensor
$S_{jk}$ whose components are given by 
\begin{equation}
S_{jk} = -2 \la \psidag i\sig_j^1 \, i\sig_k^2 \psi \ra .
\end{equation}
This object has the simple frame-free form
\begin{equation}
T = \la \psi\psidag \ra_4.
\end{equation}
Between them, $\rho$, $\bs^1_1$, $\bs^2_2$ and $T$ pick up 7 of the
possible 8 degrees of freedom in $\psi$.  The remaining freedom lies
in the phase, since all of the observables defined above are
phase-invariant.  Encoding this information caused Holland some
difficulty, but in the STA the answer is straightforward, and is
actually already contained in the above working.  The crucial
observation is that as well as containing only scalar and 4-vector
terms, the quantity $\psidag\psi$ is invariant under rotations.  So,
in addition to the scalar $\rho$, the 4-vector components of
$\psidag\psi$ must pick up important rotationally-invariant
information.  Furthermore, since $\psidag\psi$ is of the form $E\ldots
E$, the four-vector component of $\psidag\psi$ contains only two
independent terms, which can be taken as a complex combination of
the $i\sj^1\,i\sj^2$ term.  This is seen most clearly using an
explicit realisation of $\psi$.  Suppose that we write
\begin{equation}
\psi = (p - \isj^1q - i\sj^2 r + \isj^1\,\isj^2 s)E
\end{equation}
where $p$, $q$, $r$, and $s$ are complex combinations of 1 and $J$,
we then find that
\begin{equation}
\psidag \psi = [\rho + 2 i\sj^1 \, i\sj^2 (ps-qr)]E.
\end{equation}
This shows explicitly that the additional complex invariant is given
by $ps-qr$.  It is this term that picks up the phase of $\psi$, and we
therefore define the complex quantity
\begin{equation}
\alp = \la \psi i\sj^1 \, i\sj^2 \psidag\ra - \la\psi i\si^1 \,
i\sj^2 \psidag\ra \isk,   
\end{equation}
which is the STA equivalent of the complex scalar $\bar{\rho}$ defined
by Holland.  The complex scalar $\alp$ is invariant under rotations,
and under the phase change
\begin{equation}
\psi \mapsto \psi \et{J \phi}
\end{equation}
$\alp$ transforms as
\begin{equation}
\alp \mapsto  \alp \et{2 \phi \isk}.
\end{equation}
The set $\{\rho, \alp, \bs^1_1, \bs^2_2, T \}$ encode all the
information contained in the 2-particle spinor $\psi$, up to an
overall sign.  They reproduce the quantities defined by Holland, but
their STA derivation makes their properties and geometric origin much
clearer.

\subsection{Relativistic 2-Particle States}

The ideas developed for the multi-particle Pauli algebra extend
immediately to the relativistic domain.  The direct product of the two
single-particle spinor spaces (the even subalgebras) now results in a
space of $8\times 8=64$ real dimensions.  Unlike the single-particle
case, this space is not equivalent to the even subalgebra of the full
8-dimensional algebra.  The full algebra is 256-dimensional, and its
even subalgebra is therefore 128-dimensional.  It is not yet clear
whether the remaining 64-dimensional space which is not picked up by
sums of direct-product states could be of use in constructing
2-particle wavefunctions, and for the remainder of this section we
work only with the space obtained from sums of direct-product states.
Post-multiplying the direct-product space by the quantum correlator
$E$ reduces it to 32 real dimensions, which are equivalent to the 16
complex dimensions employed in standard 2-particle relativistic
quantum theory.  All the single-particle observables discussed in
Section~\ref{Ss3-Deqn} extend simply.  In particular, we define the
vectors
\begin{align}
\clj &= \la \psi (\go^1 + \go^2) \psirev \ra_1 \label{9relJ} \\
s &= \la \psi (\gk^1 + \gk^2) \psirev \ra_1 
\end{align}
which are respectively the 2-particle current and spin-vector.  (The
calligraphic symbol $\clj$ is used to avoid confusion with the
correlated bivector $J$.)  We also define the spin bivector $S$ by
\begin{equation}
S = \la \psi J \psirev \ra_2.
\end{equation}

Of particular interest are the new Lorentz-invariant quantities that
arise in this approach.  From the work of the preceding section, we
form the quantity $\psirev\psi$, which decomposes into
\begin{equation}
\psirev\psi = \la\psirev\psi\ra_{0,8} + \la\psirev\psi\ra_4.
\end{equation}
The grade-0 and grade-8 terms are the 2-particle generalisation of the
scalar + pseudoscalar combination $\psi\psirev=\rho\exp(i\beta)$ found
at the single-particle level.  Of greater interest are the 4-vector
terms.  These offer a wealth of Lorentz-invariant 2-particle
observables, the meaning of which we are only beginning to appreciate.
Such invariants are rarely seen in the traditional matrix approach.

\subsubsection*{The Relativistic Singlet State and Relativistic
Invariants} 
\label{Ss-relinvs}

Our task here is to find a relativistic analogue of the Pauli singlet
state discussed in Section~\ref{Ss-singlet}.  Recalling the definition
of $\eps$~\eqref{9psinglet}, the property that ensured that $\eps$ was a
singlet state was that
\begin{equation}
i\sigk^1 \eps = - i\sigk^2 \eps, \qquad k=1 \ldots 3.
\label{9meq1}
\end{equation}
In addition to~\eqref{9meq1} a relativistic singlet state, which we will
denote as $\eta$, must satisfy
\begin{equation}
\sigk^1 \eta = - \sigk^2 \eta, \qquad k=1 \ldots 3.
\label{9meq2}
\end{equation}
It follows that $\eta$ satisfies
\begin{equation}
i^1 \eta = \si^1\sj^1\sk^1 \eta = -\sk^2\sj^2\si^2 \eta =
i^2\eta 
\end{equation}
so that
\begin{align}
\eta &= -i^1i^2 \eta \\
\implies \eta &= \half(1-i^1i^2) \eta.
\end{align}
The state $\eta$ can therefore be constructed by multiplying $\eps$ by
the idempotent $\half(1-i^1i^2)$.  We therefore define
\begin{equation}
\eta = \epsilon \sqhalf(1-i^1i^2) = (\isji - \isjj)
\half(1-\iski \, \iskj) \half(1-i^1i^2) ,
\label{9mdefeta}
\end{equation}
which is normalised such that $(\eta,\eta)_S=1$.  The invariant
$\eta$ satisfies
\begin{equation}
i\sigk^1 \eta = i\sigk^1 \, \eps \half(1-i^1i^2) = - i\sigk^2 \eta
\qquad k=1 \ldots 3
\end{equation}
and
\begin{equation}
\sigk^1 \eta = -\sigk^1 i^1i^2 \eta = i^2 \, i\sigk^2 \eta = -
\sigk^2 \eta  \qquad k=1 \ldots 3.
\end{equation}
These results are summarised by
\begin{equation}
M^1 \eta = \tilde{M}^2 \eta,
\label{9mepsinv}
\end{equation}
where $M$ is an even multivector in either the particle-1 or
particle-2 STA.  The proof that $\eta$ is a relativistic invariant
now reduces to the simple identity
\begin{equation}
R^1R^2 \eta = R^1 \Rrev^1 \eta = \eta,
\label{9msingletinv}
\end{equation}
where $R$ is a single-particle relativistic rotor.

Equation~\eqref{9mepsinv} can be seen as arising from a more primitive
relation between vectors in the separate spaces.  Using the result
that $\go^1\go^2$ commutes with $\eta$, we can derive
\begin{align}
\gamdm^1 \eta \go^1 
&= \gamdm^1 \go^1\go^2 \eta \go^2\go^1\go^1 \nn\\ 
&= \go^2 (\gamdm \go)^1 \eta \go^2 \nn \\
&= \go^2 \go^2 \gamdm^2 \eta \go^2 \nn \\
&= \gamdm^2 \eta \go^2,
\end{align}
and hence we find that, for an arbitrary vector $a$, 
\begin{equation}
a^1 \eta \go^1 = a^2 \eta \go^2.
\label{9mveceps}
\end{equation}
Equation~\eqref{9mepsinv} now follows immediately from~\eqref{9mveceps} by
writing
\begin{align}
(ab)^1 \eta 
&= a^1 b^1 \eta \go^1 \go^1 \nn \\
&= a^1 b^2 \eta \go^2 \go^1 \nn \\
&= b^2 a^1 \eta \go^1 \go^2 \nn \\
&= b^2 a^2 \eta \go^2 \go^2 \nn \\
&= (ba)^2 \eta.
\end{align}
Equation~\eqref{9mveceps} can therefore be viewed as the fundamental
property of the relativistic invariant $\eta$.

From $\eta$ a number of Lorentz-invariant 2-particle multivectors
can be constructed by sandwiching arbitrary multivectors between
$\eta$ and $\tilde{\eta}$.  The simplest such object is
\begin{align}
\eta \tilde{\eta} 
&= \eps \half(1-i^1i^2) \tilde{\eps} \nn \\ 
&= \half(1 + i\si^1\,i\si^2 + i\sj^1\,i\sj^2 + i\sk^1\,i\sk^2)
\half(1-i^1i^2) \nn \\
&= \qrt(1-i^1i^2) - \qrt(\sigk^1\,\sigk^2 - i\sigk^1 \, i\sigk^2).
\label{9minv1}
\end{align}
This contains a scalar $+$ pseudoscalar (grade-8) term, which is obviously
invariant, together with the invariant grade-4 multivector
$(\sigk^1\,\sigk^2 - i\sigk^1\,i\sigk^2)$.  The next simplest object
is
\begin{align}
\eta \go^1 \go^2 \tilde{\eta} 
&= \half(1 + i\si^1\,i\si^2 + i\sj^1\,i\sj^2 +i\sk^1\,i\sk^2)
\half(1-i^1i^2) \go^1 \go^2  \nn \\
&= \qrt(\go^1\go^2 + i^1i^2\gam_k^1\gam_k^2 - i^1i^2\go^1\go^2 -
\gam_k^1\gam_k^2) \nn \\ 
&= \qrt(\go^1 \go^2 - \gam_k^1 \gam_k^2)(1-i^1i^2).
\label{9mKinv} 
\end{align}
On defining the bivector
\begin{equation}
K = \gamdm^1 {\gamum}^2
\label{9mdefK}
\end{equation}
and the 2-particle pseudoscalar
\begin{equation}
W = i^1i^2 = i^2 i^1
\end{equation}
the invariants from~\eqref{9mKinv} are simply $K$ and $WK$.  That $W$ is
invariant under rotations is obvious, and the invariance of $K$ under
joint rotations in the two particle spaces follows from
equation~\eqref{9mKinv}.  The bivector $K$ is of the form of a `doubling'
bivector discussed in~\cite{dor93-spin}, where such bivectors are
shown to play an important role in the bivector realisation of many
Lie algebras.

\begin{table}
\begin{center}
\begin{tabular}{ccc}
\hline \hline
           &                 Type of                  &       \\
Invariant  & \hspace{.7cm} Interaction  \hspace{.7cm} & Grade \\
\hline 
$1 $       & Scalar       & 0 \\
$K $       & Vector       & 2 \\
$K \wdg K$ & Bivector     & 4 \\
$WK $      & Pseudovector & 6 \\
$W $       & Pseudoscalar & 8 \\
\hline \hline
\end{tabular}
\end{center}
\caption{\em 2-Particle Relativistic Invariants}
\label{9t-invs}
\end{table}

From the definition of $K$~\eqref{9mdefK}, we find that
\begin{align}
K \wdg K 
&= - 2 \go^1\go^2 \gam_k^1 \gam_k^2 + (\gam_k^1 \gam_k^2) \wdg (\gam_j^1
\gam_j^2) \nn \\
&= 2(\sigk^1 \, \sigk^2 - i\sigk^1 \, i\sigk^2),
\end{align}
which recovers the grade-4 invariant from~\eqref{9minv1}.  The full set of
2-particle invariants constructed from $K$ are summarised in
Table~\ref{9t-invs}.  These invariants are well-known and have been
used in constructing phenomenological models of interacting
particles~\cite{gal92,koi82}.  The STA derivation of the invariants is
quite new, however, and the fundamental role played by the bivector
$K$ is hidden in the matrix formalism.

\subsection{Multiparticle Wave Equations}

In order to extend the local-observables approach to quantum theory
to the multiparticle domain, we need to construct a relativistic wave
equation satisfied by an $n$-particle wavefunction.  This is a
subject that is given little attention in the literature,
with most textbooks dealing solely with the field-quantised
description of an $n$-particle system.  An $n$-particle wave equation
is essential, however, if one aims to give a relativistic description
of a bound system (where field quantisation and perturbation theory on
their own are insufficient).  A description of this approach is given
in Chapter~10 of Itzykson~\& Zuber~\cite{itz-quant}, who deal mainly
with the Bethe-Salpeter equation for a relativistic 2-particle system.
Written in the STA, this equation becomes
\begin{equation}
(j \grad^1 - m_1) (j \grad^2 - m_2) \psi(r,s) = I(r,s) \psi(r,s)
\label{9bet-salp}
\end{equation}
where $j$ represents right-sided multiplication by $J$, $I(r,s)$ is an
integral operator representing the inter-particle interaction, and
\begin{equation}
\grad^1 = \gamdm^1 \deriv{}{r^\mu}, \hs{1} 
\grad^2 = \gamdm^2 \deriv{}{s^\mu},
\label{9defgrd12}
\end{equation}
with $r$ and $s$ the 4-D positions of the two particles.  Strictly, we
should have written $\grad^1_r$ and $\grad^2_s$ instead of simply
$\grad^1$ and $\grad^2$.  In this case, however, the subscripts can
safely be ignored.

The problem with equation~\eqref{9bet-salp} is that it is not first-order
in the 8-dimensional vector derivative $\grad=\grad^1+\grad^2$.  We
are therefore unable to generalise many of the simple first-order
propagation techniques discussed in Section~\ref{S-scatt}.  Clearly, we
would like to find an alternative to~\eqref{9bet-salp} which retains the
first-order nature of the single-particle Dirac equation.  Here we
will simply assert what we believe to be a good candidate for such an
equation, and then work out its consequences.  The equation we shall
study, for two free spin-1/2 particles of masses $m_1$ and $m_2$
respectively, is
\begin{equation}
\left( \frac{\nabla^1}{m_1}+\frac{\nabla^2}{m_2} \right) \psi(x) \left(
i\gk^1 + i\gk^2 \right) = 2\psi(x). 
\label{9rwveq}
\end{equation}
We can assume, \textit{a priori}, that $\psi$ is not in the correlated
subspace of the the direct-product space.  But, since $E$ commutes
with $i\gk^1 + i\gk^2$, any solution to~\eqref{9rwveq} can be reduced to a
solution in the correlated space simply by right-multiplying by $E$.
Written out explicitly, the vector $x$ in equation~\eqref{9rwveq} is
\begin{equation}
x = r^1 + s^2 = \gamdm^1 r^\mu + \gamdm^2 s^\mu
\end{equation}
where $\{r^\mu,s^\mu\}$ are a set of 8 independent components for
$\psi$.  Of course, all particle motions ultimately occur in a single
space, in which the vectors $r$ and $s$ label two independent position
vectors.  We stress that in this approach there are two time-like
coordinates, $r^0$ and $s^0$, which is necessary if our 2-particle
equation is to be Lorentz covariant.  The derivatives $\grad^1$ and
$\grad^2$ are as defined by equation~\eqref{9defgrd12}, and the
8-dimensional vector derivative $\grad=\grad_x$ is given by
\begin{equation}
\grad = \grad^1 + \grad^2.
\end{equation}
Equation~\eqref{9rwveq} can be derived from a Lorentz-invariant action
integral in 8-dimensional configuration space in which the $1/m_1$ and
$1/m_2$ factors enter via a linear distortion of the vector derivative
$\grad$. We write this as
\begin{equation}
\left( \frac{\nabla^1}{m_1}+\frac{\nabla^2}{m_2} \right) = \ho \left(
\nabla \right),
\end{equation}
where $\ho$ is the linear mapping of vectors to vectors defined by
\begin{equation}
\ho (a) = \frac{1}{m_1} \left( a \dt \gamma^{\mu \, 1} \right)
\gamdm^1 + \frac{1}{m_2} \left( a \dt \gamma^{\mu \, 2} \right)
\gamdm^2. 
\end{equation}
This distortion is of the type used in the gauge theory approach to
gravity developed
in~\cite{DGL-grav,DGL-erice,DGL-grav-bel,DGL-cosm-bel}, and it is
extremely suggestive that mass enters equation~\eqref{9rwveq} via this
route.

Any candidate 2-particle wave equation must be satisfied by factored
states of the form
\begin{equation}
\psi = \phi^1(r^1) \chi^2(s^2) E,
\label{9rwe88}
\end{equation}
where $\phi^1$ and $\chi^2$ are solutions of the separate
single-particle Dirac equations,
\begin{equation}
\grad \phi = - m_1 \phi i \gk, \hs{1} \grad \chi = -m_2 \chi i \gk.
\label{9rwe2}
\end{equation}
To verify that our equation~\eqref{9rwveq} meets this requirement, we
substitute in the direct-product state~\eqref{9rwe88} and use~\eqref{9rwe2} to
obtain 
\begin{equation}
\left( \frac{\nabla^1}{m_1}+\frac{\nabla^2}{m_2} \right) \phi^1 \chi^2
E \left(i\gk^1 + i\gk^2 \right) = - \phi^1 \chi^2 E \left(i\gk^1 +
i\gk^2 \right) \left( i\gk^1 + i\gk^2 \right), 
\end{equation}
where we have used the result that $\nabla^2$ commutes with $\phi^1$.
Now, since $i\gk^1$ and $i\gk^2$ anticommute, we have
\begin{equation}
\left( i\gk^1 + i\gk^2 \right) \left( i\gk^1 + i\gk^2
\right) = -2
\end{equation}
so that 
\begin{equation}
\left( \frac{\nabla^1}{m_1}+\frac{\nabla^2}{m_2} \right) \phi^1 \chi^2 E
\left( i\gk^1 + i\gk^2 \right) = 2 \phi^1 \chi^2 E,
\end{equation}
and~\eqref{9rwveq} is satisfied.  Equation~\eqref{9rwveq} is only satisfied by
direct-product states as a result of the fact that vectors from
separate particle spaces anticommute.  Hence equation~\eqref{9rwveq} does
not have an equivalent expression in terms of the direct-product
matrix formulation, which can only form \textit{commuting} operators
from different spaces.

\subsection{The Pauli Principle}

In quantum theory, indistinguishable particles must obey either
Fermi-Dirac or Bose-Einstein statistics.  For fermions this
requirement results in the Pauli exclusion principle that no two
particles can occupy a state in which their properties are identical.
At the relativistic multiparticle level, the Pauli principle is
usually encoded in the anticommutation of the creation and
annihilation operators of fermionic field theory.  Here we show that
the principle can be successfully encoded in a simple geometrical
manner at the level of the relativistic wavefunction, without
requiring the apparatus of quantum field theory.

We start by introducing the grade-4 multivector
\begin{equation}
I = \Gamma_0 \Gamma_1 \Gamma_2 \Gamma_3,
\end{equation}
where
\begin{equation}
\Gamma_{\mu} = \frac{1}{\sqrt{2}} \left( \gamdm^1 + \gamdm^2 \right). 
\end{equation}
It is a simple matter to verify that $I$ has the properties
\begin{equation}
I^2=-1,
\end{equation}
and
\begin{equation}
I \gamdm^1 I = \gamdm^2, \hs{1} I \gamdm^2 I = \gamdm^1.
\end{equation}
It follows that $I$ functions as a geometrical version of the particle
exchange operator.  In particular, acting on the 8-dimensional
position vector $x=r^1+s^2$ we find that
\begin{equation}
IxI = r^2 + s^1
\end{equation}
where
\begin{equation}
r^2 = \gamdm^2 r^\mu, \hs{1} s^1 = \gamdm^1 s^\mu.
\end{equation}
So $I$ can certainly be used to interchange the coordinates of
particles 1 and 2.  But, if $I$ is to play a fundamental role in our
version of the Pauli principle, we must first confirm that it is
independent of our choice of initial frame.  To see that it is,
suppose that we start with a rotated frame $\{R\gamdm\Rrev\}$ and
define
\begin{equation}
\Gamma_\mu' = \frac{1}{\sqrt{2}} \left( R^1\gamdm^1 \Rrev^1 + R^2
\gamdm^2 \Rrev^2 \right) = R^1 R^2 \Gamma_\mu \Rrev^2 \Rrev^1.
\end{equation}
The new $\Gamma_\mu'$ give rise to the rotated 4-vector
\begin{equation}
I' = R^1 R^2 I \Rrev^2 \Rrev^1.
\label{9Pexc1}
\end{equation}
But, acting on a bivector in particle space 1, we find that
\begin{equation}
I a^1 \wdg b^1 I = - (I a^1I) \wdg (I b^1 I) = - a^2 \wdg b^2,
\end{equation}
and the same is true of an arbitrary even element in either space.
More generally, $I \ldots I$ applied to an even element in one
particle space flips it to the other particle space and changes sign,
while applied to an odd element it just flips the particle space.  It
follows that
\begin{equation}
I \Rrev^2 \Rrev^1 = \Rrev^1 I \Rrev^1 = \Rrev^1 \Rrev^2 I,
\end{equation}
and substituting this into~\eqref{9Pexc1} we find that $I'=I$, so $I$
is indeed independent of the chosen orthonormal frame.

We can now use the 4-vector $I$ to encode the Pauli exchange principle
geometrically.  Let $\psi(x)$ be a wavefunction for two electrons.
Our suggested relativistic generalization of the Pauli principle is
that $\psi(x)$ should be invariant under the operation
\begin{equation}
\psi(x) \mapsto I\psi(IxI)I.
\label{9pausymm1}
\end{equation}
For $n$-particle systems the extension is straightforward: the
wavefunction must be invariant under the interchange enforced by the
$I$'s constructed from each pair of particles.

We must first check that~\eqref{9pausymm1} is an allowed symmetry of the
2-particle Dirac equation.  With $x'$ defined as $IxI$ it is simple to
verify that
\begin{equation}
\grad_{x'} = \grad_r^2 + \grad_s^1 = I \grad I,
\end{equation}
and hence that
\begin{equation}
\grad = I \grad_{x'} I.
\end{equation}
So, assuming that $\psi(x)$ satisfies the 2-particle
equation~\eqref{9rwveq} with equal masses $m$, we find that
\begin{align}
\grad [I\psi(IxI)I] \left( i\gk^1 + i\gk^2 \right)
&= - I\grad_{x'} \psi(x') I \left( i\gk^1 + i\gk^2 \right) \\ \nn 
&= m I \psi(x') \left( i\gk^1 + i\gk^2 \right) I \left( i\gk^1 + i\gk^2
\right) .
\end{align}
But $i\gk^1+i\gk^2$ is odd and symmetric under interchange of its
particle labels.  It follows that
\begin{equation}
I \left( i\gk^1 + i\gk^2 \right) I = \left( i\gk^1 + i\gk^2 \right) 
\end{equation}
and hence that
\begin{equation}
\grad [I\psi(IxI)I] \left( i\gk^1 + i\gk^2 \right) = 2m I\psi(IxI)I.
\end{equation}
So, if $\psi(x)$ is a solution of the 2-particle equal-mass Dirac
equation, then so to is $I\psi(IxI)I$.

Next we must check that the proposed relativistic Pauli principle
deals correctly with well-known elementary cases.  Suppose that two
electrons are in the same spatial state. Then we should expect our
principle to enforce the condition that they are in an antisymmetric
spin state. For example, consider $i\sigma_2^1-i\sigma_2^2$, the spin
singlet state.  We find that
\begin{equation}
I(i\sigma_2^1-i\sigma_2^2)I= -i\sigma_2^2+i\sigma_2^1 ,
\end{equation}
recovering the original state, which is therefore compatible with
our principle. On the other hand 
\begin{equation}
I(i\sigma_2^1+i\sigma_2^2)I = -(i\sigma_2^1+i\sigma_2^2),
\end{equation}
so no part of this state can be added in to the wavefunction, which
again is correct.  In conclusion, given some 2-particle solution
$\psi(x)$, the corresponding state
\begin{equation}
\psi_I = \psi(x)+I\psi(IxI) I
\label{9pp1}
\end{equation}
still satisfies the Dirac equation and is invariant
under $\psi(x)\mapsto I\psi(IxI) I$.  We therefore claim that the
state $\psi_I$ is the correct relativistic generalisation of a state
satisfying the Pauli principle.  In deference to standard quantum
theory, we refer to equation~\eqref{9pp1} as an antisymmetrisation
procedure.

The final issue to address is the Lorentz covariance of the
antisymmetrisation procedure~\eqref{9pp1}.  Suppose that we start with an
arbitrary wavefunction $\psi(x)$ satisfying the 2-particle equal-mass
equation~\eqref{9rwveq}.  If we boost this state via
\begin{equation}
\psi(x) \mapsto \psi'(x) = R^1 R^2 \psi(\Rrev^2\Rrev^1 x R^1 R^2) 
\end{equation}
then $\psi'(x)$ also satisfies the same equation~\eqref{9rwveq}.  The
boosted wavefunction $\psi'(x)$ can be thought of as corresponding to
a different observer in relative motion.  The boosted state $\psi'(x)$
can also be antisymmetrised to yield a solution satisfying our
relativistic Pauli principle.  But, for this procedure to be
covariant, the same state must be obtained if we first antisymmetrise
the original $\psi(x)$, and then boost the result.  Thus we require
that
\begin{equation}
S \psi(\Srev x S) + I S\psi(I \Srev x S I) I = S [ \psi(\Srev x S) + I
\psi (\Srev IxI S)I ] 
\label{9pp4}
\end{equation}
where $S= R^1R^2$.  Equation~\eqref{9pp4} reduces to the requirement
that
\begin{equation}
I S\psi(I \Srev x S I) = S I \psi (\Srev IxI S)
\end{equation}
which is satisfied provided that
\begin{equation}
IS = SI
\end{equation}
or
\begin{equation}
R^1R^2 I = I R^1 R^2.
\end{equation}
But we proved precisely this equation in demonstrating the
frame-invariance of $I$, so our relativistic version of the
Pauli principle is Lorentz invariant.  This is important as, rather
like the inclusion of the quantum correlator, the Pauli procedure
discussed here looks highly non-local in character.

\subsection{8-Dimensional Streamlines and Pauli Exclusion}

For a single Dirac particle, a characteristic feature of the STA
approach is that the probability current is a rotated/dilated version
of the $\gamma_0$ vector, $J=\psi\go\psirev$. This current has zero
divergence and can therefore be used to define streamlines, as
discussed in Section~\ref{S-tunn}.  Here we demonstrate how the same
idea extends to the 2-particle case.  We find that the conserved
current is now formed from $\psi$ acting on the
$\gamma_0^1+\gamma_0^2$ vector, and therefore exists in 8-dimensional
configuration space.  This current can be used to derive streamlines
for two particles in correlated motion.  This approach should
ultimately enable us to gain a better insight into what happens in
experiments of the Bell type, where spin measurements on pairs of
particles are performed over spacelike separations.  We saw in
Section~\ref{S-spinmeas} how the local observables viewpoint leads to
a radical re-interpretation of what happens in a single
spin-measurement, and we can expect an equally radical shift to occur
in the analysis of spin measurements of correlated particles.  As a
preliminary step in this direction, here we construct the current for
two free particles approaching each other head-on.  The streamlines
for this current are evaluated and used to study both the effects of
the Pauli antisymmetrisation and the spin-dependence of the
trajectories.  This work generalises that of Dewdney \textit{et
al.}~\cite{dew84,vig-causal} to the relativistic domain.

\begin{figure}
\centerline{
\hbox{\psfig{figure=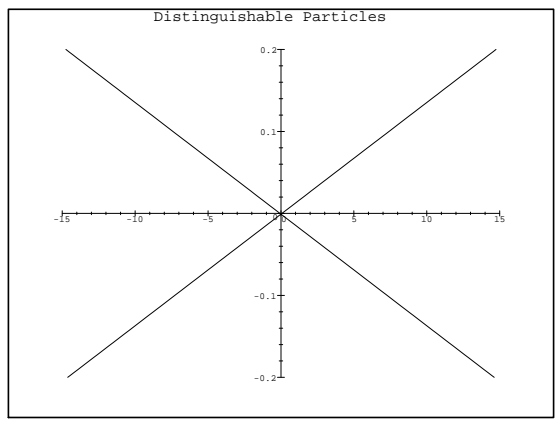,height=8cm,angle=-90}} }
\caption[dummy1]{\em Streamlines generated by the unsymmetrised 2-particle
wavefunction $\psi=\phi^1(r^1)\chi^2(s^2)E$.  Time is shown on the
vertical axis.  $\phi$ and $\chi$ are Gaussian wavepackets moving in
opposite directions and the `collision' is arranged to take place at
$t=0$.  The lack of any antisymmetrisation applied to the wavefunction
means that the streamlines pass straight through each other.}
\label{9Fig-X1}
\end{figure}

We start with the 2-particle Dirac equation~\eqref{9rwveq}, and multiply
on the right by $E$ to ensure the total wavefunction is in the
correlated subspace. Also, since we want to work with the
indistinguishable case, we assume that both masses are $m$. In this
case our basic equation is
\begin{equation}
\grad \psi E \left( i\gk^1 + i\gk^2\right) = 2 m \psi E
\label{9strpp1}
\end{equation}
and, since
\begin{equation}
E \left(i\gk^1 + i\gk^2\right) = J (\go^1+\go^2),
\end{equation}
equation~\eqref{9strpp1} can be written in the equivalent form
\begin{equation}
\grad \psi E \left(\go^1 + \go^2\right) = -2 m \psi J.
\end{equation}
Now, assuming that $\psi$ satisfies $\psi= \psi E$, we obtain
\begin{equation}
\grad \psi (\gamma_0^1+\gamma_0^2) \psirev = - 2 m \psi J \psirev, 
\end{equation}
and adding this equation to its reverse yields
\begin{equation}
\grad \psi (\gamma_0^1+\gamma_0^2) \psirev +  \psi
(\gamma_0^1+\gamma_0^2) \dot{\psirev} \dgrad = 0.
\end{equation}
The scalar part of this equation gives
\begin{equation}
\grad \dt \langle \psi (\gamma_0^1+\gamma_0^2) \tilde{\psi} \rangle_1
= 0, 
\end{equation}
which shows that the current we seek is
\begin{equation}
\clj = \langle \psi (\gamma_0^1+\gamma_0^2) \tilde{\psi} \rangle_1,
\end{equation}
as defined in equation~\eqref{9relJ}.  The vector $\clj$ has components in
both particle-1 and particle-2 spaces, which we write as
\begin{equation}
\clj=\clj_1^1+\clj_2^2.
\end{equation}
The current $\clj$ is conserved in {\em eight}-dimensional space, so
its streamlines never cross there.  The streamlines of the individual
particles, however, are obtained by integrating $\clj_1$ and $\clj_2$
in ordinary 4-d space, and these can of course cross.  An example of
this is illustrated in Figure~\ref{9Fig-X1}, which shows the
streamlines corresponding to distinguishable particles in two Gaussian
wavepackets approaching each other head-on.  The wavefunction used to
produce this figure is just
\begin{equation}
\psi = \phi^1(r^1) \chi^2(s^2) E, 
\label{multi-eqn14}
\end{equation}
with $\phi$ and $\chi$ being Gaussian wavepackets, moving in opposite
directions. Since the distinguishable case is assumed, no Pauli
antisymmetrisation is used.  The individual currents for each particle
are given by
\begin{align}
\clj_1(r,s) &= \phi(r) \go \phirev(r) \, \la\chi(s)\tilde{\chi}(s)\ra, 
\nn \\
\clj_2(r,s) &= \chi(s) \go \tilde{\chi}(s) \, \la \phi(r) \phirev(r) \ra
\label{multi-eqn15}
\end{align}
and, as can be seen, the streamlines (and the wavepackets) simply pass
straight through each other.

An interesting feature emerges in the individual currents
in~\eqref{multi-eqn15}.  One of the main problems with single-particle
Dirac theory is that the current is always positive-definite so, if we
wish to interpret it as a charge current, it fails to represent
antiparticles correctly.  The switch of sign of the current necessary
to represent positrons is put into conventional theory essentially `by
hand', via the anticommutation and normal ordering rules of fermionic
field theory. In equation~\eqref{multi-eqn15}, however, the norm
$\la\chi\tilde{\chi}\ra$ of the second state multiplies the current
for the first, and vice versa.  Since $\la\chi\tilde{\chi}\ra$ can be
negative, it is possible to obtain currents which flow backwards in
time.  This suggests that the required switch of signs can be
accomplished whilst remaining wholly within a wavefunction-based
approach.  An apparent problem is that, if only one particle has a
negative norm state --- say for example $\chi$ has
$\langle\chi\tilde{\chi}\rangle < 0$ --- then it is the $\phi$ current
which is reversed, and not the $\chi$ current.  However, it is easy to
see that this objection is not relevant to indistinguishable
particles, and it is to these we now turn.

\begin{figure}[t!]
\begin{center}
\psfig{figure=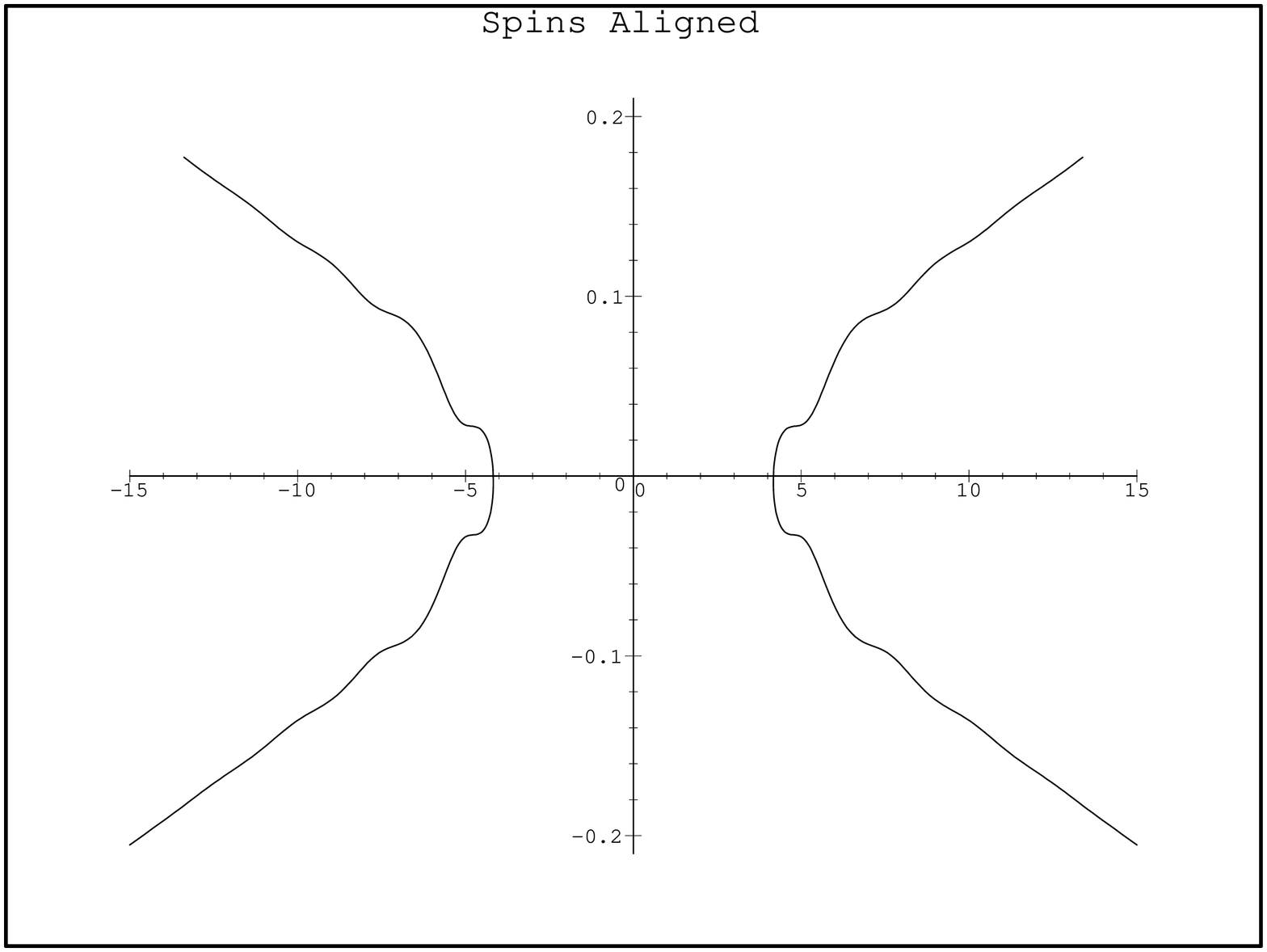,height=8cm,angle=-90} \\
\psfig{figure=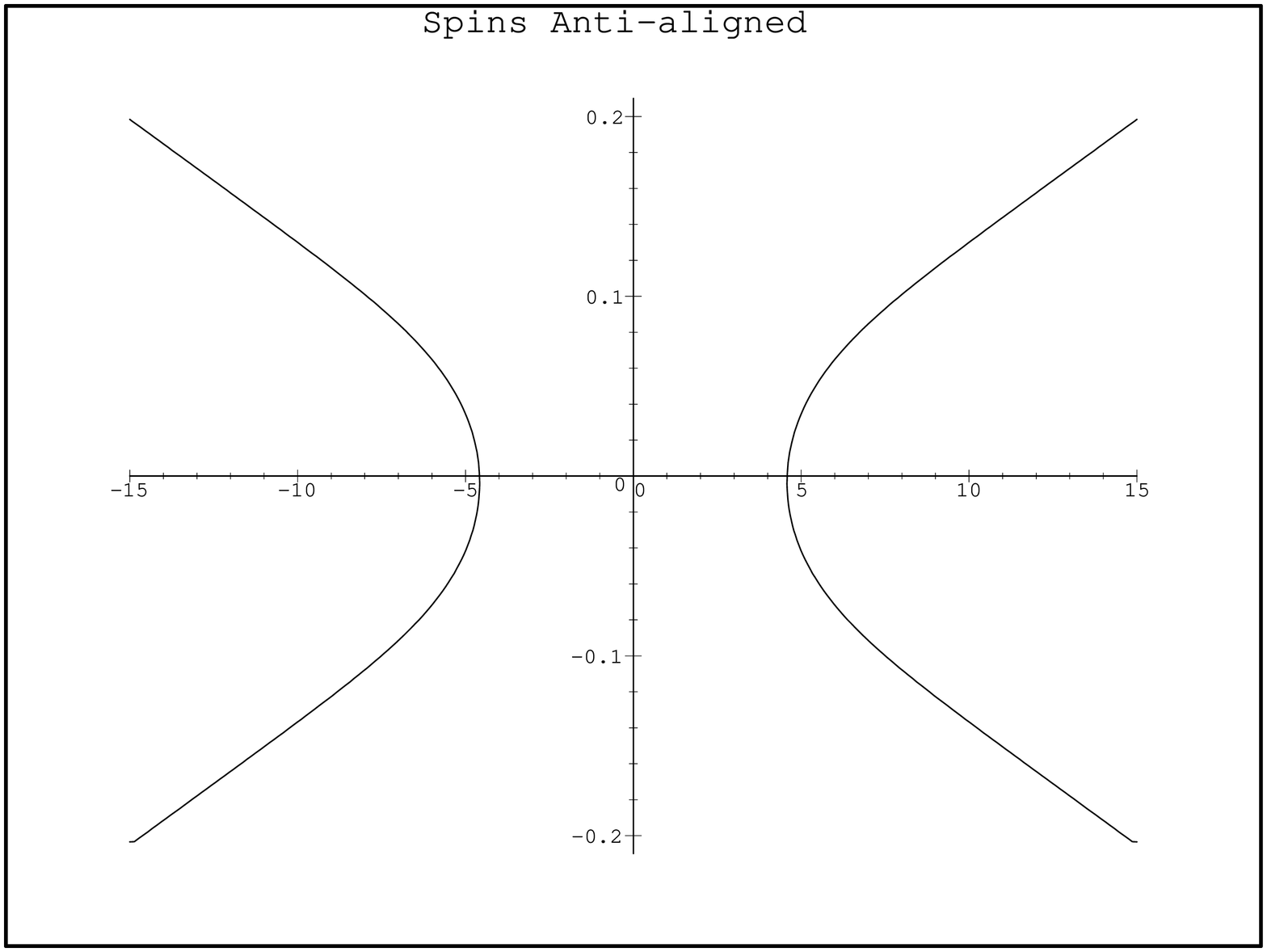,height=8cm,angle=-90} 
\end{center}
\caption[dummy1]{\em Streamlines generated by the antisymmetrised 2-particle
wavefunction $\psi=[\phi^1(r^1)\chi^2(s^2)-\chi^1(r^2)\phi^2(s^1)]E$.
The individual wavepackets pass through each other, but the
streamlines from separate particles do not cross.  The top figure has both
particles with spins aligned in the $+z$-direction, and the bottom figure
shows particles with opposite spins, with $\phi$ in the $+z$
direction, and $\chi$ in the $-z$ direction.  Both wavepackets have
energy 527KeV and a spatial spread of $\sim$ 20pm.  The spatial units
are $10^{-12}$m and the units of time are $10^{-18}$s.  The effects of
the antisymmetrisation are only important where there is significant
wavepacket overlap.}
\label{9Fig-X2}
\end{figure}

We now apply the Pauli symmetrization procedure of the previous
subsection to the wavefunction of equation~\eqref{multi-eqn14}, so as 
to obtain a wavefunction applicable to indistinguishable
particles.  This yields
\begin{equation}
\psi = \left(\phi^1(r^1) \chi^2(s^2) - \chi^1(r^2) \phi^2(s^1) \right)
E, 
\end{equation}
from which we form $\clj_1$ and $\clj_2$, as before.  We must next
decide which spin states to use for the two particles.  We first take
both particles to have their spin vectors pointing in the positive
$z$-direction, with all motion in the $\pm z$-direction.  The
resulting streamlines are shown in Figure~\ref{9Fig-X2} (top).  The
streamlines now `repel' one another, rather than being able to pass
straight through.  The corrugated appearance of the lines near the
origin is the result of the streamlines having to pass through a
region of highly-oscillatory destructive interference, since the
probability of both particles occupying the same position (the origin)
with the same spin state is zero.  If instead the particles are put in
\textit{different} spin states then the streamlines shown at the
bottom of Figure~\ref{9Fig-X2} result. In this case there is no destructive
interference near the origin, and the streamlines are smooth there.
However, they still repel!  The explanation for this lies in the
symmetry properties of the 2-particle current.  Given that the
wavefunction $\psi$ has been antisymmetrised according to our version
of the Pauli principle, then it is straightforward to show that
\begin{equation}
I \clj(IxI) I = \clj(x).
\end{equation}
It follows that at the same spacetime position, encoded by $IxI=x$ in
the 2-particle algebra, the two currents $\clj_1$ and $\clj_2$ are
equal.  Hence, if two streamlines ever met, they could never separate
again.  For the simulations presented here, it follows from the
symmetry of the set-up that the spatial currents at the origin are
both zero and so, as the particles approach the origin, they are
forced to slow up. The delay means that they are then swept back in
the direction they have just come from by the wavepacket travelling
through from the other side.  We therefore see that `repulsion' as
measured by streamlines has its origin in indistinguishability, and
that the spin of the states exerts only a marginal effect.

\section{Further Applications}
\label{S-further}

In this section we briefly review two further applications of
spacetime algebra to important areas of electron physics.  The first
of these is classical and semiclassical mechanics.  As well as
simplifying many calculations in quantum mechanics, spacetime algebra
is well suited to handling problems in classical mechanics where
electrons are treated as point charges following a single trajectory.
In recent years there has been considerable interest in finding
modifications to the simple classical equations to include the effects
of spin, without losing the idea of a definite
trajectory~\cite{hes90,bar84}.  One of the aims behind this work is to
find a suitable classical model which can be quantised via the
path-integral route~\cite{bar-dur89}.  One of the more promising
candidates is discussed here, and we outline some improvements that
could repair some immediate defects.

The second application discussed here is to Grassmann algebra and the
associated `calculus' introduced by Berezin~\cite{ber-meth}. Grassmann
quantities are employed widely in quantum field theory, and the
Berezin calculus plays a crucial role in the path-integral
quantisation of fermionic systems.  Here we outline how many of the
calculations can be performed within geometric algebra, and draw
attention to some work in the literature.  We do not attempt a more
detailed analysis of path-integral quantisation here.

\subsection{Classical and Semiclassical Mechanics}

The Lorentz force law for a point-particle with velocity $v$, mass $m$
and charge $q$ is
\begin{equation}
\vdot = \frac{q}{m} F \dt v
\label{10Lfl}
\end{equation}
where $\vdot$ denotes differentiation with respect to the affine
parameter and $F$ is the external electromagnetic field bivector.  
Any future-pointing unit timelike vector can be written in terms of a
rotor acting on a fixed vector $\go$,
\begin{equation}
v(\tau) = R(\tau) \go \Rrev(\tau)
\end{equation}
from which we find
\begin{equation}
\vdot = (2\Rdot \Rrev) \dt v.
\end{equation}
The quantity $\Rdot\Rrev$ is a bivector, so we can recover
equation~\eqref{10Lfl} by setting
\begin{align}
2\Rdot \Rrev &= \frac{q}{m} F \\
\implies \hs{0.5} \Rdot &=  \frac{q}{2m} F R.
\label{10rotfrm}
\end{align}
This is not the only equation for $R$ that is consistent
with~\eqref{10Lfl}, since any bivector that commutes with $v$ could be
added to $F$.  However, \eqref{10rotfrm} is without doubt the simplest
equation available.

It turns out that equation~\eqref{10rotfrm} is often easier to analyse
than~\eqref{10Lfl}, as was first shown by Hestenes~\cite{hes74}.
Furthermore, we can extend this approach to include a classical notion
of spin.  Let us suppose that, as well as describing the tangent vector
$v$, the rotor $R$ determines how a frame of vectors is transported
along the curve.  We can then define the spin vector as the unit
spatial vector 
\begin{equation}
s = R \gk \Rrev,
\end{equation}
which matches the definition given for the quantum observable.  If we
now assume that equation~\eqref{10rotfrm} is valid, we find that the
spin-vector satisfies the equation
\begin{equation}
\sdot = \frac{q}{m} F \dt s,
\end{equation}
which gives the correct precession equation for a particle of
gyromagnetic ratio 2~\cite{hes74}.  It follows that $g=2$ can be
viewed as the natural value from the viewpoint of the relativistic
classical mechanics of a rotating frame --- a striking fact that
deserves to be more widely known.  We can use the same approach to
analyse the motion of a particle with a $g$-factor other than 2 by
replacing~\eqref{10rotfrm} with
\begin{equation}
\Rdot =  \frac{q}{2m} [F R + (g/2-1) R \bB],
\end{equation}
which reproduces the Bargmann-Michel-Telegdi equation employed in the
analysis of spin-precession measurements~\cite{hes-geom82}.

A remarkable aspect of the Dirac theory is that the current
$\psi\go\psirev$ and the momentum (which is defined in terms of the
momentum operator) are not necessarily colinear.  This suggests that a
more realistic classical model for an electron should employ an
independent quantity for the momentum which is not necessarily related
to the tangent vector to the spacetime trajectory.  Such a model was
proposed by Barut~\& Zanghi~\cite{bar84}, who did not employ the STA,
and was analysed further in~\cite{DGL93-lft} (see also~\cite{rod93}).
Written in the STA, the action proposed by Barut~\& Zanghi takes the
form
\begin{equation}
S = \int \! d\lam \, \la \psidot \isk \psirev + p(\xdot -
\psi\go\psirev) + q A(x) \psi\go\psirev \ra 
\label{10BZ}
\end{equation}
where the dynamical variables are $x(\lam)$, $p(\lam)$ and
$\psi(\lam)$.  Variation with respect to these variables yields the
equations~\cite{DGL93-lft}
\begin{align}
\xdot &= \psi\go\psirev \label{10BZ1}\\
\dot{P} &= q F\dt \xdot \\
\psidot\isk &= P \psi \go \label{10BZ3}
\end{align}
where
\begin{equation}
P = p-qA.
\end{equation}
These constitute a set of first-order equations so, with $x$, $p$ and
$\psi$ given for some initial value of $\lam$, the future evolution is
uniquely determined.  Equations \eqref{10BZ1}--\eqref{10BZ3} contain a number
of unsatisfactory features.  One does not expect to see $P$ entering
the Lorentz force law~\eqref{10BZ3}, but rather the dynamical variable
$p$.  This problem is simply addressed by replacing~\eqref{10BZ} with

\begin{equation}
S_1 = \int \! d\lam \, \la \psidot \isk \psirev + p(\xdot -
\psi\go\psirev) + q \xdot A(x) \ra 
\label{10BZalt1}
\end{equation}
so that the $\dot{p}$ and $\psidot$ equations become
\begin{align}
\dot{p} &= q F\dt \xdot \\
\psidot\isk &= p \psi\go.
\end{align}
The quantity $p\dt\xdot$ is a constant of the motion, and can be viewed
as defining the mass.

A more serious problem remains, however.  If we form the spin
bivector $S=\psi\isk\psirev$ we find that
\begin{equation}
\Sdot = 2 p \wdg \xdot
\end{equation}
so, if $p$ and $\xdot$ are initially colinear, the spin bivector does
not precess, even in the presence of an external
$\bB$-field~\cite{gul-steps}.  To solve this problem an extra term
must be introduced into the action. The simplest modification is
\begin{equation}
S_2 = \int \! d\lam \, \la \psidot \isk \psirev + p(\xdot -
\psi\go\psirev) + q \xdot A(x) -\frac{q}{2m} F \psi \isk \psirev \ra  
\label{10BZalt2}
\end{equation}
which now yields the equations
\begin{align}
\xdot &= \psi\go\psirev \\
\dot{p} &= q F\dt \xdot - \frac{q}{2m} \grad F(x) \dt S \\
\psidot\isk &= p \psi\go - \frac{q}{2m}F \psi.
\end{align}
The problem with this system of equations is that $m$ has to be
introduced explicitly, and there is nothing to identify this quantity
with $p\dt\xdot$.  If we assume that $p=m\xdot$ we can recover the
pair of equations 
\begin{align}
\dot{S} &= \frac{q}{m} F \crs S \\
\vdot &= \frac{q}{m}  F\dt v - \frac{q}{2m^2} \grad F(x) \dt S ,
\end{align}
which were studied in~\cite{hol91}.

Whilst a satisfactory semiclassical mechanics for an electron still
eludes us, it should be clear that the STA is a very useful tool in
constructing and analysing candidate models.

\subsection{Grassmann Algebra}

Grassmann algebras play an essential role in many areas of modern
quantum theory.  However, nearly all calculations with Grassmann
algebra can be performed more efficiently with geometric algebra.  A
set of quantities $\{\zeta_i\}$ form a Grassmann algebra if their
product is totally antisymmetric
\begin{equation}
\zeta_i \zeta_j = - \zeta_j \zeta_i.
\end{equation}
Examples include fermion creation operators, the fermionic generators
of a supersymmetry algebra, and ghost fields in the path integral
quantisation of non-abelian gauge theories.  Any expression involving
the Grassmann variables $\{\zeta_i\}$ has a geometric algebra
equivalent in which the $\{\zeta_i\}$ are replaced by a frame of
independent vectors $\{e_i\}$ and the Grassmann product is replaced by
the outer (wedge) product~\cite{DGL93-gras,DGL-polmvl}.  For example,
we can make the replacement
\begin{equation}
\zeta_i \zeta_j \trans e_i \wdg e_j.
\end{equation}
This translation on its own clearly does not achieve a great deal, but
the geometric algebra form becomes more powerful when we consider the
`calculus' defined by Berezin~\cite{ber-meth}.  This calculus is
defined by the rules
\begin{align}
\deriv{\zeta_{j}}{\zeta_{i}} &= \delta_{ij} , \\
\zeta_{j} \backderiv{}{\zeta_{i}} &= \delta_{ij},
\end{align}
together with the `graded Leibniz' rule',
\begin{equation}
\deriv{}{\zeta_{i}} (f_{1} f_{2}) = \deriv{f_{1}}{\zeta_{i}} f_{2} +
(-1)^{[f_{1}]} f_{1} \deriv{f_{2}}{\zeta_{i}},
\end{equation}
where $[f_{1}]$ is the parity (even/odd) of $f_{1}$.  In geometric
algebra, the operation of the Grassmann derivatives can be replaced by
inner products of the reciprocal frame vectors
\begin{equation}
\deriv{}{\zeta_{i}} (  \trans  e^{i} \dt (
\end{equation} 
so that
\begin{equation}
\deriv{\zeta_{j}}{\zeta_{i}} \trans e^{i} \dt e_{j} = \delta^{i}_{j}.
\end{equation}
Some consequences of this translation procedure were discussed
in~\cite{DGL93-gras}, where it was shown that the geometric product
made available by the geometric algebra formulation simplifies many
computations.  Applications discussed in~\cite{DGL93-gras} included
`Grauss' integrals, pseudoclassical mechanics, path integrals and
Grass\-mann--Fourier transforms.  It was also shown that super-Lie
algebras have a very simple representation within geometric algebra.
There seems little doubt that the systematic replacement of Grassmann
variables with geometric multivectors would considerably enhance our
understanding of quantum field theory.

\section{Conclusions}

There is a growing realisation that geometric algebra provides a
unified and powerful tool for the study of many areas of mathematics,
physics and engineering.  The underlying algebraic structure (Clifford
algebra) appears in many key areas of physics and
geometry~\cite{law-spin}, and the geometric techniques are finding
increasing application in areas as diverse as gravitation
theory~\cite{DGL-erice} and robotics~\cite{hes94I,hes94II}.  The only
impediment to the wider adoption of geometric algebra appears to be
physicists' understandable reluctance to adopt new techniques.  We
hope that the applications discussed in this paper make a convincing
case for the use of geometric algebra, and in particular the STA, in
electron physics.  Unfortunately, in concentrating on a single area of
physics, the unifying potential of geometric algebra does not
necessarily come across.  However, a brief look at other applications
should convince one of the wider utility of many of the techniques
developed here.

Further work in this field will centre on the multiparticle STA.  At
various points we have discussed using the multiparticle STA to
analyse the non-locality revealed by EPR-type experiments.  This is
just one of many potential applications of the approach outlined here.
Others include following the streamlines for two particles through a
scattering event, or using the 3-particle algebra to model pair
creation.  It will also be of considerable interest to develop
simplified techniques for handling more complicated many-body
problems.  Behind these goals lies the desire to construct an
alternative to the current technique of fermionic field quantisation.
The canonical anticommutation relations imposed there remain
mysterious, despite forty years of discussion of the spin-statistics
theorem.  Elsewhere, there is still a clear need to develop the
wavepacket approach to tunnelling.  This is true not only of fermions,
but also of photons, on which most of the present experiments are
performed.

Looking further afield, the approach to the Dirac equation described
in Section~\ref{S-opers} extends simply to the case of a gravitational
background~\cite{DGL-grav}.  The wavepacket and multiparticle
techniques developed here are essentially all that is required to
address issues such as superradiance and pair creation by black holes.
Closer to home, the STA is a powerful tool for classical relativistic
physics.  We dealt briefly with the construction of classical models
for the electron in Section~\ref{S-further}.  Elsewhere, similar
techniques have been applied to the study of radiation reaction and
the Lorentz-Dirac equation~\cite{gul-steps}.  The range of
applicability of geometric algebra is truly vast.  We believe that all
physicists should be exposed to its benefits and insights.

\appendix

\section{The Spherical Monogenic Functions}
\label{App-monos}

We begin by assuming that the spherical monogenic is an eigenstate of the
$\bx\wdg\bgrad$ and $J_3$ operators, where all operators follow the
conventions of Section~\ref{4S-angmonos}.  We label this state as
$\psi(l,\mu)$, so
\begin{equation}
-\bx\wdg\bgrad \psi(l,\mu) = l \psi(l,\mu), \hs{1} J_3 \psi(l,\mu) =
\mu \psi(l,\mu).
\end{equation}
The $J_i$ operators satisfy
\begin{align}
J_iJ_i &= - [(i\sigi)\dt(\bx\wdg\vec{\bgrad}) - \half i\sigi]
[(i\sigi)\dt(\bx\wdg\bgrad) - \half i\sigi]  \nn \\
&= 3/4 - \bx\wdg\bgrad + \la \bx\wdg\vec{\bgrad} \bx\wdg\bgrad \ra 
\end{align}
where the $\vec{\bgrad}$ indicates that the derivative acts on
everything to its right.  Since
\begin{equation}
\la \bx\wdg\vec{\bgrad} \bx\wdg\bgrad \ra \psi = \bx\wdg\bgrad
(\bx\wdg\bgrad \psi) - \bx\wdg\bgrad \psi
\end{equation}
we find that
\begin{align}
J_iJ_i \psi(l,\mu) &= (3/4 +2l +l^2) \psi(l,\mu) \nn \\
&= (l+1/2)(l+3/2) \psi(l,\mu).
\end{align}
With the ladder operators $J_+$ and $J_-$ defined by
\begin{align}
& J_+ = J_1 + jJ_2  & \\
& J_- = J_1 - jJ_2, &
\end{align}
it is a simple matter to prove the following results:
\begin{equation}
\begin{array}{ccc}
[J_+, J_-] = 2 J_3 & \hs{0.4} & J_iJ_i = J_-J_+ +J_3 + {J_3}^2 \\
{[J_\pm, J_3]} = \mp J_\pm & \hs{0.4} & J_iJ_i = J_+J_- -J_3 + {J_3}^2.
\end{array}
\label{A1-1}
\end{equation}
The raising operator $J_+$ increases the eigenvalue of $J_3$ by an
integer.  But, for fixed $l$, $\mu$ must ultimately attain some
maximum value.  Denoting this value as $\mu_+$, we must reach a state
for which
\begin{equation}
J_+ \psi(l,\mu_+) = 0.
\end{equation}
Acting on this state with $J_iJ_i$ and using one of the results
in~\eqref{A1-1} we find that
\begin{equation}
(l+1/2)(l+3/2) = \mu_+(\mu_+ +1)
\end{equation}
and, as $l$ is positive and $\mu_+$ represents an upper bound, it
follows that
\begin{equation}
\mu_+ = l +1/2.
\end{equation}
There must similarly be a lowest eigenvalue of $J_3$ and a
corresponding state with
\begin{equation}
J_- \psi(l,\mu_-) = 0.
\end{equation}
In this case we find that
\begin{equation}
(l+1/2)(l+3/2) = \mu_-(\mu_- -1) 
\end{equation}
\begin{equation}
\implies \mu_- = -(l+1/2).
\end{equation}
The spectrum of eigenvalues of $J_3$ therefore ranges from $(l+1/2)$ to
$-(l+1/2)$, a total of $2(l+1)$ states.  Since the $J_3$ eigenvalues
are always of the form (integer $+1/2$), it is simpler to label
the spherical monogenics with a pair of integers.  We therefore write
the spherical monogenics as $\psi_l^m$, where
\begin{align}
-\bx\wdg\bgrad \psi_l^m = l \psi_l^m  \qquad & l \geq 0 \\
J_3 \psi_l^m = (m+\half) \psi_l^m  \qquad & -1-l \leq m \leq l.
\end{align}

To find an explicit form for the $\psi^m_l$ we first construct the
highest-$m$ case.  This satisfies 
\begin{equation}
J_+ \psi_l^l = 0
\end{equation}
and it is not hard to see that this equation is solved by
\begin{equation}
\psi_l^l \propto \sin^l\!\theta\, \et{-l\phi\isk}.
\end{equation}
Introducing a convenient factor, we write
\begin{equation}
\psi_l^l = (2l+1) P^l_l(\cos\!\theta) \, \et{l\phi\isk}.
\end{equation}
Our convention for the associated Legendre polynomials follows
Gradshteyn~\& Ryzhik~\cite{gr-tables}, so
\begin{equation}
P_l^m(x) = \frac{(-1)^m}{2^l l!} (1-x^2)^{m/2}
\frac{d^{l+m}}{dx^{l+m}} (x^2-1)^l
\end{equation}
and we have the following recursion relations:
\begin{equation}
(1-x^2) \frac{dP_l^m(x)}{dx} + mx P_l^m(x) = -(1-x^2)^{1/2}
P_l^{m+1}(x) 
\end{equation}
\begin{equation}
(1-x^2) \frac{dP_l^m(x)}{dx} - mx P_l^m(x) = (1-x^2)^{1/2}
(l+m)(l-m+1) P_l^{m-1}(x) .
\end{equation}
The lowering operator $J_-$ has the following effect on $\psi$:
\begin{equation}
J_- \psi = [-\partial_\theta \psi + \cot\!\theta\, \partial_\phi \psi
\isk] \et{-\phi\isk} - i\sj \, \half(\psi + \sk\psi\sk).
\end{equation}
The latter term just projects out the $\{1,\isk\}$ terms and
multiplies them by $-i\sj$.  This is the analog of the lowering
matrix in the standard formalism.  The derivatives acting on
$\psi_l^l$ form
\begin{align}
\lefteqn{[-\partial_\theta \psi_l^l + \cot\!\theta\, \partial_\phi
\psi_l^l  \isk] \et{-\phi\isk}} \hs{2} \nn \\  
&= (2l+1) [-\partial_\theta P_l^l(\cos\!\theta) - l \cot\!\theta\,
P_l^l(\cos\!\theta) ]\et{(l-1)\phi\isk}  \nn \\
&= (2l+1) 2l P_l^{l-1}(\cos\!\theta) \et{(l-1)\phi\isk},
\end{align}
and, if we use the result that
\begin{equation}
\sig_\phi = \sj \et{\phi i\sk},
\end{equation}
we find that
\begin{equation}
\psi_l^{l-1} \propto [2l P_l^{l-1}(\cos\!\theta) -
P_l^l(\cos\!\theta) i\sig_\phi ] \et{(l-1)\phi\isk}.
\end{equation}
Proceeding in this manner, we are led to the following formula for the
spherical monogenics:
\begin{equation}
\psi_l^m = [ (l+m+1)P_l^m(\cos\!\theta) -
P_l^{m+1}(\cos\!\theta)i\sig_\phi ] \et{m\phi\isk},
\end{equation}
in which $l$ is a positive integer or zero, $m$ ranges from $-(l+1)$
to $l$ and the $P_l^m$ are taken to be zero if $|m|>l$.  The positive-
and negative-$m$ states can be related using the result that
\begin{equation}
P_l^{-m}(x) = (-1)^m \frac{(l-m)!}{(l+m)!} P_l^m(x),
\end{equation}
from which it can be shown that
\begin{equation}
\psi_l^m (-i\sj) = (-1)^m \frac{(l+m+1)!}{(l-m)!} \psi_l^{-(m+1)}.
\end{equation}
The spherical monogenics presented here are unnormalised.  Normalisation
fact\-ors are not hard to compute, and we find that
\begin{equation}
\int_0^\pi \! d\theta \int_0^{2\pi} \! d\phi \, \sin\!\theta\, \psi_l^m
{\psi_l^m}^\dagger = 4 \pi \frac{(l+m+1)!}{(l-m)!}.
\end{equation}

\end{document}